\documentclass{appolb}
\usepackage{graphicx}
\usepackage{amsmath}
\usepackage{amssymb}
\usepackage{enumitem}
\usepackage{dsfont}

\usepackage[usenames, dvipsnames]{color}

\usepackage[normalem]{ulem}
\newcommand{\be}{\begin{equation}}
\newcommand{\ee}{\end{equation}}
\newcommand{\ba}{\begin{eqnarray}} 
\newcommand{\ea}{\end{eqnarray}} 
\newcommand{\ban}{\begin{eqnarray*}}
\newcommand{\ean}{\end{eqnarray*}}
\newcommand{\nn}{\nonumber}

\newcommand{\bea}{\begin{eqnarray}}
\newcommand{\eea}{\end{eqnarray}}
\newcommand{\ww}{{\rm w}}

\begin{document}
\eqsec  

\title{The earliest phase of relativistic heavy-ion collisions\footnote{Presented at the 63rd Cracow School of Theoretical Physics {\it Nuclear Matter at Extreme Densities and High Temperatures}, Zakopane, Poland, September 17-23, 2023.}}

\author{Margaret E. Carrington
\address{Department of Physics, Brandon University, Brandon, Manitoba, Canada,
\\ Winnipeg Institute for Theoretical Physics, Winnipeg, Manitoba, Canada}
\\[3mm]
{Stanis\l aw Mr\' owczy\' nski
\address{National Centre for Nuclear Research, Warsaw, Poland,
\\ Institute of Physics, Jan Kochanowski University, Kielce, Poland }}
}

\maketitle

\begin{abstract}
According to the Color Glass Condensate approach to relativistic heavy-ion collisions, the earliest phase of the collision is a glasma which is made of highly populated gluon fields that can be treated classically. Using a proper time expansion we study analytically various properties of the glasma. In particular, we compute the glasma energy-momentum tensor which allows us to obtain the energy density, longitudinal and transverse pressure, collective flow, and angular momentum. We also study the role of the glasma in jet quenching by computing collisional energy loss and transverse momentum broadening. 
\end{abstract}

\newpage
\tableofcontents

\newpage
\section{Introduction}

The earliest phase of relativistic heavy-ion collisions is the least understood. The phenomena occurring during this phase are largely `forgotten' due to the subsequent temporal evolution of the system, and consequently experiments provide very limited information about its properties. What happens during this phase is usually paramterized using a few of its primary characteristics, such as an energy density profile, and used only to provide initial conditions for the later, much better understood, hydrodynamic phase. The earliest phase, however, is of special interest for several reasons. At very early times the  system is strongly anisotropic, far from thermodynamic equilibrium, and the energy density reaches its maximal values. The processes which take place in this phase can significantly affect the subsequent evolution of the system and its final-state characteristics. 

Several different strategies have been used to understand and describe the earliest phase of relativistic heavy-ion collisions. The framework of the Color Glass Condensate effective theory \cite{McLerran:1993ni, McLerran:1993ka, McLerran:1994vd}, see also the review \cite{Gelis:2012ri}, is commonly applied. The theory is based on a separation of scales between hard valence partons and soft gluons. The system that exists at very early times is called a `glasma'. It consists of large occupation number, coherent chromodynamic fields that are essentially classical. The dynamics of the glasma fields is determined by the classical Yang-Mills (YM) equations with sources provided by the valence partons. To calculate observables one performs averaging over a Gaussian distribution of colour charges within each colliding nucleus. The original McLerran-Venugopalan (MV) model assumed a source charge density that was  homogeneous in the transverse plane and infinitely Lorentz contracted. A description of real finite-size nuclei requires including the effects of varying nuclear density in the transverse plane. It is also necessary to take into account a finite width of the sources across the light-cone \cite{JalilianMarian:1996xn, Kovchegov:1999yj}. 

Properties of the glasma have been studied for over two decades using more and more advanced numerical simulations, see Refs.~\cite{Sun:2019fud,Boguslavski:2021buh,Ipp:2021lwz,Avramescu:2023qvv,Matsuda:2023gle} as examples of recent works in this direction. There are also analytic approaches, but they are usually very limited in their applicability. We use a method designed to study the earliest phase of relativistic heavy-ion collisions that involves an expansion of the YM equations in powers of the proper time $\tau$, which is treated as a small parameter\footnote{The dimensionless small parameter is the proper time times the saturation scale.}. This method, which is sometimes called a `near field expansion', was proposed in \cite{Fries:2005yc} and further developed in \cite{Fukushima:2007yk,Fujii:2008km,Chen:2015wia,Fries:2017ina,Li:2017iat}. Results are valid only for small values of $\tau$ but they are analytic and free of artifacts of numerical computation like those caused by taking a continuous limit in the case of lattice calculations. 

Some important characteristics of glasma can be calculated from the energy-momentum tensor expressed through the classical chromodynamic fields and averaged over the colour configurations of the colliding nuclei. One can obtain in this way the glasma energy density, and the transverse and longitudinal pressures, which give us insight into the process of the system's equilibration. The energy and momentum fluxes, which are also calculated from the energy-momentum tensor, tell us how the system expands and how collective flow develops. The angular momentum carried by the glasma can also be found from the energy-momentum tensor.

Partons with high transverse momenta and heavy quarks are produced in relativistic heavy-ion collisions only through hard interactions with large momentum transfer at the earliest phase of the collision. The production mechanisms of these hard probes are thus described by perturbative QCD. They propagate through the evolving medium probing QCD matter at different phases throughout the whole system's evolution. During this propagation heavy quarks and high-$p_T$ partons lose a substantial fraction of their initial energy which causes a significant suppression of final high-$p_T$ hadrons, commonly known as jet quenching. The suppression of high-$p_T$ hadrons is treated as a signal of the formation of quark-gluon plasma, because only a deconfined state of matter could produce such significant braking of hard partons.  

The energy loss of a probe is caused by collisions and/or radiation and depends on the medium content and its dynamics. The medium produced in heavy-ion collisions quickly approaches equilibrium, within about 1~fm, and the long lasting equilibrium phase, with lifetime approximately 10 fm, is expected to be mostly responsible for jet quenching. However, the energy density of the transient non-equilibrium phase is significantly higher than that of the equilibrium phase and it is therefore important to consider the possibility that the non-equilibrium phase plays an important role. Two distinct pre-equilibrium phases can be identified: one just before the thermal quark-gluon plasma is formed, when the medium consists of quasi-particles with non-equilibrium distributions of momenta, and the strict earliest phase, called glasma, when the medium is described in terms of strong classical gluon fields rather than partons. We are interested in the glasma phase. 

In a series of papers \cite{Carrington:2020ssh,Carrington:2021qvi,Carrington:2020sww,Carrington:2022bnv,Carrington:2021dvw,Carrington:2023nty}, we have used the small $\tau$ expansion extensively to describe the earliest phase of relativistic heavy-ion collisions. The papers \cite{Carrington:2020ssh,Carrington:2021qvi,Carrington:2023nty} are devoted to various characteristics of glasma which are calculated from the energy-momentum tensor. These characteristics include: the energy density, longitudinal and transverse pressure, collective flow, and angular momentum. In the articles \cite{Carrington:2020sww,Carrington:2022bnv,Carrington:2021dvw,Carrington:2023nty} we studied the role of the glasma in jet quenching, computing the collisional energy loss and the transverse momentum broadening. The idea behind this series of lectures is to present the whole project in a systematic and coherent way. We also update some of our results and add some discussion on the regime of validity of our calculation. 

In Sec.~\ref{sec-formalism} we present the formalism we use to describe the glasma. We discuss the equations of motion, the boundary conditions, and the proper time expansion, and we explain how the energy-momentum tensor is calculated. Correlators of pre-collision potentials of incoming nuclei, which are the building blocks of our computational method, are discussed in detail. We pay special attention to their regularization in both the infrared and ultraviolet domains. We also discuss how the correlators are modified when the effect of the finite size of the colliding nuclei is taken into account.

Sec.~\ref{sec-numerical} is devoted to glasma characteristics obtained from the energy-momentum tensor. We present  our numerical results for the energy density, pressures and energy fluxes. We discuss the evolution of glasma anisotropy and show that the time dependence of the radial and azimuthally asymmetric flow of the glasma resembles hydrodynamic behaviour. We compute the glasma angular momentum and demonstrate that only a small fraction of the angular momentum of the incoming nuclei is transferred to glasma, and consequently the system does not rotate significantly. 

In Sec.~\ref{sec-jet-quenching} the role of glasma in  jet quenching is studied. We first discuss the Fokker-Planck equation, which is our main theoretical tool, and a physical picture of parton transport across the glasma. Our numerical results on collisional energy loss $dE/dx$ and momentum broadening $\hat{q}$ are shown, and their dependence on the adopted regularization procedure is analyzed. The gauge dependence of our results is also briefly discussed. We finally show that the glasma has a sizable impact on jet quenching, comparable to that of the long lasting equilibrium phase. 

Our conclusions are collected in Sec.~\ref{sec-conclusions}.

Throughout the article we use the natural system of units with $c = \hbar = k_B =1$. We neglect henceforth the prefix `chromo' when referring to chromoelectric or chromomagnetic fields. Since we study QCD only, this should not cause confusion. Notational details and a collection of some useful formulas can be found in Appendix~\ref{AppendixA}. 

\section{Formalism}
\label{sec-formalism}

In this section we present the formalism of the CGC effective theory that we use and our computational method. 

We consider a collision of two heavy ions moving towards each other along the $z$-axis and colliding at $t=z=0$. 
The transverse coordinates are denoted by the two-vector $\vec x_\perp$. The time and longitudinal coordinates ($t,z)$ can be written in two different combinations which will both be  useful: in different situations we use either light-cone coordinates, $x^\pm=(t\pm z)/\sqrt{2}$, or Milne coordinates, $\tau=\sqrt{t^2-z^2}=\sqrt{2x^+ x^-}$ and $\eta=\ln(x^+/x^-)/2$.

Tensor equations, like equation (\ref{YMeqn}), are valid in any coordinate system. However, in some parts of our calculation it will be easier to use a particular basis. All vectors and tensors can be written in the Minkowski, light-cone or Milne basis. 
For example: we can write either 
\ban
A_{\rm mink}^\mu(x) &\equiv& (A^0(x),A^z(x),\vec A(x)), 
\\
A_{\rm lc}^\mu(x) &\equiv& (A^+(x),A^-(x),\vec A(x)), 
\\
A_{\rm milne}^\mu(x) &\equiv& (A^\tau(x),A^\eta(x),\vec A(x)) .
\ean
Transformations from one basis to another are performed using the appropriate general coordinate transformation (see equation (\ref{milne2mink})). The  transverse components of any vector or tensor are the same in all three bases, and we will use indices $(i,j,k,l\dots)$ to denote  transverse components. Individual components are sometimes written with letter indices, using an obvious notation (for example, $A^{i=1} \equiv A^x$ and $A^{i=2} \equiv A^y$). In most equations it is obvious which basis is being used (for example in equation (\ref{ansatz}) the superscripts on the left side make it clear that the potential is written in the light-cone basis). In any situation where the basis is not clear, we include a subscript stating explicitly which basis is used. 

\subsection{Equations of motion} 
\label{sec-EoM}

In the formulation of the CGC effective theory that we use, the dynamics of the small x gluons is determined from the classical YM equation
\be
\label{YMeqn}
[\nabla_\mu,F^{\mu\nu}]=J^\nu ,
\ee
where $\nabla_\mu$ is the covariant derivative. In Minkowski and light-cone coordinates $\nabla_\mu = D_\mu \equiv \partial_\mu - i gA_\mu$ but in Milne coordinates the covariant derivative also includes  Christoffel symbols. The field-strength tensor is 
\be
\label{field-strength}
F_{\mu\nu}=\frac{i}{g}[\nabla_\mu, \nabla_\nu] .
\ee
Both $J^\mu$ and $A^\mu$ are SU($N_c$) valued functions that can be written as, for example, $A^\mu=A^\mu_a t_a$. 

The two ions moving towards each other along the $z$-axis contain large x valence partons that provide the source on the right side of equation (\ref{YMeqn})
\be
\label{J-LC}
\begin{aligned}
& ~~~~~~~~~~~~~~~~~~~~
J^\mu(x) = J^{\mu}_1(x)+J^{\mu}_2(x) ,
\\
& J^{\mu}_1(x) = \delta^{\mu +} g \rho_1(x^-,\vec{x}_\perp),
~~~~~~~
J^{\mu}_2(x) = \delta^{\mu - } g \rho_2(x^+,\vec{x}_\perp) ,
\end{aligned}
\ee
where the indices 1 and 2 indicate the ions moving to the right (the positive $z$ direction) and the left (the negative $z$ direction), respectively, and $\rho_1$ and $\rho_2$ are the densities of the colour charges. The valence partons are assumed to remain ultra-relativistic throughout the collision, and the currents are static (independent of the light-cone time). Physically this means that the lifetime of the valence partons is much greater than that of the small x degrees of freedom. We refer to the path of ion 1 as the positive light-cone, and ion 2 moves along the negative light-cone. Because of Lorentz contraction, both ions have a very small but finite region of support across the light-cone over $-\ww/2\le x^\mp \le \ww/2$. The limit ${\rm w}\to 0$ will be taken at the end of the calculation (see Appendix B of Ref.~\cite{Carrington:2020ssh} for details), but in intermediate steps of the calculation it is necessary to keep w non-zero. 

Our goal is to find the gauge field in the forward light-cone, which corresponds to the post-collision part of spacetime. It is natural to describe this region using Milne coordinates. We will work in the axial gauge $A^\tau=0$, which is also called Fock-Schwinger gauge. The gauge potential in the forward light-cone has the form
\bea
\label{ansatz2}
A^\mu_{\rm milne} =\theta(\tau) \big(0,\alpha(\tau, \vec x_\perp),\vec\alpha_\perp(\tau, \vec x_\perp)\big) ,
\eea
where the functions $\alpha(\tau,\vec x_\perp)$ and $\vec\alpha_\perp(\tau,\vec x_\perp)$ are independent of rapidity, as is appropriate for a boost invariant system. The YM equation and the energy-momentum tensor are simplified in these coordinates. As will be explained in detail below, our method is to expand the gauge potential in $\tau$, solve the YM equation order by order in the expansion, and obtain expressions that depend on the initial potentials $\alpha(0, \vec x_\perp)$ and $\vec\alpha_\perp(0, \vec x_\perp)$. 

The initial potentials must be connected to the source terms that represent the currents of the two colliding ions. For this purpose, one writes the gauge potential in light-cone coordinates, where the regions of spacetime that correspond to the pre- and post-collision fields are separated. In light cone coordinates the gauge condition $A^\tau=0$ is
\be
\label{foch}
x^+ A^- + x^- A^+  = 0 
\ee
and the gluon potential is given by the ansatz \cite{Kovner:1995ts,Kovner:1995ja}
\be
\label{ansatz}
\begin{aligned}
A^+(x) &= \Theta(x^+)\Theta(x^-) x^+ \alpha(\tau,\vec x_\perp)  ,
\\[2mm]
A^-(x) &=  -\Theta(x^+)\Theta(x^-) x^- \alpha(\tau,\vec x_\perp) ,
\\[2mm]
A^i(x) &= \Theta(x^+)\Theta(x^-) \alpha_\perp^i(\tau,\vec x_\perp)
\\
& ~ +\Theta(-x^+)\Theta(x^-) \beta_1^i(x^-,\vec x_\perp)
      +\Theta(x^+)\Theta(-x^-) \beta_2^i(x^+,\vec x_\perp) ,
\end{aligned}
\ee
which satisfies Eq.~(\ref{foch}) in all regions of spacetime. The step functions separate the glasma potentials in the post-collision part of space-time, determined by the  functions $\alpha$ and $\vec\alpha_\perp$, from the pre-collision potential of each ion, denoted $\vec\beta_1$ and $\vec\beta_2$. In the post-collision part of space-time the four-component vector potential is represented in terms of three independent scalar functions, and in each of the pre-collision regions there are two independent functions. 

\subsection{Outline of the computational method} 
\label{sec-method}

The first step is to calculate the gauge potential in the post-collision region of spacetime. These potentials can then be used to obtain electric and magnetic fields which are directly related to observables. There are five main steps in the procedure.

\begin{enumerate}[label=\Alph*.]

\item Using Milne coordinates with the ansatz (\ref{ansatz}), we expand the functions $\alpha(\tau,\vec x_\perp)$ and $\vec\alpha_\perp(\tau,\vec x_\perp)$ in $\tau$, and solve the YM equation for the coefficients of these expansions. 

\item The solutions obtained in step A are rewritten in terms of the initial potentials $\alpha(0,\vec x_\perp)$ and $\vec\alpha_\perp (0,\vec x_\perp)$ and their derivatives. 

\item Applying the boundary conditions that connect the  potentials $\alpha(0,\vec x_\perp)$ and $\vec\alpha_\perp(0,\vec x_\perp)$ to the potentials $\vec\beta_1(x^-,\vec x_\perp)$ and $\vec\beta_2(x^+,\vec x_\perp)$ in the pre-collision region, 
the potentials obtained in step B are rewritten in terms of pre-collision potentials for each ion, and their derivatives. 

\item The pre-collision potentials $\beta^i_1(x^-,\vec x_\perp)$ and $\beta^i_2(x^+,\vec x_\perp)$, which are generated by the colour charge distributions of the individual incoming nuclei $\rho_1(x^-,\vec x_\perp)$ and $\rho_2(x^+,\vec x_\perp)$, are expressed in terms of these distributions by solving the YM equation in the pre-collision region. 

\item The physical observables we consider are nonlinear functions of derivatives of the gauge potentials obtained in steps A, B, C and D. To obtain physical, colour neutral, results we use an averaging procedure that is explained below.
 
\end{enumerate}

The procedure to average over colour in step E is complicated. An observable of interest is expressed through products of colour charge  distributions and their derivatives. An important assumption of the CGC approach is that the distribution of colour charges is Gaussian within each nucleus. This means that the average of a product of colour charge distributions can be written as a sum of terms that combine the averages of all possible pairs in full analogy to Wick's theorem. The average of a product of pre-collision potentials, which depend on the colour charges that created them in a nontrivial way, is much more difficult to calculate, and the calculation becomes more and more complicated as the number of potentials increases \cite{Blaizot:2004wv,Fukushima:2007dy,FillionGourdeau:2008ij,Lappi:2017skr,Albacete:2018bbv}. We use the Glasma Graph Approximation \cite{Lappi:2017skr} which has been used in other near field expanded calculations and is equivalent to the application of Wick's theorem to light-cone potentials directly. For the simple case of homogeneous ions that are infinite in the transverse plane, we have found a method \cite{Carrington:2020ssh}, discussed in Sec.~\ref{sec-glasma-graph}, which shows that the effect of this approximation is small. We stress however that the range of validity of the Glasma Graph Approximation has not been carefully studied and this is an open and important issue. In the context of our calculation, the use of the Glasma Graph Approximation means that an observable can be written in terms of the two-point correlator of the pre-collision potentials. This correlator was originally calculated in Ref. \cite{JalilianMarian:1996xn} and generalized to include effects of varying nuclear density in Refs. \cite{Chen:2015wia,Carrington:2020ssh}.

\subsection{Proper time expansion}
\label{sec-proper-time-expansion}

In this section we solve the sourceless YM equation in the forward light-cone in Milne coordinates. The sourceless YM equation (\ref{YMeqn}) can be written
\be
\label{YM-0} 
g^{\sigma \mu}\big[\nabla_\sigma[\nabla_\mu,\nabla_\nu]\big]=0 ,
\ee
where the covariant derivative in this equation includes both the gauge field contribution, and the Christoffel symbols that are necessary in Milne coordinates (see Eqs.~(\ref{nabla}, \ref{connect})). The ansatz for the gauge potential (\ref{ansatz}) contains only three functions $\alpha$ and $\vec\alpha_\perp$ in the forward light-cone (due to the gauge condition), and therefore the four components of the YM equation are not all independent. We use the three equations obtained from setting $\nu\in(1,2,3)$ and find analytic solutions for the three ansatz functions by expanding in the proper time and solving for the coefficients of the expansion. We write
\be
\label{exp1}
\alpha(\tau,\vec x_\perp) = \sum_{n=0}^\infty \tau^n \alpha^{(n)}(\vec x_\perp)
=\alpha^{(0)}(\vec x_\perp) + \tau \alpha^{(1)}(\vec x_\perp) + \tau^2 \alpha^{(2)}(\vec x_\perp) + \cdots  ,
\ee
and similarly
\be
\label{exp2}
\vec\alpha_\perp(\tau,\vec x_\perp) = \sum_{n=0}^\infty \tau^n \vec\alpha^{(n)}_\perp(\vec x_\perp) .
\ee
It can be shown using a recursive procedure that all of the coefficients multiplying odd powers of $\tau$ are zero \cite{Chen:2015wia}.
The coefficients $\alpha^{(2)}(\vec x_\perp)$ and $\vec\alpha_\perp^{(2)}(\vec x_\perp)$ are found in terms of $\alpha^{(0)}(\vec x_\perp)$ and $\vec\alpha^{(0)}(\vec x_\perp)$, the coefficients $\alpha^{(4)}(\vec x_\perp)$ and $\vec\alpha_\perp^{(4)}(\vec x_\perp)$ are then found in terms of $\alpha^{(2)}(\vec x_\perp)$, $\vec\alpha_\perp^{(2)}(\vec x_\perp)$, $\alpha^{(0)}(\vec x_\perp)$ and $\vec\alpha_\perp^{(0)}(\vec x_\perp)$, etc. The process is tedious but perfectly straightforward and can be carried out algorithmically to any order. As a consistency check, we verify that the solution obtained satisfies the YM equation with $\nu=0$.

The results can be written in compact form in terms of the fields at lowest order in the $\tau$ expansion. The only non-zero components of the electric and magnetic fields at $\tau=0$ are
\ba
\label{Ez-initial}
&& E \equiv E^z(0,\vec x_\perp) = -2 \alpha(0,\vec x_\perp) ,
\\[2mm]  \nn
&& B \equiv B^z(0,\vec x_\perp) = \partial^y \alpha_\perp^x(0,\vec x_\perp) 
- \partial^x \alpha_\perp^y(0,\vec x_\perp) 
\\ \label{Bz-initial}
&&~~~~~~~~~~~~~~~~~~~~~~~~~~~~~~~~
-ig[\alpha_\perp^y(0,\vec x_\perp),\alpha_\perp^x(0,\vec x_\perp)].
\ea
To fourth order the even coefficients in the series in Eqs.~(\ref{exp1}, \ref{exp2}) are (omitting all arguments)
\be
\label{al-4}
\begin{aligned}
& \alpha^{(2)} = \frac{1}{8} \big[{\cal D}^j,[{\cal D}^j,\alpha^{(0)}]\big] ,
\\
& \alpha^{(4)} = \frac{ig}{48} \epsilon^{ij} \big[[{\cal D}^i, \alpha^{(0)}],[{\cal D}^j, B]\big] 
+\frac{1}{192} \bigg[{\cal D}^k,\Big[{\cal D}^k, \big[{\cal D}^j,[{\cal D}^j,\alpha^{(0)}]\big]\Big]\bigg] ,
\\ 
& \alpha^{i(2)}_{\perp} = \frac{1}{4} \epsilon^{ij} [{\cal D}^j, B] ,
\\
& \alpha^{i(4)}_{\perp} = \frac{ig}{64}  \left[\left[{\cal D}^i,B\right],B\right] 
+ \frac{1}{64}\epsilon^{ij} \Big[{\cal D}^j,\big[{\cal D}^k, [{\cal D}^k, B]\big]\Big]
\\ 
&~~~~~~~~~~~~~~~~~~~~~~~~~~~~~~~~~~~~~~~~~~~~~~~~~~~~~~~~~
 + \frac{ig}{16}\big[\alpha^{(0)},[{\cal D}^i_,\alpha^{(0)}]\big] ,
\end{aligned}
\ee
where we define
\be
{\cal D}^i \equiv \partial^i - ig \, \alpha_\perp^i(0,\vec x_\perp) ,
\ee
and the notation $\epsilon^{ij}$ represents a matrix with values $\epsilon^{11}=\epsilon^{22}=0$ and $\epsilon^{12}=-\epsilon^{21}=1$.

\subsection{Boundary conditions}
\label{sec-boundary-conditions}

The initial potentials $\alpha^{(0)}(\vec x_\perp)$ and $\vec\alpha^{(0)}_\perp (\vec x_\perp)$ are related to the pre-collision potentials $\vec\beta_1(x^-,\vec x_\perp)$ and $\vec\beta_2(x^+,\vec x_\perp)$ through a set of boundary conditions. These conditions were originally obtained by matching terms from the pre- and post-collision regions that are singular on the light-cone \cite{Kovner:1995ts, Kovner:1995ja}. We work with sources with small but finite longitudinal width, and the limit that this width goes to zero cannot be taken until after the boundary conditions have been used. The boundary conditions should therefore be obtained by integrating the YM equation across the light-cone. This procedure is presented in Appendix C of our work \cite{Carrington:2020ssh} and the boundary conditions are
\ba
\label{cond1}
&& \alpha^{i}_\perp(0,\vec{x}_\perp) = \alpha^{i(0)}_\perp(\vec{x}_\perp) 
= \lim_{\text{w}\to 0}\big(\beta^i_1 (x^-,\vec{x}_\perp) + \beta^i_2
(x^+,\vec{x}_\perp)\big) ,
\\ \label{cond2}
&& \alpha(0,\vec{x}_\perp) = \alpha^{(0)}(\vec{x}_\perp) 
= -\frac{ig}{2}\lim_{\text{w}\to 0}\;[\beta^i_1 (x^-,\vec{x}_\perp),\beta^i_2 (x^+,\vec{x}_\perp)] ,
\ea
where the notation $\lim_{\text{w}\to 0}$ indicates that the width of the sources across the light-cone is taken to zero. As explained in section \ref{sec-corr}, the pre-collision potentials depend only on transverse coordinates in this limit.

\subsection{Energy momentum tensor}
\label{sec-energy-momentum-tensor}

Various observables can be obtained from the energy-momentum tensor. We choose the following definition of the tensor
\be
\label{Tmunu}
T^{\mu\nu} = 2{\rm Tr}\big[F^{\mu\rho}F_\rho^{~\nu}+\frac{1}{4}g^{\mu\nu}F^{\rho\sigma}F_{\rho\sigma}\big] 
= F_a^{\mu\lambda}F_{\lambda\,a}^{~\nu}+\frac{1}{4}g^{\mu\nu}F_a^{\rho\sigma}F_{\rho\sigma\,a} ,
\ee
where $F^{\mu\nu}$ is the field-strength tensor (\ref{field-strength}). The definition is obtained by adding a total divergence to the canonical result to produce an expression that is gauge invariant, conserved, symmetric and traceless. We note that the two field-strength tensors in equation (\ref{Tmunu}) are actually calculated at different points, denoted $x=(x^+,x^-,\vec x_\perp)$ and $y=(y^+,y^-,\vec y_\perp)$, and we take $(x-y)\to 0$ at the end of the calculation. 

Using equations (\ref{ansatz2}, \ref{exp1}, \ref{exp2}) and the iterative solutions found in Sec.~\ref{sec-proper-time-expansion}, we obtain a lengthy expression for the field-strength tensor in Milne coordinates that depends only on the initial potentials $\alpha^{(0)}(\vec x_\perp)$ and $\vec\alpha^{(0)}_\perp (\vec x_\perp)$ and the expansion parameter $\tau$. We further express the tensor in terms of only pre-collision potentials using the boundary conditions (\ref{cond1}, \ref{cond2}). 

We have checked that the resulting energy-momentum tensor is symmetric and has zero divergence
\be
\nabla_\mu T^{\mu\nu} =
\partial_\mu T^{\mu\nu} - \Gamma^\mu_{\mu\rho}T^{\rho\nu}-\Gamma^\nu_{\mu\rho}T^{\mu\rho} = 0 .
\ee
The energy-momentum tensor in Minkowski space can be found as $ T^{\mu\nu}_{\rm mink} = M^\mu_{~\rho} M^\nu_{~\sigma} T^{\rho\sigma}_{\rm milne}$ with the transformation matrix given by Eq.~(\ref{milne2mink}).

\subsection{Correlation functions of pre-collision potentials}
\label{sec-corr}

All components of the energy-momentum tensor have the form of sums of products of pre-collision potentials. A generic term has the form 
\be
\label{core-ex-1}
\beta_{1a}^i(x^-,\vec x_\perp) \beta_{1b}^k(x^-,\vec x_\perp)\beta_{2c}^k(x^+,\vec x_\perp) \dots \beta_{1d}^m(y^-,\vec y_\perp)\beta_{2e}^l(y^+,\vec y_\perp) .
\ee
The pre-collision potentials can be expressed in terms of the charge distributions of the ion sources by solving the YM equation in the pre-collision region. These colour charge distributions are not known, and an important input to the CGC approach is the use of an averaging procedure based on the assumption of a Gaussian distribution of colour charges within each nucleus. A product of colour charges is replaced by its average over this Gaussian distribution (which will be denoted with angle brackets). We make the assumption that sources from different ions are uncorrelated, or equivalently that correlation functions of products of sources from different ions can be set to zero. This means that we need to consider only averages of the form $\langle \rho_1 \rho_1 \dots \rho_1\rangle$ and $\langle \rho_2 \rho_2 \dots \rho_2\rangle $ where all sources are from the same ion. This approximation is justified because the pre-collision distributions are independent from each other due to causality, and they remain uncorrelated post-collision because the CGC approach we are using assumes that the colour sources are static. This assumption is a standard component of most CGC calculations, but it necessarily means that the approach does not capture the QCD dynamics of a heavy ion collision completely.

The correlation function of the colour charge density of the first ion $\rho_1(x^-,\vec x_\perp)$ is assumed to have  the form
\be
\label{big-in}
\langle \rho_1(x^-,\vec x_\perp) \, \rho_1(y^-,\vec y_\perp)\rangle 
\equiv g^2 \lambda_1(x^-,\vec x_\perp) \, \delta(x^--y^-) \, \delta^{(2)}(\vec x_\perp - \vec y_\perp) ,
\ee
where $\lambda_1(x^-,\vec x_\perp) = h(x^-) \mu_1(\vec x_\perp) $ with $h(x^-)$ a sharply peaked non-negative function  normalized to one with width ${\rm w}$  around $x^-=0$ and $\mu_1(\vec x_\perp)$ is a surface colour charge density. The integration of $\lambda_1(x^-,\vec x_\perp) $ over $x^-$ obviously gives $\mu_1(\vec x_\perp)$. We make the analogous definitions for the second ion, and the width w is taken to zero at the end of the calculation. Since the average over a Gaussian distribution of colour densities which are independent random variables can be rewritten as a sum over the averages of all possible pairs, the average over any product of colour sources can be written in terms of the fundamental correlator (\ref{big-in}). 
 
As discussed in Sec.~\ref{sec-method}, we use the approximation that Wick's theorem can be applied to light-cone potentials directly, which is  called the Glasma Graph Approximation \cite{Lappi:2015vta}. Correlations of any even number of potentials from the same ion are written as products of correlators of pairs of potentials. The only correlator that must be calculated is the two-point correlator of pre-collision potentials from the same ion. We define
\be
\label{core5-20}
\begin{aligned}
& \delta_{ab} B_1^{ij}(\vec{x}_\perp,\vec y_\perp) \equiv  
\lim_{{\rm w} \to 0} \langle \beta_{1\,a}^i(x^-,\vec x_\perp) \, \beta_{1\,b}^j(y^-,\vec y_\perp)\rangle ,
\\  
& \delta_{ab} B_2^{ij}(\vec{x}_\perp,\vec y_\perp) \equiv  
\lim_{{\rm w} \to 0} \langle \beta_{2\,a}^i(x^+,\vec x_\perp) ] \, \beta_{2\,b}^j(y^+,\vec y_\perp)\rangle .
\end{aligned}
\ee
Our calculation of the functions $B_n^{ij}(\vec x_\perp,\vec y_\perp)$ with $n\in\{1,2\}$ can be found in Appendix D of Ref.~\cite{Carrington:2020ssh} and the result is
\ba
\nn
&& B_n^{ij}(\vec{x}_\perp,\vec y_\perp) = \frac{1}{g^2 N_c \Delta\tilde\gamma_n(\vec x_\perp,\vec y_\perp)}  
\bigg(\exp\Big[g^4 N_c \Delta\tilde\gamma_n(\vec x_\perp,\vec y_\perp) \Big]-1\bigg)
\\[2mm] \label{B-res}
&&~~~~~~~~~~~~~~~~~~~~~~~~~~~~~~~~~~~~~~~~~\times 
\partial_x^i \partial_y^j \tilde\gamma_n(\vec x_\perp,\vec y_\perp) ,
\ea
with 
\be
\label{Gamma-tilde-def}
\Delta\tilde\gamma_n(\vec x_\perp,\vec y_\perp) \equiv 
\tilde\gamma_n(\vec x_\perp,\vec y_\perp) 
- \frac{1}{2}\tilde\gamma_n(\vec x_\perp,\vec x_\perp) 
- \frac{1}{2}\tilde\gamma_n(\vec y_\perp,\vec y_\perp)  
\ee
and
\be
\label{gamma-tilde-def}
\tilde \gamma_n(\vec x_\perp,\vec y_\perp) \equiv 
\int d^2 z_\perp \mu_n(\vec z_\perp) \, G(\vec x_\perp-\vec z_\perp) \, G(\vec y_\perp-\vec z_\perp) ,
\ee
where $G(\vec x_\perp)$ is the Green's function of the two-dimensional Poisson equation and equals 
\be
\label{Gf-def}
G(\vec x_\perp) = \frac{1}{2\pi} K_0(m |\vec x_\perp|) .
\ee
The function $K_0$ is a modified Bessel function of the second kind, and $m$ is an infrared regulator whose definition will be discussed in Sec.~\ref{sec-ir-reg}. We note that the correlators $B_1^{ij}(\vec{x}_\perp,\vec y_\perp)$ and $B_2^{ij}(\vec{x}_\perp,\vec y_\perp)$ will be different only if the two ions, and their corresponding  surface colour charge densities $\mu_1(\vec z_\perp)$ and $\mu_2(\vec z_\perp)$, are different. 

If the surface density of colour charges is uniform and $\mu_1(\vec x_\perp) =\mu_2(\vec x_\perp) = \bar\mu$, the functions $\tilde\gamma_1(\vec x_\perp, \vec y_\perp)$ and $\tilde\gamma_2(\vec x_\perp, \vec y_\perp)$ are equal to each other and depend only on the magnitude of the relative coordinate $r = |\vec x_\perp - \vec y_\perp|$, and will  both be denoted $\bar\gamma(r)$. In this case the correlator (\ref{B-res}) becomes
\be
B^{ij}(\vec x_\perp,\vec y_\perp) = g^2\,\left(\frac{e^{g^4 N_c \left(\bar\gamma(r)-\bar\gamma(0)\right)}-1}{g^4 N_c \left(\bar\gamma(r)-\bar\gamma(0)\right)} \right)\partial_x^i \partial_y^j \bar\gamma(r) \label{B-constant}
\ee
with
\be
\label{gamma-k-const}
\bar \gamma(r)  = \frac{\bar\mu}{4 \pi  m} r K_1(m r) . 
\ee
Substituting (\ref{gamma-k-const}) into (\ref{B-constant}) one finds that the correlator $B^{ij}$ diverges logarithmically when $r\to 0$ and therefore has to be regularized. This will be discussed in Sec.~\ref{sec-uv-reg}.

All higher order correlators are expressed through the correlators (\ref{core5-20}, \ref{B-res}). For example, the average of four potentials, two from each ion, is
\ba
\nn
&& \lim_{{\rm w} \to 0} \langle\beta^i_{1\,a}(x^-,\vec x_\perp) \, \beta^j_{1\,b}(y^-,\vec y_\perp) \,
\beta^l_{2\,c}(x^+,\vec x_\perp)\beta^m_{2\,d}(y^+,\vec y_\perp) \rangle 
\\ \label{wick}
&&~~~~~~~~~~~~~~~~~~~~~~~~~~~~~~
= \delta_{ab}\delta_{cd} B^{ij}(\vec x_\perp,\vec y_\perp) \, B^{lm}(\vec x_\perp,\vec y_\perp) .
\ea 
When one of the potentials is differentiated with respect to a transverse coordinate we have, for example,
\be
\lim_{{\rm w} \to 0} \langle\partial^k_x \beta^i_{1\,a}(x^-,\vec x_\perp)
\,\beta^j_{1\,b}(y^-,\vec y_\perp)\rangle = \delta_{ab}  \partial^k_x B^{ij}(\vec x_\perp,\vec y_\perp) .
\ee

\subsection{Accuracy of the  Glasma Graph Approximation}
\label{sec-glasma-graph}

We apply the Glasma Graph Approximation which allows us to use Wick's theorem to calculate products of light-cone gauge potentials.  Although the approximation is consistently used throughout, we can still obtain a quantitative measure of its validity in our calculation of the energy-momentum tensor. 

The initial longitudinal magnetic field in equation (\ref{Bz-initial}) is rewritten using the initial condition (\ref{cond2}) as 
\be
\label{B-long}
B = B^z(0,\vec x_\perp) =  F_1^{21}(\vec x_\perp) + F_2^{21}(\vec x_\perp) 
+i g \epsilon^{ij}[\beta_1^i(\vec x_\perp),\beta_2^j(\vec x_\perp)] .
\ee
If we use the fact that the pre-collision potentials are pure gauge ($ F_1^{ij}= F_2^{ij}=0$), Eq.~(\ref{B-long}) can be written as 
\be
\label{B-short}
B_{\rm pg} =  i g \epsilon^{ij}[\beta_1^i(\vec x_\perp),\beta_2^j(\vec x_\perp)] .
\ee
We expect that the averaging procedure gives $\langle B\rangle = \langle B_{\rm pg}\rangle$ or $\langle F^{21}_1\rangle = \langle F^{21}_2\rangle= \langle F^{12}_1\rangle = \langle F^{12}_2\rangle =0$. We consider the field-strength from the first ion (suppressing the subscript 1). Expanding out all terms we have
\be
F^{12}(\vec x_\perp) = \partial_x^1 \beta^2(\vec x_\perp) - \partial_x^2 \beta^1(\vec x_\perp)  - i g \beta^1(\vec x_\perp)\beta^2(\vec x_\perp) + i g\beta^2(\vec x_\perp)\beta^1(\vec x_\perp) \,.
\ee 
The expectation value of this expression is obviously zero because the averaging procedure involves a trace over colour indices. In fact it is zero even before the colour trace is taken, since the first two terms have only one potential (all terms with an odd number of potentials are set to zero), and the third and fourth term give zero using the expression for the two-point correlator that is derived below (see Eq.~(\ref{corr-res-basic})). 

Next we consider the expectation value
\be
\lim_{r\to 0}\langle F^{12}(\vec x_\perp) \partial^2_y \beta^1(\vec y_\perp)\, \rangle\,.
\ee
This should also vanish, since $F^{12}=0$. The terms with three potentials are set to zero, and the terms with two potentials give
\ba
\nn
&&\lim_{r\to 0}\langle F^{12}(\vec x_\perp) \partial^2_y \beta^1(\vec y_\perp)\rangle  
\\ [2mm] \label{pg-bad}
&&~~~~
= \frac{1}{2}(N_c^2-1)  \lim_{r\to 0} \big[\partial^1_x\partial^2_y B^{21}(\vec x_\perp,\vec y_\perp) - \partial^2_x\partial^2_y B^{11}(\vec x_\perp,\vec y_\perp)\big].
\ea
Equation (\ref{alina-116}), which is derived in Sec.~\ref{sec-uv-reg}, shows that the right side of (\ref{pg-bad}) is not zero. This contradiction is related to the Glasma Graph Approximation, which sets all correlators with odd numbers of potentials to zero. Therefore, we can test the approximation by comparing the energy-momentum tensor obtained with the two different forms of the lowest order magnetic field in Eqs.~(\ref{B-long}) and (\ref{B-short}). 

\subsection{Regularization}
\label{sec-reg}

To calculate the observables we are interested in we need correlators of  potentials at the same point, which should be obtained from the correlator (\ref{B-res}) taking the limit $\vec r=\vec x_\perp-\vec y_\perp~\to~0$. This limit produces a divergence, and an ultraviolet regulator is needed to control it. We note that this divergence is expected since the CGC model we are using is classical and breaks down at small distances. An infrared regulator is also needed to properly define the Green's functions (\ref {Gf-def}) from which the correlators of pre-collision potentials are constructed.  In this section we discuss the parameters that are needed to control these ultraviolet and infrared divergences. 

\subsubsection{Infrared regulator}
\label{sec-ir-reg}

In Eq.~(\ref{Gf-def}) there is introduced an infrared regulator denoted $m$. To obtain more insight into how this regulator should be chosen, we expand the Green's function in (\ref{Gf-def}) around $m=0$ which gives
\bea
\label{Gf-exp}
G(\vec{r}\,) \approx \frac{1}{2\pi}\ln\left(\frac{L}{r}\right)
~~ \text{with}~~
L = \frac{2 e^{-\gamma_E}}{m} \approx \frac{1.12}{m},
\eea
where $\gamma_E\approx 0.577$ is  Euler's constant. Since the valence parton sources come from individual nucleons, confinement tells us that their effects should die off at transverse length scales larger than $1/\Lambda_{\rm QCD}$, and the Green's function should therefore be defined with boundary conditions so that it vanishes at $r \gtrsim 1/\Lambda_{\rm QCD}$. From Eq.~(\ref{Gf-exp}) we see that we should choose $m\sim \Lambda_{\rm QCD}$.

\subsubsection{Ultraviolet regulator}
\label{sec-uv-reg}

Substituting (\ref{gamma-k-const}) into (\ref{B-constant}) one finds that the correlator $B^{ij}$ diverges logarithmically when $r\to 0$. One approach to regulate this divergence, and the divergences that appear in derivatives of the two-point correlator, is to expand in $r$ and regulate any factors involving inverse powers of $r$, or logarithms of $r$, by making the replacement $r\to 1/Q_s$.  A different method  \cite{Fujii:2008km} is more suitable for a calculation of local quantities like the energy-momentum tensor, where we want to take the relative coordinate $r$ strictly to zero. We rewrite the function $\bar\gamma(r)$ as a momentum integral 
\be
\label{gamma-k-const-2}
\bar \gamma(r)  = \frac{\bar\mu}{4 \pi  m} r K_1(m r) 
= \bar \mu \int \frac{d^2k}{(2\pi)^2} \frac{e^{i\vec r\cdot \vec k} }{\left(k^2+m^2\right)^2} ,
\ee
and then substituting (\ref{gamma-k-const-2}) into (\ref{B-constant}) we  obtain\footnote{In Ref.~\cite{Carrington:2020ssh} there was a misprint in the equation corresponding to (\ref{log-div}) (see Eq.~(32)) where the expression in the middle has the wrong power of $k$.}
\be
\label{log-div}
\lim_{r\to 0}B^{ij}(\vec x_\perp,\vec y_\perp) 
= g^2 \bar \mu \int \frac{d^2k}{(2\pi)^2} \frac{\hat k^i \hat k^j k}{\left(k^2+m^2\right)^2} 
= \delta^{ij}\frac{g^2\bar\mu}{4\pi} \int_0^\infty \frac{dk \,k^3}{(k^2+m^2)^2}\,.
\ee
In the last line we have used  $\hat k^i \hat k^j\to \delta^{ij}/2$ which follows from the integration over the angular variables. We introduce a momentum cutoff that we call $\Lambda$ to regulate the logarithmically divergent integral in Eq.~(\ref{log-div}), which gives 
\be
\label{corr-res-basic}
\lim_{r\to 0}B^{ij}(\vec x_\perp,\vec y_\perp) 
=\delta^{ij}g^2\frac{\bar \mu}{8\pi}\left({\rm LN}-\frac{\Lambda^2}{\Lambda^2+m^2}\right) ,
\ee
where we have defined 
\be
{\rm LN} \equiv \ln\left(\frac{\Lambda^2}{m^2}+1\right) .
\ee

Derivatives of the correlator $B^{ij}(\vec x_\perp,\vec y_\perp)$ can be calculated in the same way. After integrating over the angular variables, products of odd numbers of unit vectors $\hat k$ give zero and products of four and six unit vectors are replaced with sums of products of delta functions using
\ba
\label{khat-4}
\hat k^i \hat k^j \hat k^k \hat k^l 
&\to& \frac{1}{8}( \delta_{il} \delta_{jk}+\delta_{ik} \delta_{jl}+\delta_{ij} \delta_{kl}) 
\\[2mm] \nn
\hat k^i \hat k^j \hat k^k \hat k^l \hat k^m \hat k^n 
&\to& \frac{1}{48}(\delta _{in} \delta _{jm} \delta _{kl} +\delta _{im} \delta _{jn} \delta_{kl}
                     +\delta _{ij} \delta _{kl} \delta _{mn} +\delta _{in} \delta _{jl} \delta_{km}
\\ \nn
 &&~\,
    +\delta _{il} \delta _{jn} \delta _{km} +\delta _{im} \delta _{jl} \delta_{kn} 
    +\delta _{il} \delta _{jm} \delta _{kn}+\delta _{in} \delta _{jk} \delta_{lm}
\\ \nn &&~\,
    +\delta _{ik} \delta _{jn} \delta _{lm} +\delta _{ij} \delta _{kn} \delta_{lm}
    +\delta_{im} \delta _{jk} \delta _{ln}  +\delta _{ik} \delta _{jm} \delta _{ln}
\\  &&~\,
    +\delta _{ij} \delta _{km} \delta _{ln}+\delta _{il} \delta _{jk} \delta_{mn}
    +\delta _{ik} \delta _{jl} \delta _{mn}).
\label{khat-6}
\ea
Formulas similar to (\ref{khat-4}) and (\ref{khat-6}) can be obtained for any even number of unit vectors. 

When two derivatives act on the two-point correlator we obtain 
\ba
\nn 
&& \lim_{r\to 0} \partial^l_{n_1} \partial^m_{n_2} B^{ij}(\vec{x}_\perp, \vec{y}_\perp)
=(-1)^{n_1+n_2+1} \frac{g^2 \bar \mu }{16\pi} 
\\ \nn
&&~~~~\times
\Big[(\delta^{ij}\delta^{lm} + \delta^{il}\delta^{jm} + \delta^{im}\delta^{jl})
\Big( \frac{\Lambda^4 + 2\Lambda^2 m^2}{2(\Lambda^2+m^2)} - m^2 {\rm LN} \Big) 
\\  \label{alina-116} 
&& ~~~~~~~~~~~~~~ + \delta^{lm}\delta^{ij}  \frac{3g^4 \bar \mu}{8\pi} 
\Big({\rm LN}^2 + \frac{\Lambda^4}{(\Lambda^2+m^2)^2} - \frac{2{\rm LN}\Lambda^2}{\Lambda^2+m^2}\Big) \Big] ,
\ea
where we use the notation that the index $n_i$ is 1 if the transverse derivatives are with respect to $\vec x_\perp$ and 2 if the transverse derivatives are with respect to $\vec y_\perp$. We note that the leading order contributions to equation (\ref{alina-116}) should agree with Eq.~(27) in Ref. \cite{Fujii:2008km}, but we have a different sign for the last term in that equation.

Although one might expect that the alternative regularization scheme described above equation (\ref{gamma-k-const-2}) would give very similar results, this is not always true. For some correlators the only difference is a redefinition of the mass scale which appears in the argument of a logarithmic factor, but in some cases the two regularization methods give parametrically different results. We consider as an example the biggest contribution to (\ref{alina-116}) with $n_1=1$ and $n_2=2$ which is 
\be
\label{116-lo}
\partial^l_x \partial^m_y B^{ij}(\vec x_\perp,\vec y_\perp)\Big|_{\rm lo} 
= \frac{g^2 \bar\mu \Lambda^2}{32\pi}
(\delta^{ij}\delta^{lm} + \delta^{il}\delta^{jm} + \delta^{im}\delta^{jl}) . 
\ee
If we start from equation (\ref{B-constant}), take the derivatives, expand in $r$, and regulate divergences using $r\to 1/Q_s$, the leading order contribution is 
\be
\label{116-no}
\partial^l_x \partial^m_y B^{ij}(\vec x_\perp,\vec y_\perp)\Big|_{\rm alternate} 
= \frac{g^2\bar\mu  m^2 {\rm LG}}{16\pi}
(\delta^{ij}\delta^{lm} + \delta^{il}\delta^{jm} + \delta^{im}\delta^{jl}) ,
\ee
where we have introduced the definitions
\be
\label{mhat-def}
{\rm LG} \equiv \ln[(\hat m /Q_s)^2]  \text{~~and~~} \hat m \equiv \frac{m}{2}e^{\gamma_E+\frac{1}{2}} .
\ee
Comparing equations (\ref{116-lo}, \ref{116-no}) we see that when we write the correlator as a momentum integral with an ultraviolet cutoff the regulated result is of order $\Lambda^2$, but if we expand in $r$ and make the replacement $r\to 1/Q_s$ the regulated expression is of order $m^2$. The expansion method would therefore replace the ultraviolet regulator with the infrared one. In Ref.~\cite{Chen:2015wia} the energy-momentum tensor is calculated using both of these regularization methods, depending on which correlator is being calculated. 
In our calculation of the energy-momentum tensor we consistently regularize by writing each correlator as a momentum integral, differentiating as needed, taking the limit $r\to 0$, and calculating the regulated momentum integral. We note also that we treat both the one-point correlator $B^{ij}(\vec x_\perp,\vec x_\perp)$ and the two-point correlator $B^{ij}(\vec x_\perp,\vec y_\perp)$ in the same way. 

In our calculation of transport coefficients the regularization is done in a completely different way, because the physics of the calculation is different. In this case, the variable $r$ is an integration variable that goes to zero at the lower limit of the integration. The regularization is done using a step function in coordinate space, and is discussed in Sec.~\ref{sec-res-c}.

\subsubsection{The MV scale}
\label{sec-MV-scale}

The colour charge density provides a dimensionful scale $\sqrt{\bar\mu}$ which is usually called the  McLerran-Venugopalan (MV) scale and is defined in our notation as $g^2 \sqrt{\bar\mu}$. This scale is related to the saturation scale $Q_s$, although the exact relationship between them cannot be determined within the CGC approach, as repeatedly discussed, see for example \cite{Iancu:2003xm, Lappi:2007ku}. Expanding the function $\Delta\tilde\gamma(\vec x_\perp,\vec y_\perp)$ in Eq.~(\ref{Gamma-tilde-def}) in $mr\ll 1$  gives
\be
\label{MVc}
\Delta\tilde\gamma(\vec x_\perp,\vec y_\perp) \Big|_{\rm MV} = \frac{\bar\mu r^2}{16\pi}
\Big[2\ln\left(\frac{mr}{2}\right)+2\gamma_E -1  + {\cal O}\left((mr)^2\right) \Big] 
\ee
which is rewritten using (\ref{mhat-def}) in the form
\be
\label{MVd}
\Delta\tilde\gamma(\vec x_\perp,\vec y_\perp) \Big|_{\rm MV}  = \frac{\bar\mu r^2}{16\pi}
\left({\rm LG}-2  + {\cal O}\left((\hat m\, r)^2\right) \right) .
\ee
Substituting the expression (\ref{MVd}) into the correlator (\ref{B-res}), we obtain 
\be
\label{MVh}
B^{ij}(\vec x_\perp,\vec y_\perp) \Big|_{\rm MV}  \approx
 \frac{2}{g^2 N_c r^2} \, \delta^{ij}  \big(1 - \delta^{\epsilon}\big) ,
\ee
where we have defined two dimensionless parameters
\be
\delta \equiv \hat m r \ll1 \text{~~~~ and ~~~~}  \epsilon \equiv \frac{g^4 N_c}{8\pi} \bar\mu r^2 .
\ee

The result in (\ref{MVh}) can be used to argue that the MV scale is proportional to the saturation scale \cite{JalilianMarian:1996xn}. For $\epsilon\ll 1$  we have $1-\delta^{\epsilon} \approx -\epsilon\ln\delta$ 
and the correlator (\ref{MVh}) becomes
\be
\label{Ai0-Aj0-approx-1}
B^{ij}(\vec x_\perp,\vec y_\perp)\Big|_{\rm MV}     
\approx -\frac{g^2\bar \mu}{8 \pi}   \delta^{ij}  \ln (\hat m^2 r^2)  .
\ee
For $\epsilon>1$ we have $1-\delta^{\epsilon} \approx 1$ and (\ref{MVh}) takes the form
\be
\label{Ai0-Aj0-approx-2}
B^{ij}(\vec x_\perp,\vec y_\perp)\Big|_{\rm MV}     
\approx \frac{2 \delta^{ij} }{g^2 N_c} \frac{1}{r^2}  .
\ee
The condition $\epsilon<1$ corresponds to small transverse distances, or momentum scales $k_\perp \gtrsim \bar\alpha \sqrt{\bar\mu}$ with $\bar\alpha = g^2\sqrt{N_c/(8\pi)}$. The opposite case $\epsilon>1$ means transverse distances that are large (but still much smaller than $1/\Lambda_{\rm QCD}$), or momentum scales that satisfy $\bar\alpha \sqrt{\bar \mu} > k_\perp > \Lambda_{\rm QCD}$. From the Fourier transforms of the expressions (\ref{Ai0-Aj0-approx-1}, \ref{Ai0-Aj0-approx-2}) we see that at small transverse distances, or large momentum scales, the correlation function falls like $1/k_\perp^2$ (perturbative behaviour) and at large transverse distances, or small momentum scales, the correlation function rises like $\sim \ln(k_\perp)$. The number of gluons in a range $dk_\perp$ about some $k_\perp$ is related to the trace of the gluon propagator in equation (\ref{MVh}), multiplied by the phase space factor $k_\perp$. The peak occurs approximately at the momentum scale that corresponds to $\epsilon=1$, which divides the growing and falling regions of the distribution function. Therefore the typical transverse momenta of gluons, or the saturation scale, satisfies
\be
\label{A-def}
\epsilon\big|_{r\sim 1/Q_s}  = 2\pi \alpha_s^2 N_c r^2\bar \mu\big|_{r \sim 1/Q_s}  = 1
\ee 
which gives $Q_s^2 \sim g^4\bar\mu$. 
As mentioned above, the proportionality factor cannot be determined within the CGC approach. We define
\be
\label{myUbar}
Q_s^2 = g^4  \bar\mu. 
\ee
Due to the ambiguity associated with the value of the MV scale, our numerical results for quantities like the energy density and pressure should be regarded as order of magnitude estimates. Quantities that depend on ratios of different elements of the energy-momentum tensor, like Fourier coefficients of the azimuthal flow, will have much weaker dependence on the MV scale.

\subsection{Correlators of pre-collision potentials of finite nuclei}
\label{sec-corr-finite}

Until now we have considered only the simple case of nuclei that are infinite and uniform in the transverse plane. In this section we consider more physically realistic collisions where the nuclei are finite and the nuclear area densities are not constant. 

The correlator of pre-collision potentials given by Eqs.~(\ref{core5-20} - \ref{Gf-def}) is not only applicable to nuclei which are transversally uniform, which corresponds to taking $\mu(\vec x_\perp)=\bar\mu$, but can also be used for non-uniform charge densities. We now consider the situation where the surface density of colour charges for each nucleus is assumed to be the two-dimensional projection of a Woods-Saxon distribution
\be
\mu(\vec x_\perp) \label{def-mu2-2} 
 = \Big(\frac{A}{207}\Big)^{1/3}\frac{\bar\mu}{2a\ln(1+e^{R_A/a})} 
\int^\infty_{-\infty} \frac{dz}{1 + \exp\big[(\sqrt{(\vec x_\perp)^2 + z^2} - R_A)/a\big]}.
\ee
The parameters $R_A$ and $a$ give the radius and skin thickness of a nucleus of mass number $A$, and their numerical values are discussed in Sec.~\ref{sec-finite-nuclei}. The integral in (\ref{def-mu2-2}) is normalized so that for a lead nucleus $\mu(\vec 0)=\bar\mu$.  

The correlator of pre-collision potentials can be obtained from Eqs.~(\ref{core5-20} - \ref{Gf-def}) by substituting the distribution (\ref{def-mu2-2}) into Eq.~(\ref{gamma-tilde-def}) and performing a gradient expansion, following the method developed in Ref.~\cite{Chen:2015wia}\footnote{We remind the reader that for a realistic nucleus, which is made up of individual nucleons, the transverse charge distribution is not a very smooth function. It is possible that the transverse charge distribution of a real nucleus could be sufficiently irregular that a Woods-Saxon distribution is not a good representation.}. The coordinates $\vec x_\perp$ and $\vec y_\perp$ are rewritten in terms of  relative and average coordinates, which are defined as $\vec x_\perp-\vec y_\perp$ and $(\vec x_\perp+\vec y_\perp)/2$, respectively. To consider collisions with non-zero impact parameter we expand the distribution $\mu_1(\vec z_\perp)$ around $\vec R-\vec b/2$, and $\mu_2(\vec z_\perp)$ around $\vec R+\vec b/2$. We will keep terms up to second order in gradients of the distribution. The parameter that we assume to be small is
\be
\label{delta-def}
\delta = \frac{|\nabla^i\mu(\vec R \pm \frac{\vec b}{2})|}{m \mu(\vec R \pm \frac{\vec b}{2})} ,
\ee
where the gradient operator indicates differentiation with respect to the argument of the function. The region of validity of this expansion is discussed in Sec.~\ref{sec-finite-nuclei}.

In the rest of this section we drop the subscript that indicates which ion is being considered and set $\vec b=0$. Performing the gradient expansion and keeping terms up to second order in gradients of $\mu$,  equation (\ref{gamma-tilde-def})  becomes
\ba
\nn
&&
\tilde\gamma(\vec x_\perp,\vec y_\perp) = \frac{\mu(\vec R)r}{4\pi m}K_1(mr)
\\  \label{gamma-def-2}
&&~~~\times
\frac{1}{2}\nabla^i \nabla^j \mu(\vec R) 
\Big(\delta^{ij}\frac{r^2}{24\pi m^2}K_2(mr)+\frac{r^i r^j}{r^2}\frac{r^3}{48\pi m}K_1(mr)\Big).
\ea
We can rewrite Eq.~(\ref{gamma-def-2}) in the form 
\be
\tilde \gamma(\vec x_\perp,\vec y_\perp) = 
\mu(\vec R) \int \frac{d^2k}{(2\pi)^2} \frac{e^{i\vec r\cdot \vec k} }{\left(k^2+m^2\right)^2} 
+ \frac{m^2}{2} \nabla ^2\mu(\vec R)  
\int \frac{d^2k}{(2\pi)^2} \frac{e^{i\vec r\cdot \vec k} }{\left(k^2+m^2\right)^4},
\label{gamma-def-3}
\ee
where we have made the replacement $\hat r^i \hat r^j \to \delta^{ij}/2$ because in the limit $\vec r\to 0$ we know $\tilde\gamma$ must be independent of the direction of the vector $\hat r$. We note that we are able to make this replacement before performing any derivatives with respect to $\vec x_\perp$ and $\vec y_\perp$,  since $\lim_{r\to 0}\partial_x^i \dots \partial_y^j \dots \hat r^k \hat r^l = 0$, where the dots indicate any number of derivatives. 

The correlator $B^{ij}(\vec x_\perp, \vec y_\perp)$ and its derivatives  have ultraviolet divergences that must be regulated. We use a modified version of the method proposed in Ref.~\cite{Fujii:2008km} which is discussed in section \ref{sec-uv-reg} in the context of uniform nuclei. To illustrate it we consider, as an example, the calculation of 
\ba
\nn
&& \partial_x^i\partial_y^j \tilde\gamma(\vec x_\perp,\vec y_\perp) = \mu(\vec R) \int \frac{d^2k}{(2\pi)^2} \frac{k^i k^j  e^{i\vec r\cdot \vec k} }{\left(k^2+m^2\right)^2} 
\\ \label{bxx-example}
&&~~~~~~~~~~~~~~~~~~~~~~~
+ \frac{m^2}{2} \nabla ^2\mu(\vec R)  \int \frac{d^2k}{(2\pi)^2} \frac{k^i k^j  e^{i\vec r\cdot \vec k} }{\left(k^2+m^2\right)^4} ,
\ea
which appears in the expression for $B^{ij}$ in Eq.~(\ref{B-res}). The integration over angular variables gives $k^i k^j \to \delta^{ij} k^2/2$. The second term in (\ref{bxx-example}) is finite, but the first term is logarithmically divergent and we regulate it using an ultraviolet momentum cutoff $\Lambda$. This cutoff will be set to the saturation scale $Q_s$. 

Now we consider the contribution from the factor in round brackets in Eq.~(\ref{B-res}). Expanding this factor we have
\be
\label{coef-exp}
\frac{e^{g^4 N_c \,\Delta\tilde\gamma_n(\vec x_\perp,\vec y_\perp)}-1}
{g^4 N_c \, \Delta\tilde\gamma_n(\vec x_\perp,\vec y_\perp)} = 1
+\frac{1}{2} g^4 N_c \, \Delta\tilde\gamma(\vec x_\perp,\vec y_\perp) 
+ \frac{1}{6} \left(g^4 N_c \, \Delta\tilde\gamma(\vec x_\perp,\vec y_\perp)\right)^2 + \dots
\ee 
When we calculate derivatives of the correlator $B^{ij}(\vec x_\perp,\vec y_\perp)$, the derivatives operate on all terms in the expansion in Eq.~(\ref{coef-exp}). At sixth order in the $\tau$ expansion the energy-momentum tensor includes terms with six derivatives acting on the correlator in Eq.~(\ref{B-res}). Naively it would seem that we need to expand the exponent in Eq.~(\ref{coef-exp}) to seventh order, since each of the six derivative operators will have a piece proportional to $\partial/\partial r^i$ which could act separately on each of the six factors in the term $(\Delta\tilde\gamma(\vec x_\perp,\vec y_\perp))^6$. For example, if we differentiate six times with respect to $r^1$ we obtain an expression of the form
\be
\lim_{\vec r\to 0} \Big(\frac{\partial}{\partial r^1}\Big)^6 
\big(\Delta\tilde\gamma(\vec x_\perp,\vec y_\perp)\big)^6 
= \lim_{\vec r\to 0} \Big(\frac{\partial}{\partial r^1}\tilde\gamma(\vec x_\perp,\vec y_\perp)\Big)^6 
+ \dots ,
\ee
where the dots represent additional terms that give zero when $\vec r$ is taken to zero. However, it is easy to see from Eq.~(\ref{gamma-def-3}) that if we differentiate $\tilde\gamma$ an odd number of times with respect to $r^1$ or $r^2$, the integration over momentum variables gives zero. This means that terms with more than three factors of $\Delta\tilde\gamma$, which are operated on with a maximum of six derivatives with respect to components of $\vec r$, can be set to zero. Equivalently, we have to expand the exponential only to fourth order. All correlators and their derivatives can be obtained using the method described above. 

\section{Numerical results for the energy-momentum tensor}
\label{sec-numerical}

We remind the reader of the geometry of the collision we are considering. The two ions approach each other along the $z$-axis and collide at the origin, at time $t=0$. Post collision, the first ion moves outward along the positive $z$-axis, and the second ion moves along the negative $z$-axis. We will consider collisions with non-zero impact parameter, which we denote $b$. The displacement vector for the first ion is $\vec b_1 = (b/2,0)$ and for the second ion we use $\vec b_2 = (-b/2,0)$.  Energy and pressure are given in GeV  and lengths in fm. We use  $N_c=3$, $m=0.2$ GeV, $Q_s=2$ GeV and $g=1$, unless stated otherwise. We consider lead-lead collisions, which corresponds to mass numbers $A_1=A_2=207$, except for a few situations where we will explicitly specify different mass numbers.

\subsection{Glasma from collisions of transversally uniform nuclei}
\label{sec-infinite-nuclei}

We start the presentation of our results with the simplified case of traversally infinite and uniform nuclei. 

\begin{figure}[t]
\centering
\includegraphics[scale=0.3]{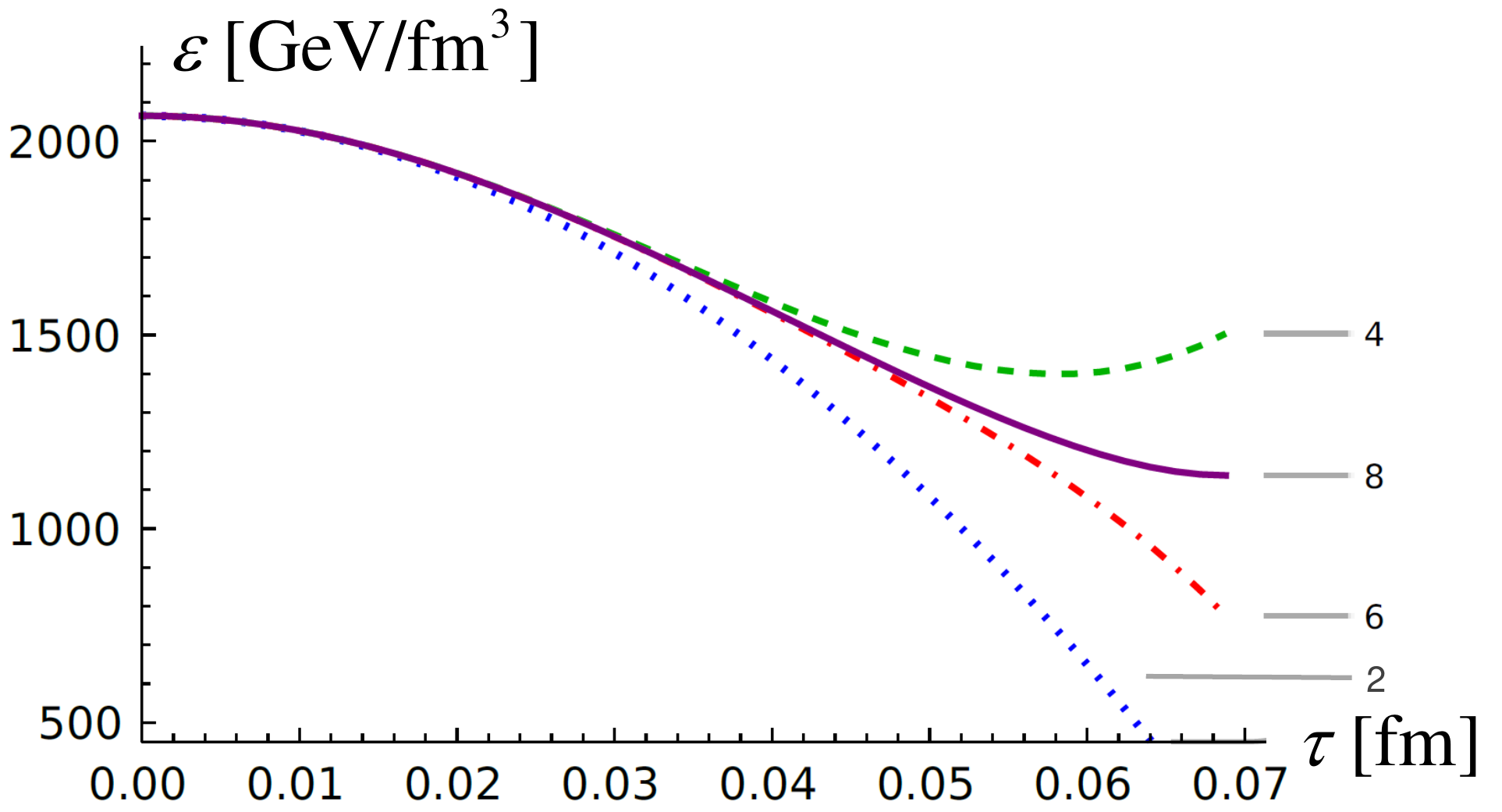}
\vspace{-1mm}
\caption{The energy density as a function of $\tau$ at $\eta=0$ at different orders in the proper time expansion.  
\label{energy-plot}}
\end{figure}

In Fig.~\ref{energy-plot} we show the energy density as a function of $\tau$ at 2nd, 4th, 6th and 8th order in the proper time expansion. The figure shows that the expansion converges well for $\tau \lesssim 0.05$ fm. The initial energy density is ${\cal E}_0$ = 2080 GeV/fm$^3$. As discussed under equation (\ref{myUbar}), the precise numerical value of the MV scale cannot be determined with a CGC approach and numerical results for quantities like the energy density should therefore be interpreted as order of magnitude estimates only. We note however that our result is not far from the estimate given in Ref.~\cite{Mrowczynski:2017kso} for the fraction of the collision energy that goes into particle production at the LHC.  

\begin{figure}
\centering
\includegraphics[scale=0.33]{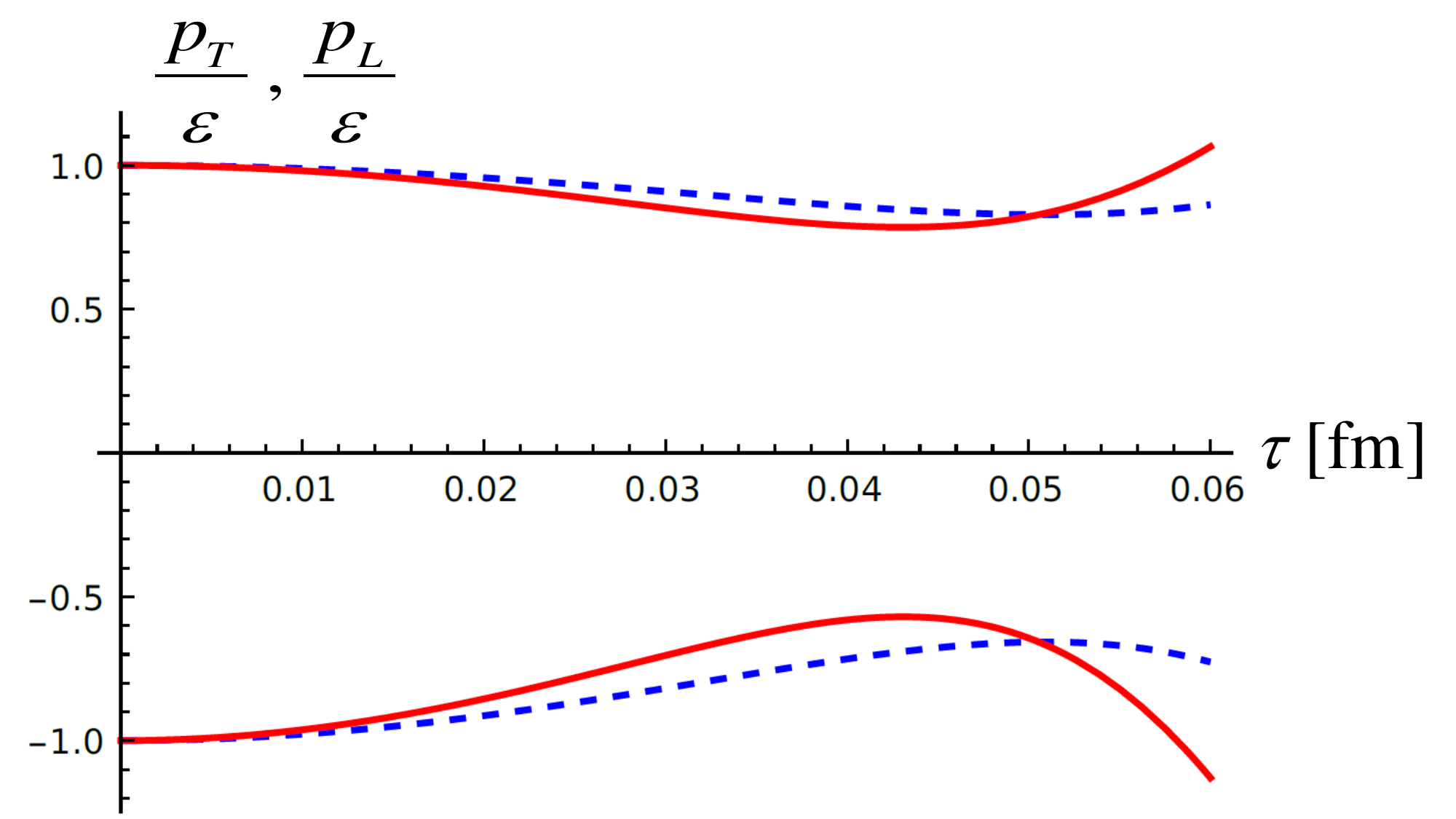}
\vspace{-2mm}
\caption{The normalized longitudinal and transverse pressures at order $\tau^4$ as functions of $\tau$ at $\eta=0$. The blue (dashed) lines show the result obtained using equation (\ref{B-short}) and the red (solid) lines are the results from equation (\ref{B-long}). In each case the lower line is $p_L/{\cal E}$ and the upper line is $p_T/{\cal E}$. \label{isot}}
\end{figure}

Next we consider the possible equilibration of the system. At $\tau=0^+$ the energy-momentum tensor has the diagonal form 
\ba
\label{diag}
T_{\rm mink}^{\rm initial} = \left(
\begin{array}{cccc}
{\cal E}_0  & 0 & 0 & 0 \\
0 & - {\cal E}_0  & 0 & 0  \\
0 & 0 & {\cal E}_0  & 0  \\
0 & 0 & 0 & {\cal E}_0  
\end{array}
\right).
\ea
The initial longitudinal pressure is large and negative. The system is therefore far from equilibrium, and also far from the regime where a quasi-particle picture would be valid. We look at the evolution of the energy density and pressure as functions of time. We define the normalized longitudinal and transverse pressures as
\be
\frac{p_L}{{\cal E}} = \frac{T_{\rm mink}^{11}}{T_{\rm mink}^{00}} 
\text{~~ and ~~} 
\frac{p_T}{{\cal E}} = \frac{1}{2}\frac{(T_{\rm mink}^{22} + T_{\rm mink}^{33})}{T_{\rm mink}^{00}} .
\ee 
If the system approaches equilibrium, the longitudinal pressure must grow as the system evolves. The energy-momentum tensor is traceless at all times ($T^\mu_{~\mu}=0$) and therefore the normalized transverse pressure must decrease as the normalized longitudinal pressure increases. 

\begin{figure}[t]
\centering
\includegraphics[scale=0.28]{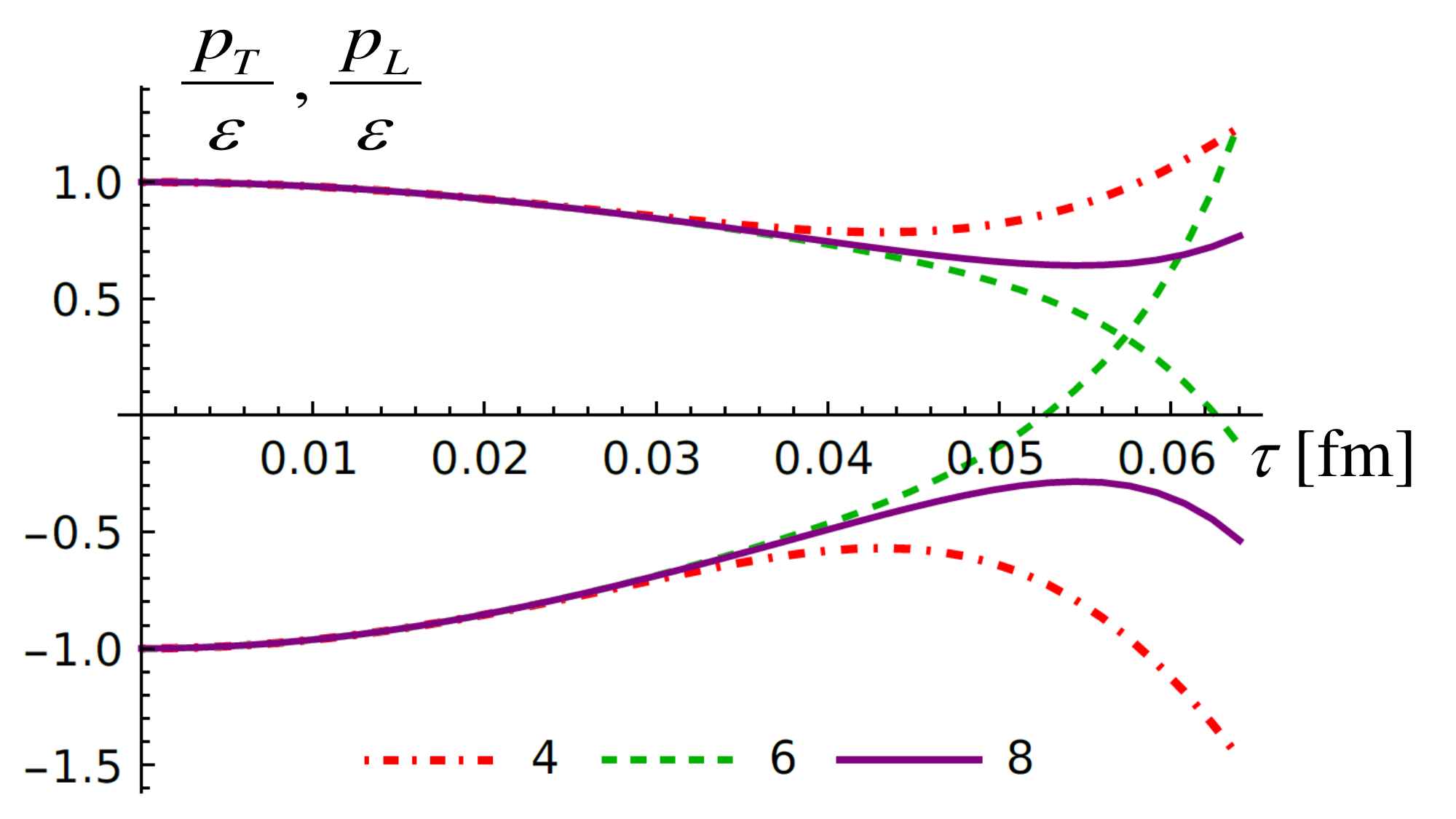}
\vspace{-2mm}
\caption{The normalized longitudinal and transverse pressures as functions of $\tau$ at $\eta=0$ computed at 4th, 6th and 8th orders in the proper time expansion. The lower lines are  $p_L/{\cal E}$ and the upper lines are $p_T/{\cal E}$.
\label{isot6}}
\end{figure}

In Fig.~\ref{isot} we show the normalized pressures to order $\tau^4$ as functions of $\tau$. The red (dashed) and blue (solid) lines are the results obtained using, respectively, equations (\ref{B-long}) and (\ref{B-short}), and the closeness of these results is an indication of the validity of the Glasma Graph Approximation. One sees that  the system starts to equilibrate, but the normalized pressures move apart again at $\tau\sim 0.05$ fm, which is consistent with the breakdown of the $\tau$ expansion observed in Fig.~\ref{energy-plot}. 

In Fig.~\ref{isot6} we show the normalized longitudinal and transverse pressures to order $\tau^4$, $\tau^6$ and $\tau^8$. One sees that the expansion breaks down at later times when the order of the expansion is increased. 

\begin{figure}[t]
\centering
\includegraphics[scale=0.55]{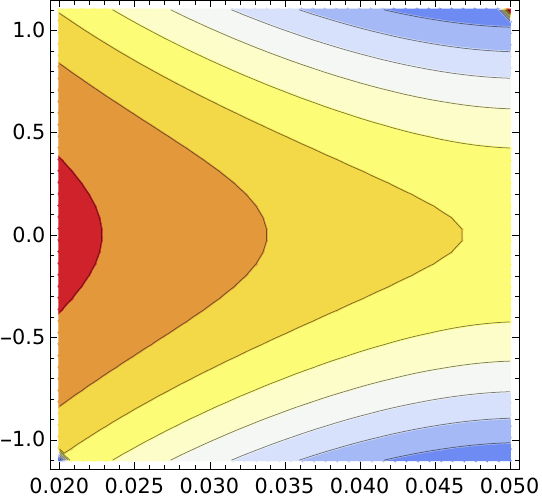}
\includegraphics[scale=0.55]{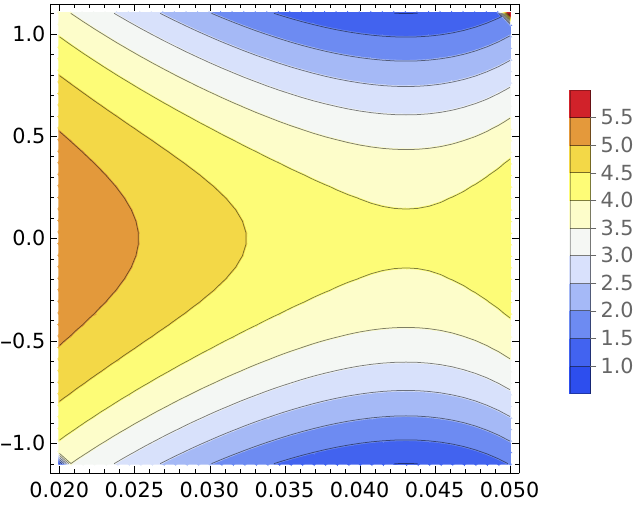}
\caption{The anisotropy measure $A_{TL}$ (\ref{AA-def}) at order $\tau^4$. The vertical axis shows $\eta$ and the horizontal axis is $\tau$. The left panel shows the result obtained using equation (\ref{B-short}), and the right panel is the result using equation (\ref{B-long}).
\label{plotA-one}}
\end{figure}

\begin{figure}[b]
\centering
\includegraphics[scale=0.55]{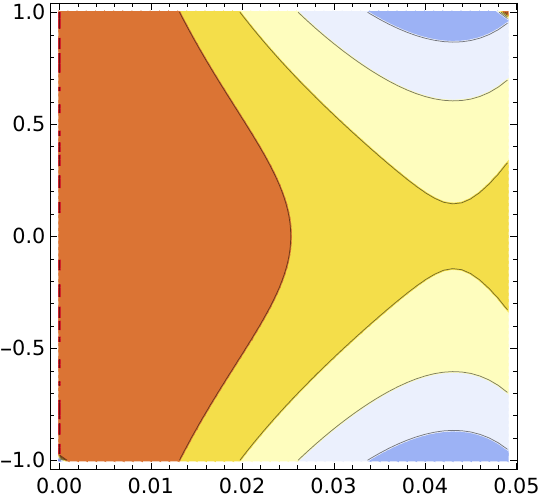}
\includegraphics[scale=0.55]{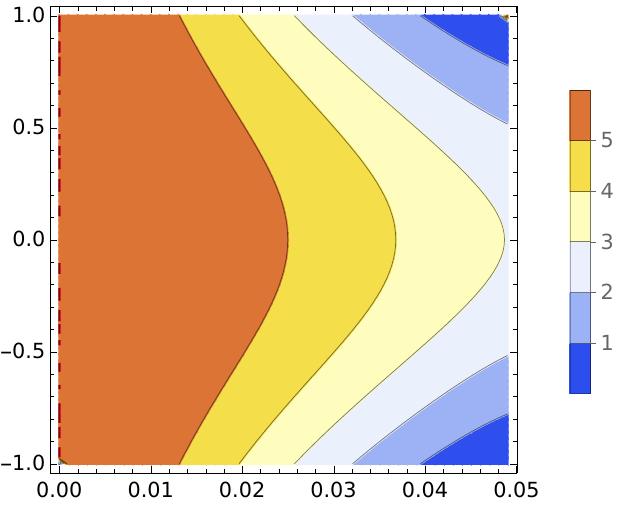}
\caption{The anisotropy measure $A_{TL}$ (\ref{AA-def}) at order $\tau^4$ (left panel) and order $\tau^8$ (right panel). The vertical axis shows $\eta$ and the horizontal axis is $\tau$. 
\label{plotA-two}}
\end{figure}

The evolution of the glasma can be studied using the anisotropy measure \cite{Jankowski:2020itt}
\be
\label{AA-def}
A_{TL} \equiv \frac{3(p_T-p_L)}{2p_T+p_L},
\ee
which takes the value $A_{TL}=6$ at $\tau=0$ (see Eq.~(\ref{diag})) and would be zero in an equilibrated plasma. 
In Fig.~\ref{plotA-one} we show $A_{TL}$ as a function of $\tau$ and $\eta$ at order $\tau^4$ using the two different results for the magnetic field in equations (\ref{B-long}) and (\ref{B-short}).  One sees that the results obtained using equations (\ref{B-long}) and (\ref{B-short}) are fairly close to each other which shows the validity of the Glasma Graph Approximation. We note that the appearance of the saddle structure in the right panel indicates the breakdown of the near field expansion. In Fig.~\ref{plotA-two} we show $A_{TL}$ at order $\tau^4$ and $\tau^8$. In the left panel we see, from the appearance of the saddle, that the fourth order calculation breaks down at $\tau\sim 0.04$ fm. The right panel shows clearly that when eighth order terms are included the region for which the expansion is valid is extended, and the system evolves significantly closer to an equilibrium state.  

\subsection{Validity of the classical approach}
\label{sec-class-approx}

Our calculation is based on a classical description and we can estimate the regime of validity of this description by looking at the constraint imposed by the uncertainty principle. The classical description requires $\Delta E \Delta t \gg 1$. Since the energy released in the collision is extremely large, as seen in Fig.~\ref{energy-plot}, the lower bound for the range of times that satisfy the constraint  will be very small, which is the idea that justifies the near field expansion. To obtain a quantitative approximation for this lower bound we estimate the initial energy as $\Delta E = {\cal E}_0 S \Delta t$ where ${\cal E}_0\approx 2000$ GeV/fm$^3$ is the initial energy density, and $S\approx 150$ ${\rm fm}^2$ is the transverse area of overlap of the colliding nuclei. From these numbers we obtain $\Delta t \gg 1/\sqrt{{\cal E}_0 S} \sim 8 \times 10^{-4}$ fm. From Figs.~\ref{energy-plot} and \ref{isot6}, and from the results presented in the next section where we consider nuclei with transverse structure,  we estimate that in our calculation the $\tau$ expansion breaks down for values $\tau \gtrsim 0.06 - 0.07$ fm. We see therefore that the region of validity of the near field expansion reaches far beyond the lower bound at which we no longer trust the classical description we are using.  

\subsection{Glasma from collisions of finite nuclei}
\label{sec-finite-nuclei}

\begin{figure}[t]
\centering
\includegraphics[scale=0.22]{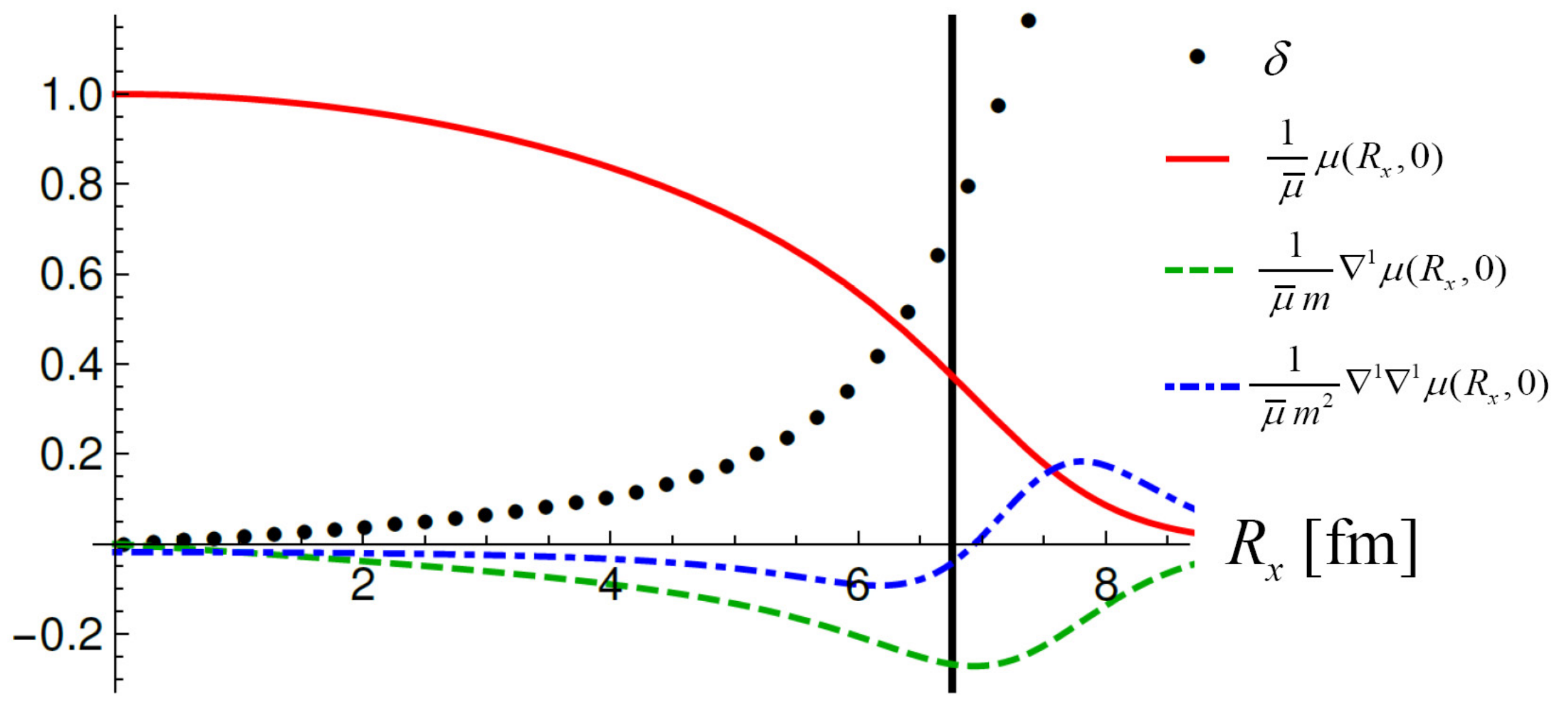}
\caption{The red (solid), green (dashed) and blue (dot-dashed) curves show the density  $\mu(\vec x_\perp)$  with $\vec x_\perp = (R_x,0)$ and its first and second derivatives. The quantity $\delta$ in equation (\ref{delta-def}) is shown by the black dots. For illustration the figure shows a vertical line that indicates the value of $R_x$ for which $\delta=0.75$. 
\label{MUd-plot}}
\end{figure}

In the previous section we presented some results for the simple case of nuclei that are infinite and uniform in the transverse plane. In this section we consider more physically realistic collisions of finite nuclei. The surface density of colour charges is no longer constant but depends on the transverse coordinates. The function $\mu(\vec x_\perp)$ which is chosen in the form of a projected Woods-Saxon distribution (\ref{def-mu2-2}). We use $r_0=1.25$ fm and $a=0.5$ fm so that the radius of a nucleus with $A=207$ is $R_A=r_0 A^{1/3}= 7.4$ fm. 

We allow for non-homogeneous nuclear densities of colliding nuclei by performing a gradient expansion around the coordinate that gives the position of the center of each nucleus in the transverse plane. In Fig.~\ref{MUd-plot} we show the density $\mu(\vec x_\perp)$ with $\vec x_\perp = (R_x,0)$, its first and second derivatives with respect to $R_x$, and the quantity $\delta$ in Eq.~(\ref{delta-def}) which must be small for the gradient expansion to converge. The condition $\delta < 0.75$ is satisfied in the region to the left of the vertical line in the figure.

Fig.~\ref{MUd-plot} shows clearly that the derivatives of the density function are appreciable only in a very small region at the edges of the nucleus. This means that if we calculate a quantity for which the dominant part of the integrand is not close to the edges of the nuclei, the gradient expansion will converge well, but the contributions of the derivative terms will likely be so small that they are negligible. On the other hand, if we calculate a quantity for which the region of the transverse plane close to the edges of the nuclei is important, the contribution from the derivative terms can be large and the convergence of the gradient expansion must be studied carefully.

In the rest of this section  we study several quantities that involve an integration over the transverse plane. These include: transverse pressure anisotropy, in Sec.~\ref{sec-energy-density}, the Fourier coefficients of the flow, in Sec.~\ref{sec-Fourier}, and the angular momentum of the glasma, in Sec.~\ref{sec-angular}. These calculations are potentially sensitive to the gradient expansion and results are reliable if they are largely insensitive to the choice of the integration limits. This condition restricts us to the consideration of fairly small impact parameters. The reason is that when the centers of the two ions are separated, the inner edge of the first/second ion, where the density changes rapidly, will be closer to the center of the second/first ion, where integrand can be large.

\subsubsection{Energy density and pressures}
\label{sec-energy-density}

We look at the initial energy density ${\cal E} = T_{\rm mink}^{00}$ at mid-spatial-rapidity ($\eta=0$) for four different configurations of the colliding ions which are defined in Table \ref{config-table}. 
The last row of the table shows the maximum initial energy density. 

\begin{table}[t]
\centering 
\begin{tabular}{|c | c | c|c|c|c|c|c|c|c|c|} 
\hline                        
~~~~~~~~ & ~~~~ $A$ ~~~~ &~~~~ $B$ ~~~~ &~ ~~ ~ $C$ ~~~~ &~ ~~~ $D$ ~~~~   \\ [0.5ex]   
\hline
$A_1$ & 207 & 207 & 207 & 207  \\
$A_2$ & 207 & 207 & 40  & 40   \\
$b_1/2$ & 0 & 3 &  3 & 0  \\
$b_2/2$ & 0 &-3 & -3 & 0     \\
\hline
 ${\cal E}^{\rm max}_0$  GeV/fm$^3$ &  2080 & 1715 & 722 & 1202 \\  
\hline
\end{tabular}
\caption{Four configurations of colliding ions}
\label{config-table}
\end{table}

\begin{figure}[t]
\centering
\includegraphics[scale=1.3]{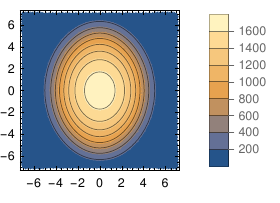} 
\includegraphics[scale=1.3]{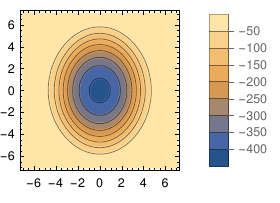} 
\caption{Energy densities in the transverse plane for case $B$. The left panel shows the energy density at $\tau=0$ and the right panel shows the difference between the energy densitiy at $\tau=0.04$ fm and the initial energy density, at sixth order in the proper time expansion. The units are GeV/fm$^3$ and the axes show $R_x$ and $R_y$ in fm. \label{fig-alm} }
\end{figure}

In Fig.~\ref{fig-alm} we show the initial energy density and the difference between the energy density at $\tau=0.04$ fm and the initial energy density at sixth order in the proper time expansion for case $B$. The energy density drops fastest at the centre and more slowly at the edges of the almond shaped interaction region. 

When we calculate the energy density we restrict to the region of the transverse plane for which $-5 \text{~fm} <|\vec R|<  5 \text{~fm}$ and the gradient expansion converges well. Within this region, the inhomogeneity of the energy density in the transverse plane is almost entirely due to the asymmetry created by the non-zero impact parameter, which produces an almond shaped region of overlap. The gradients of the individual charge distributions are small and mostly irrelevant. 

\begin{figure}[b]
\centering
\includegraphics[scale=0.3]{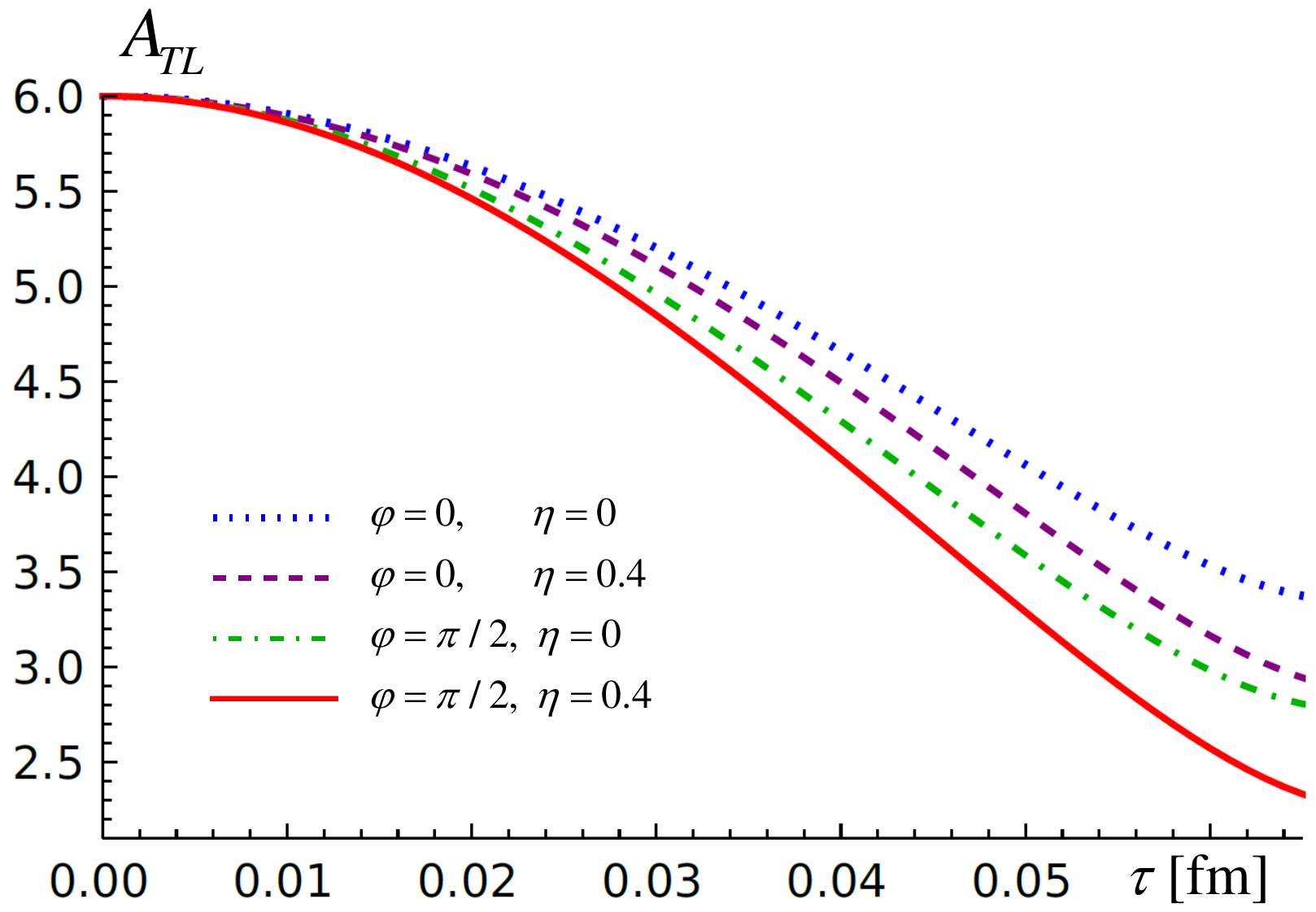} 
\vspace{-1mm}
\caption{The anisotropy measure $A_{TL}$ defined in Eq.~(\ref{AA-def}) at $R=5$ fm and $b=6$ fm.
\label{MartB}}
\end{figure}

The anisotropy measure defined in Eq.~(\ref{AA-def}) was shown in Fig.~\ref{plotA-two} for the case of uniform nuclei. The figure shows that $A_{TL}$ decreases as $\tau$ grows, up until the point at which the proper time expansion breaks down. When colliding nuclei are finite and of varying density we can study how the behaviour of $A_{TL}$ depends on azimuthal angle (denoted $\phi$), spatial rapidity, and impact parameter. As expected, $A_{TL}$ moves towards the equilibrium value more quickly when the impact parameter is smaller, and the region where the two ions overlap is greater. In Fig.~\ref{MartB} we show the measure $A_{TL}$ at eighth order in the proper time expansion  as a function of $\tau$, for different values of $\eta$ and $\phi$.  We consider $\phi=0$, which corresponds to $\vec R$ in the reaction plane, and $\phi=\pi/2$, where $\vec R$ is perpendicular to the reaction plane. The graph shows that $A_{TL}$ drops more quickly when either the azimuthal angle or the spatial rapidity increases. 

\begin{figure}[t]
\centering
\includegraphics[scale=0.385]{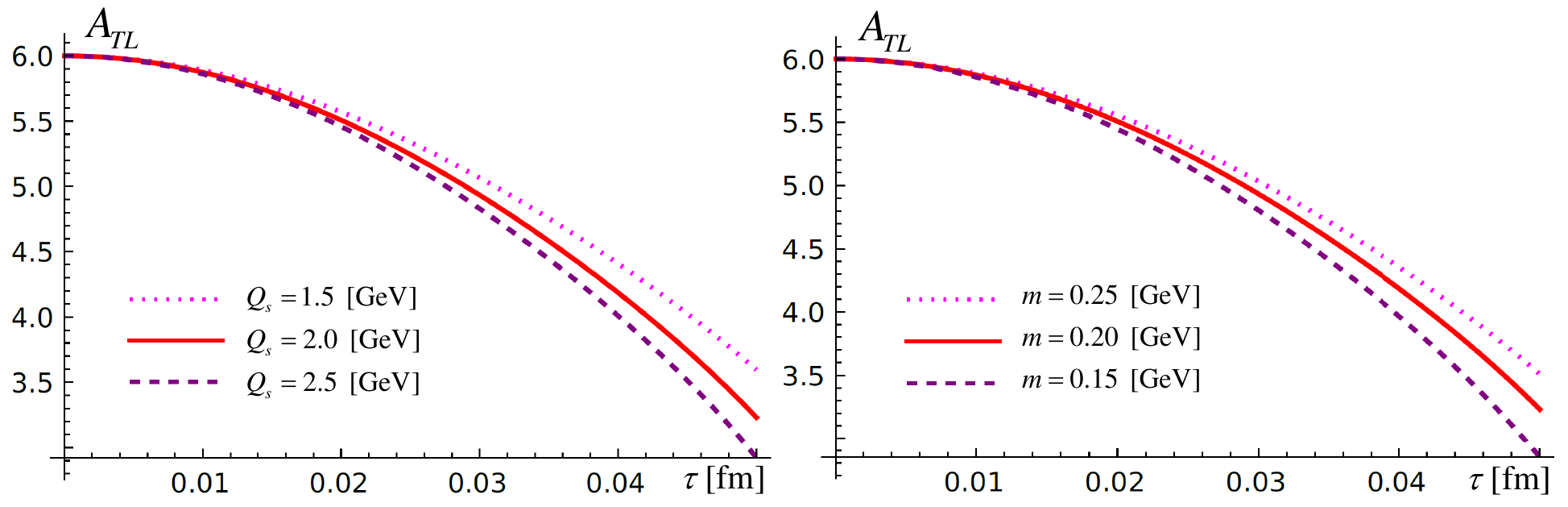} 
\caption{The anisotropy $A_{TL}$ with $R=5$ fm, $b=0$ and $\eta=0$ at order $\tau^6$ for three different values of the saturation scale $Q_s$ (left panel) and mass parameter $m$ (right panel). \label{Qs_mass}}
\end{figure}

We can also use the anisotropy measure $A_{TL}$ to demonstrate that our results are not strongly dependent on the UV and IR scales that enter the calculation ($Q_s$ and $m$ in our notation). This is important because the exact values of these scales are not known, and also because the way they enter the calculation depends on the method chosen to perform the regularization. In most of our calculations we have used $Q_s=2.0$ GeV and $m=0.2$ GeV.  In Fig.~\ref{Qs_mass} we show $A_{TL}$ at order $\tau^6$ as a function of time for three different values of $Q_s$ with $m=0.2$ GeV (left panel) and for three different values of $m$ with $Q_s=2.0$ GeV (right panel). The graphs show that within the range of validity of the $\tau$ expansion, the dependence on the value of these scales is weak. 

We can also study the anisotropy of the glasma in the transverse plane. We consider the quantity \cite{Krasnitz:2002ng}
\be
\label{vareps}
\{A_{xy}\} \equiv \frac{\{T^{yy}-T^{xx}\}}{\{ T^{xx}+T^{yy}\}} ,
\ee
where the angular brackets indicate integration over the transverse plane. For comparison we also calculate 
\be
\label{AA-int}
\{ A_{TL} \}  \equiv \frac{3\{ p_T-p_L \}}{\{ 2p_T+p_L \} }.
\ee

The leading-order contribution to $\{A_{xy}\}$ comes from the first order term in the gradient expansion and therefore this quantity, in contrast to $\{A_{TL}\}$, will be sensitive to the region of the transverse plane that is close to the edges of the nuclei. We must verify that the integral is largely independent of the upper limit that is used to perform the two dimensional integration over the transverse plane, which we call $R_{\rm max}$. 

\begin{figure}[t]
\centering
\includegraphics[scale=0.3]{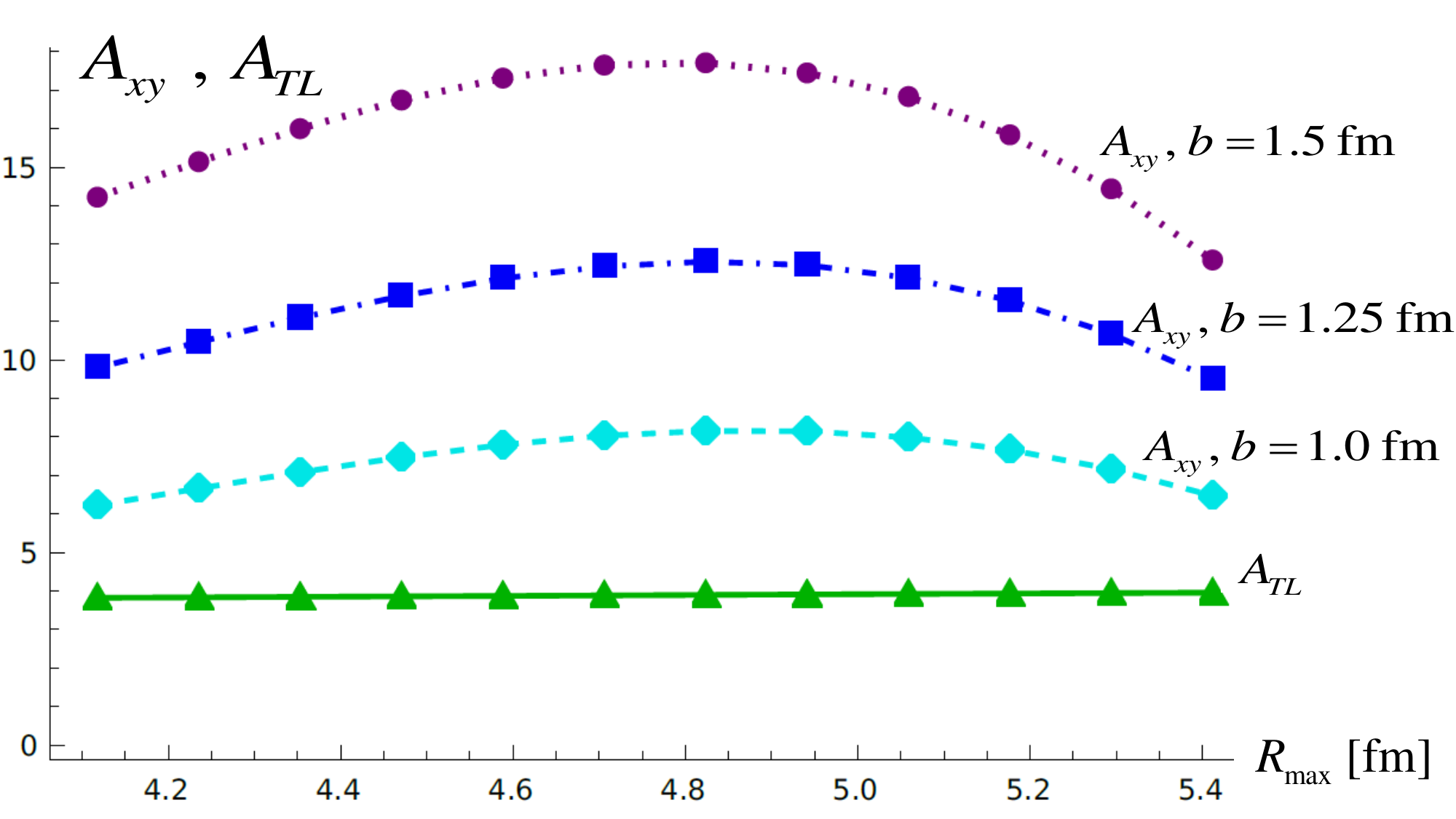} 
\caption{Results for $\{A_{TL}\}$ and $7 \times 10^5 \times \{A_{xy}\}$  at $\tau=0.04$ fm and $\eta=0$ as functions of $R_{\rm max}$ for three different impact parameters. All values of $A_{xy}$ have been multiplied by $7.5\times 10^5$. The green line shows $A_{TL}$ which is much bigger and almost completely insensitive to impact parameter.  \label{Ahat-plot}}
\end{figure}

In Fig.~\ref{Ahat-plot} we show $\{A_{xy}\}$ at $\tau=0.04$ fm for three different values of impact parameter, as a function of $R_{\rm max}$\footnote{There was a misprint in the caption of figure 10 of Ref. \cite{Carrington:2021qvi}, where the blue line was labelled incorrectly (it was multiplied by $7\times 10^4$ and not $7\times 10^5$).}. On the same graph we show the result for $\{A_{TL}\}$ at $\tau=0.04$ fm and $\eta=0$. For $\{A_{TL}\}$ the change with impact parameter is too small to be seen on the graph, and the result is almost six orders of magnitude larger than $\{A_{xy}\}$ and nearly completely independent of $R_{\rm max}$. Figure \ref{Ahat-plot} shows that as long as the integration over the transverse plane is restricted to a fairly central region, results for the pressure anisotropy are largely insensitive to the upper limit of the integration. 

\subsubsection{Radial flow}
\label{sec-flow}

\begin{figure}[t]
\begin{center}
\includegraphics[width=8cm]{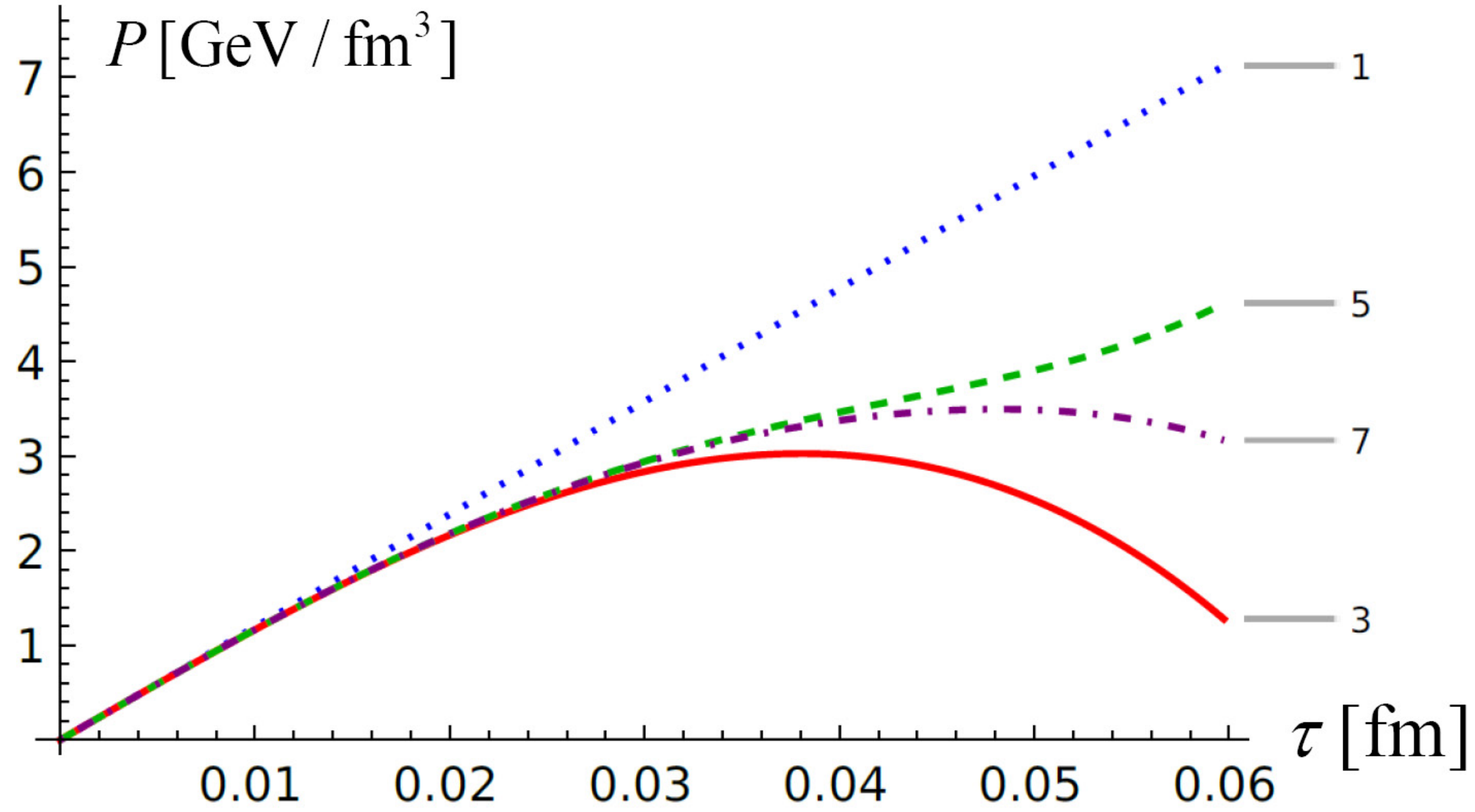}
\end{center}
\vspace{-4mm}
\caption{Radial flow to seventh order in the proper time expansion at $R=3$ fm and $\phi=\pi/2$ in collisions with $b=6$ fm.}
\label{fig-P-time}
\end{figure}

\begin{figure}[b]
\begin{center}
\includegraphics[width=8cm]{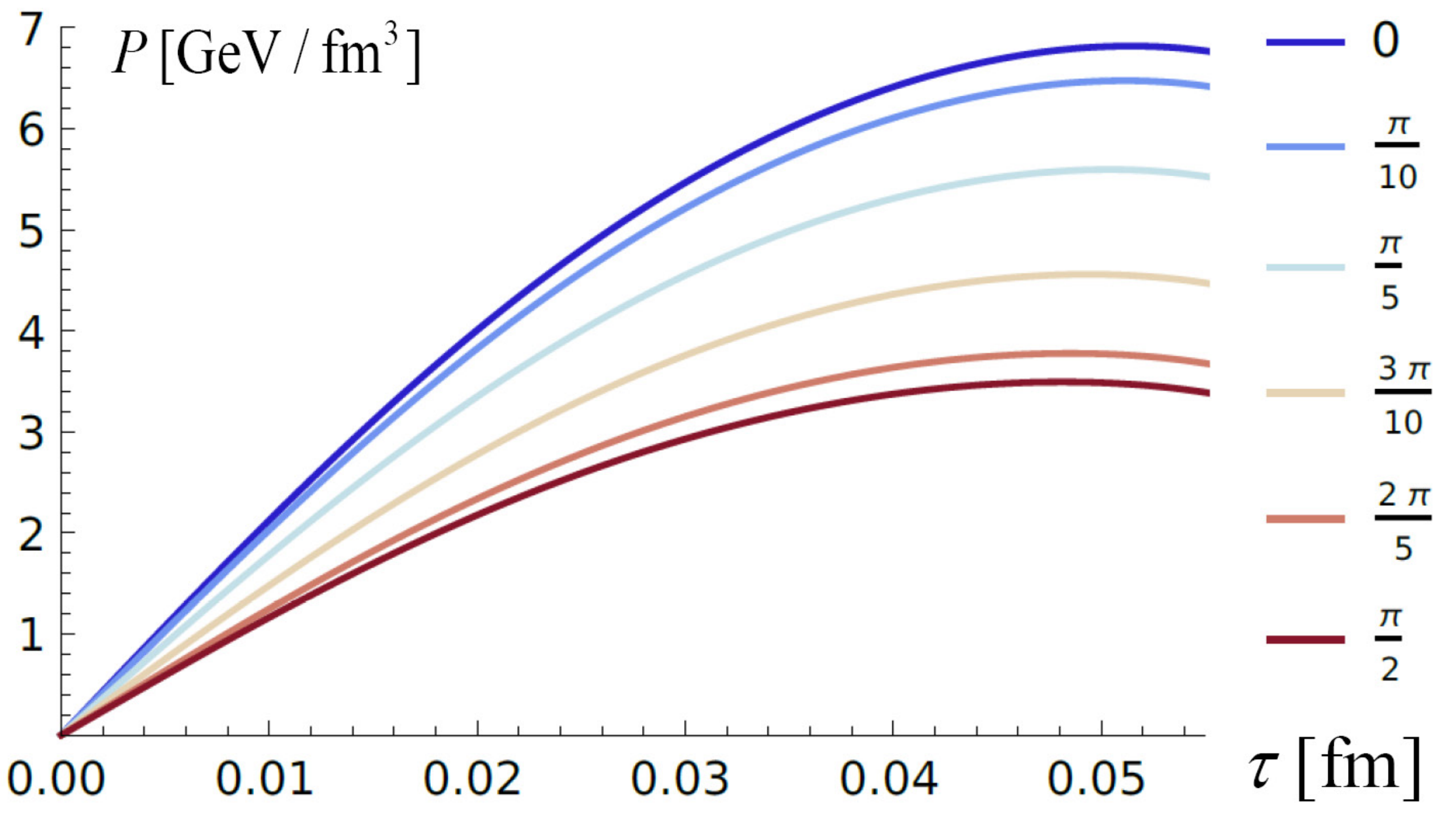}
\end{center}
\vspace{-4mm}
\caption{Radial flow to seventh order in the proper time expansion at $R=3$ fm for a range of azimuthal angles $\phi$ in collisions with $b=6$ fm.}
\label{fig-P-time-phi}
\end{figure}

To characterize the radial flow of the expanding glasma we compute the radial projection of the transverse Poynting vector $P \equiv \hat R^i T^{i0}$ where $\hat R^i \equiv R^i/|\vec{R}|$. In Fig.~\ref{fig-P-time} we show this quantity for fairly peripheral collisions with $b=6$ fm at $R=3$ fm and $\phi=\pi/2$, at different orders in the $\tau$ expansion. One observes that at seventh order our result for radial flow can be trusted to $\tau \lesssim 0.06$ fm. Figure \ref{fig-P-time-phi} shows the same quantity $P$ at $R=3$ fm for a range of azimuthal angles $\phi$ in collisions with $b=6$ fm. The flow is seen to be significantly stronger in the reaction plane ($\phi=0$) than in the direction perpendicular to it ($\phi=\pi/2$).

\begin{figure}
\begin{center}
\includegraphics[width=5.3cm]{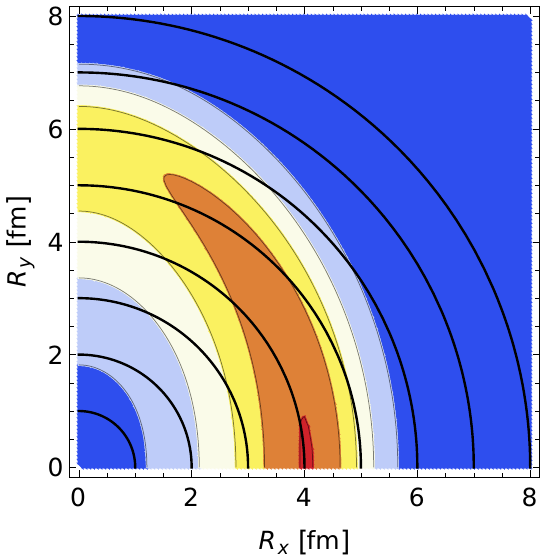}
\includegraphics[width=6.13cm]{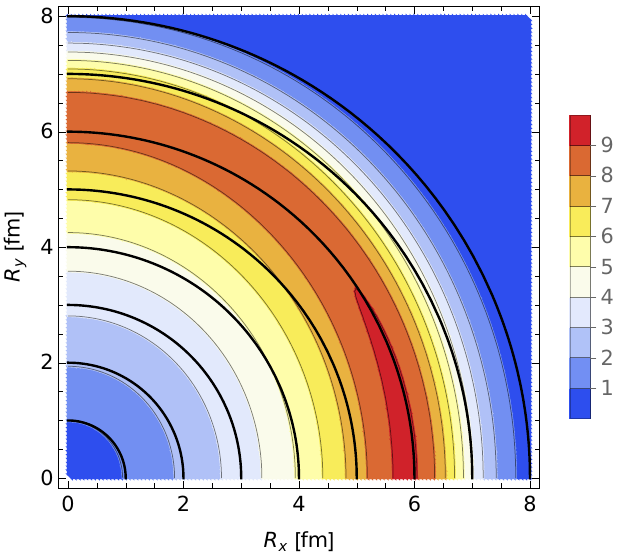}
\end{center}
\vspace{-3mm}
\caption{Radial flow in the transverse plane at $\tau = 0.05$ fm at seventh order in the proper time expansion  in collisions with $b=6$ fm  (left panel) and $b=2$ fm (right panel). The black curves mark lines of constant radius. See text for further explanation.}
\label{fig-P-trans}
\end{figure}

When the impact parameter is non-zero, we expect that the radial flow in the plane transverse to the beam direction will not be azimuthally symmetric. In our coordinate system the $x$-$y$ plane is transverse to the beam axis, and we always choose the  impact parameter along the $x$-axis. The left panel of Fig.~\ref{fig-P-trans}  shows the radial flow of the glasma for a peripheral collision with $b = 6$ fm, and the right panel is a more central collision with $b = 2$ fm. The flow is greater in the $x$ than in the $y$ direction, as expected, up to $R \approx 5$ fm in the peripheral collision and up to $R  \approx 7$ fm in the more central collision. At bigger distances there is a slight increase in the radial flow at larger azimuthal angles, but since the gradient expansion is not reliable at distances comparable to the nuclear radii, the accuracy of the calculation is lower in this region. The effect is difficult to see from the figures and the black arcs that represent quarter circles are intended to make it more easily visible.

\subsubsection{Fourier coefficients of azimuthally asymmetric flow}
\label{sec-Fourier}

The azimuthal asymmetry of the collective flow is usually quantified in terms of Fourier coefficients $v_1,v_2,v_3 \dots$. In Appendix~\ref{sec-phi-distribution} we define these coefficients treating the transverse Poynting vector $(T^{0x},T^{0y})$ as the transverse momentum of a final-state particle.  Below we discuss only the elliptic flow coefficient, $v_2$, and the eccentricity of the energy density distribution, $\varepsilon$, which are defined as
\be
v_2 = \frac{\int d^2 R \, \frac{T_{0x}^2 - T_{0y}^2}{\sqrt{T_{0x}^2+T_{0y}^2}}}
{\int d^2 R \, \sqrt{T_{0x}^2+T_{0y}^2}}
\text{~~~~~and~~~~~} 
\varepsilon = - \frac{\int d^2 R\,  \frac{R_x^2-R_y^2}{\sqrt{R_x^2+R_y^2}} \,T^{00}}
{\int d^2 R\, \sqrt{R_x^2+R_y^2} \; T^{00}} \,. 
\ee

\begin{figure}[t]
\begin{center}
\includegraphics[width=9cm]{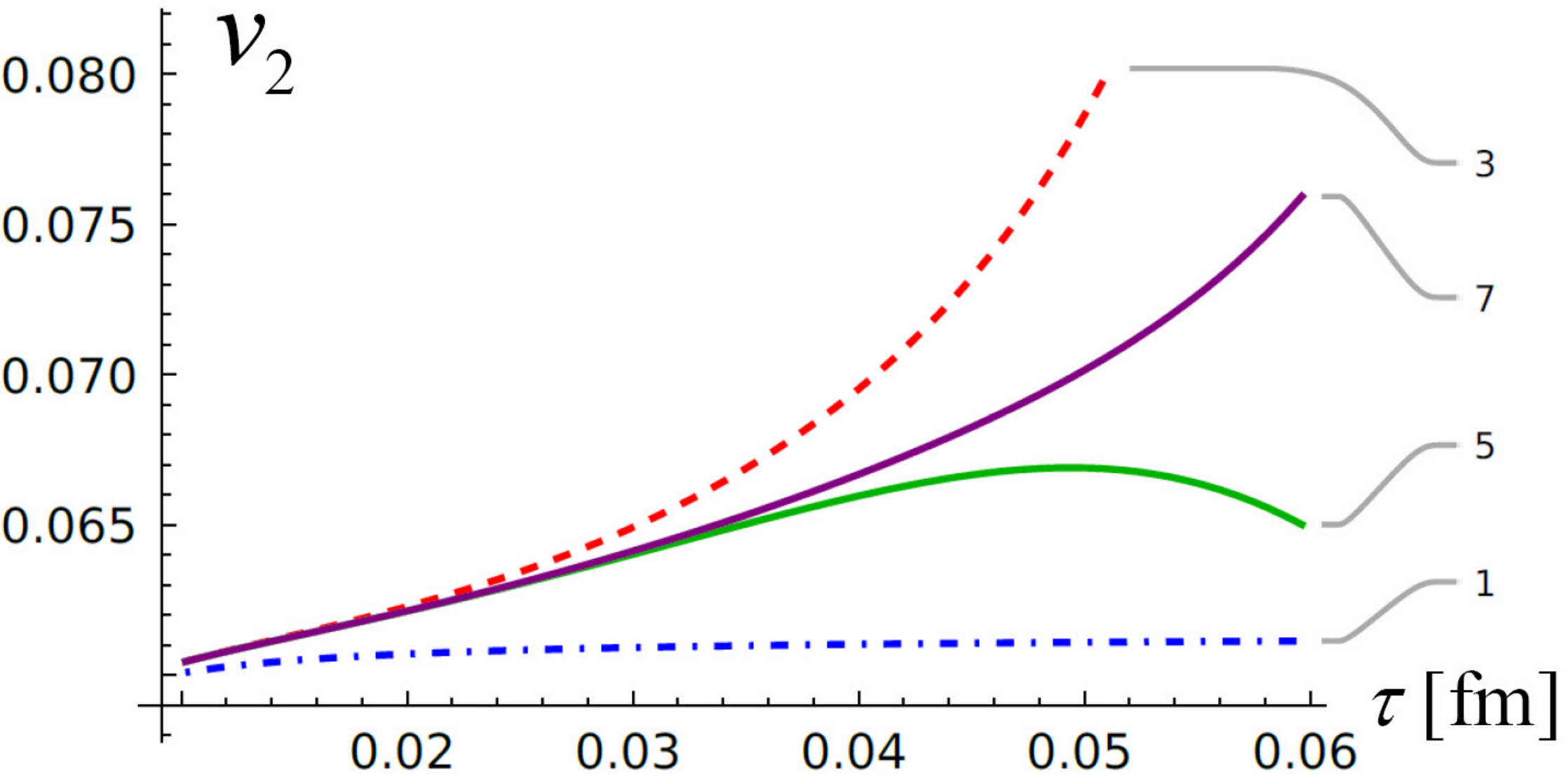}
\end{center}
\vspace{-4mm}
\caption{Elliptic flow coefficient $v_2$ versus proper time at different orders in the proper time expansion in collisions with $b=2$ fm. }
\label{fig-v2-vs-tau}
\end{figure}

In Fig.~\ref{fig-v2-vs-tau} we show the coefficient $v_2$ as a function of $\tau$ at orders one, three, five and seven of the expansion, for collisions with impact parameter $b=2$ fm. The coefficient $v_2$ is constant in time at first order in the expansion, since both the numerator and denominator are linear in $\tau$. The seventh order result clearly shows that $v_2$ does not saturate at $\tau \gtrsim 0.05$ fm, as the fifth order result might suggest, but continues to grow with time.  We note that the calculation of $v_2$ at very small times is numerically difficult because the numerator and denominator both approach zero as $\tau\to 0$. The numerical values of $v_2$ shown in Fig.~\ref{fig-v2-vs-tau} are of the same order as experimental values \cite{Heinz:2013th}. This is surprising as one expects that collective flow develops mostly during the hydrodynamic evolution that follows the glasma phase.  

\begin{figure}
\begin{center}
\includegraphics[width=10cm]{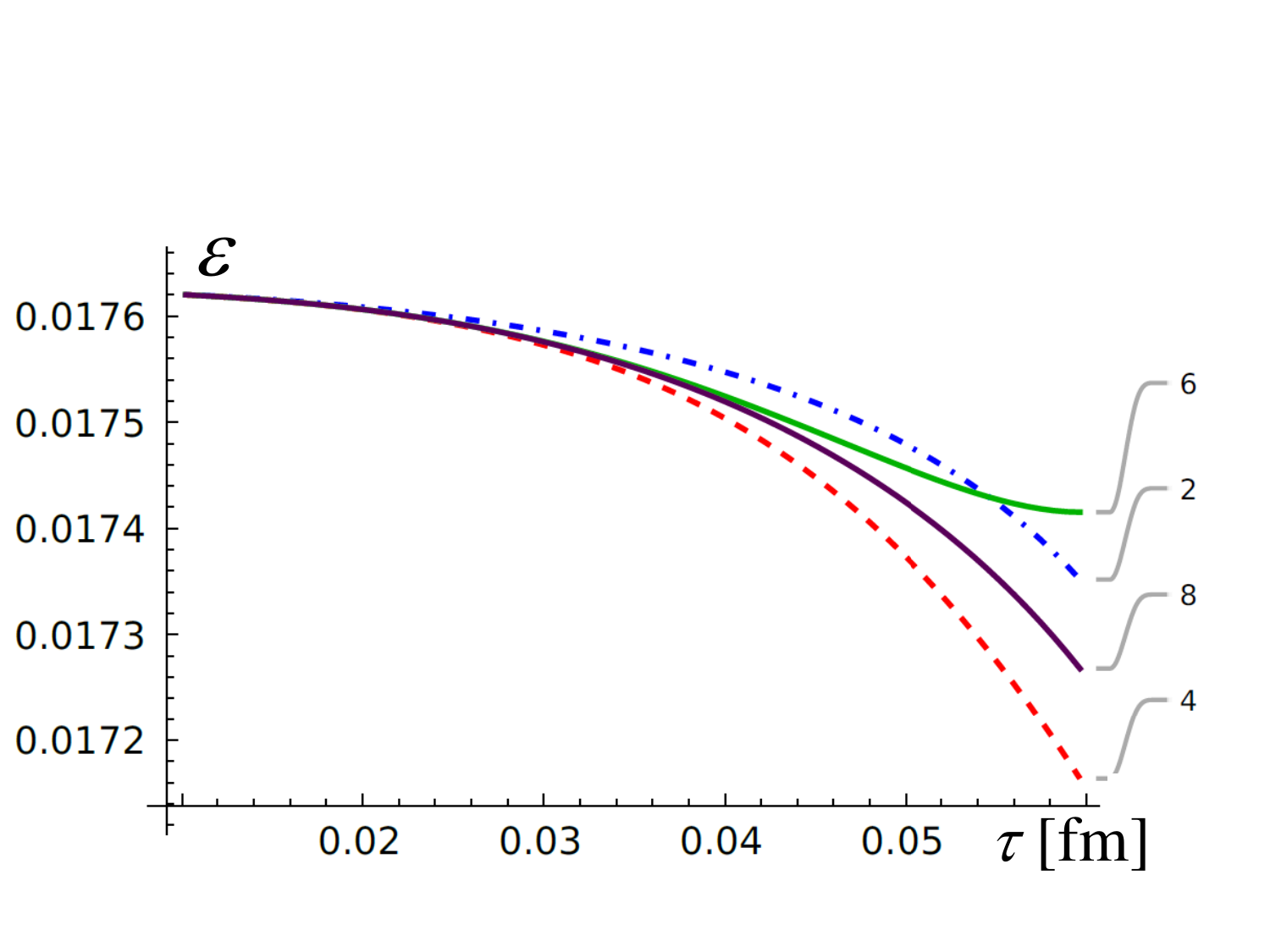}
\end{center}
\vspace{-7mm}
\caption{Eccentricity $\varepsilon$ versus proper time at different orders in the proper time expansion in collisions at $b=2$ fm.}
 \label{fig-e2-vs-tau}
\end{figure}

It is usually assumed that the experimentally observed azimuthal aniso-tropy in momentum space of a hadronic final state is caused by the azimuthal anisotropy in coordinate space of the energy density and pressure of the initial state. Physically the idea is that the final state momentum anisotropy is generated by pressure gradients. To investigate if this behaviour is seen in our calculation, we have computed the eccentricity $\varepsilon$ as a function of $\tau$ at orders two, four, six and eight of the expansion, for collisions with impact parameter $b=2$ fm. The results are presented in Fig.~\ref{fig-e2-vs-tau} which together with Fig.~\ref{fig-v2-vs-tau} show that the collective elliptic flow increases in time while the spatial eccentricity decreases, which resembles hydrodynamical behaviour even though the glasma is far from a local equilibrium state. One also sees also that the eccentricity changes much more slowly than the elliptic flow coefficient.

\begin{figure}[b]
\begin{center}
\includegraphics[width=8.0cm]{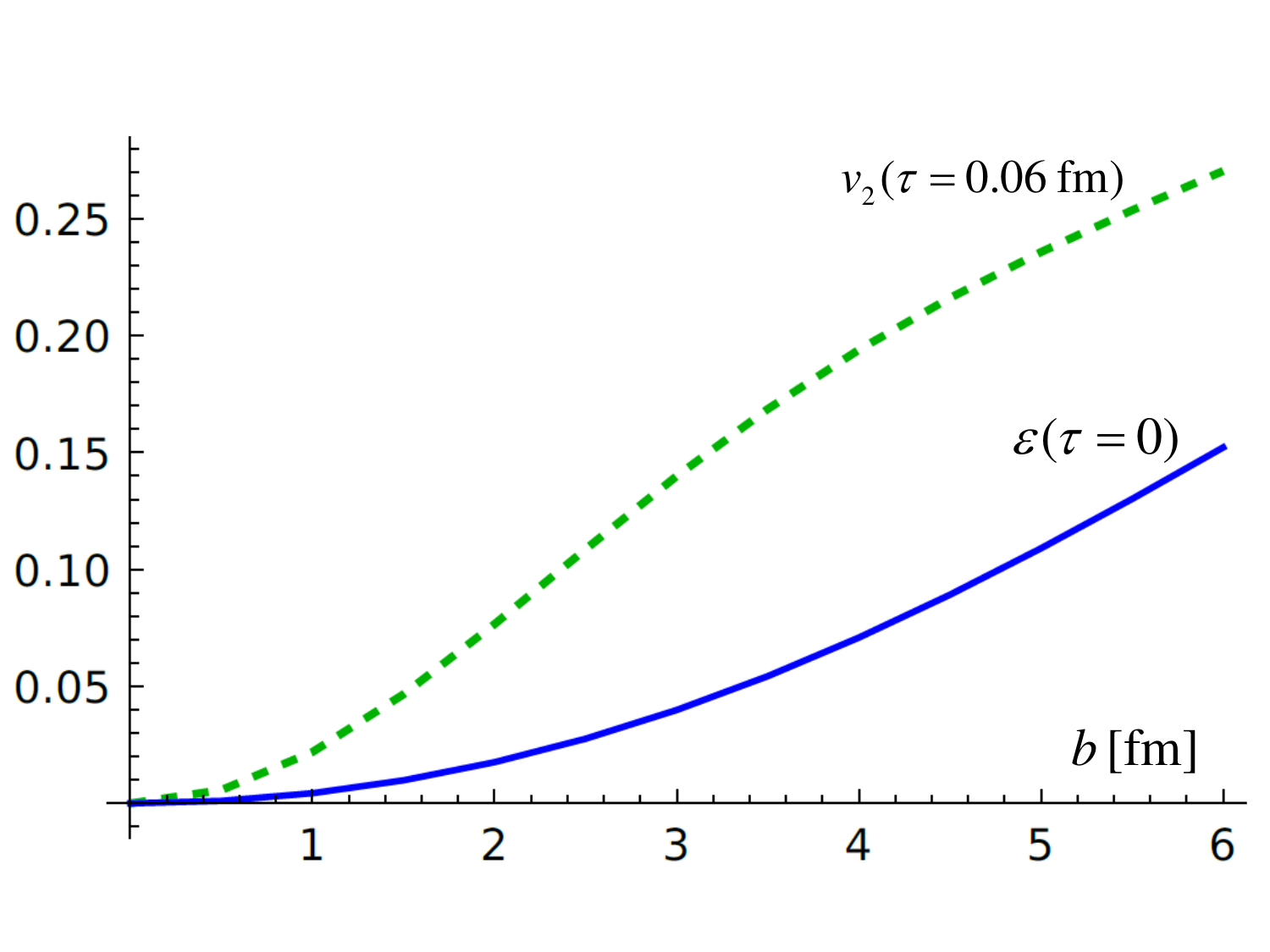}
\end{center}
\vspace{-11mm}
\caption{$v_2$ at $\tau=0.06$ fm and $\varepsilon$ at $\tau=0$ versus impact parameter.}
 \label{fig-v2-and-e2}
\end{figure}

\begin{figure}[t]
\begin{center}
\includegraphics[width=7.5cm]{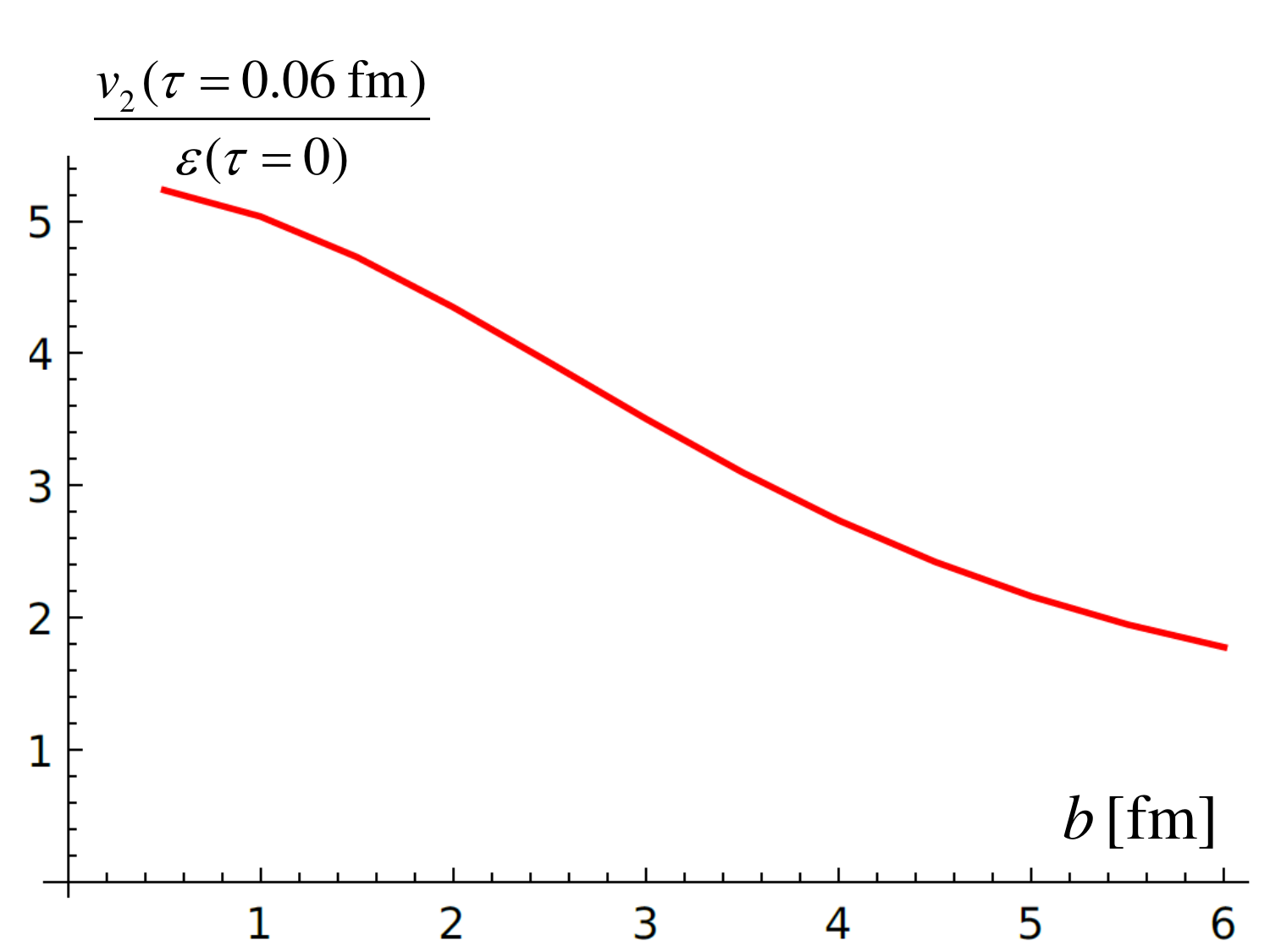}
\end{center}
\vspace{-5mm}
\caption{Ratio of $v_2$ at $\tau=0.06$ fm over $\varepsilon$ at $\tau=0$ versus impact parameter.}
 \label{fig-v2:e2}
\end{figure}

It is interesting to consider the dependence of the glasma elliptic flow on the system's initial eccentricity, which in turn depends on impact parameter. We show in Fig.~\ref{fig-v2-and-e2} the coefficient $v_2$ at $\tau=0.06$ fm computed at seventh order of the proper time expansion and the initial eccentricity, both as  functions of impact parameter. In Fig.~\ref{fig-v2:e2} the coefficient $v_2$ is divided by the initial eccentricity $\varepsilon$. One sees that the relative change in $v_2$ when the impact parameter grows from 1 fm to 6 fm is much greater than the relative change in the ratio $v_2/\varepsilon$. This behaviour indicates that  the initial spatial asymmetry of the glasma is transmitted to the momentum asymmetry of the system, which mimics the behaviour of hydrodynamics. 

\subsubsection{Angular momentum}
\label{sec-angular}

A system of relativistic heavy ions colliding at a finite impact parameter has initially a huge angular momentum perpendicular to the reaction plane. The value of the initial angular momentum carried by the nucleons that will participate in the collision is of order $10^5$ at maximum RHIC energies \cite{Gao:2007bc,Becattini:2007sr} and even larger at LHC energies. We would like to know how much of this initial angular momentum is transferred to the glasma that is produced in the collision. Since the glasma in our approach is boost invariant, we cannot compute the total angular momentum of the system which, strictly speaking, extends in rapidity from minus to plus infinity. Instead we compute the angular momentum per unit rapidity which can be obtained in a way similar to that of Ref.~\cite{Fries:2017ina}. 

One defines the tensor 
\be
\label{M-def}
M^{\mu\nu\rho}=T^{\mu\nu}R^\rho - T^{\mu\rho}R^\nu ,
\ee
where $R^\mu$ denotes a component of the position vector. The energy-momentum tensor is divergenceless and therefore $\nabla_\mu M^{\mu\nu\rho} = 0$. Using Stokes' theorem one obtains a set of six conserved quantities
\be
\label{J1}
J^{\nu\rho} = \int_\Sigma d^3 y \sqrt{|\gamma|} \, n_\mu M^{\mu\nu\rho},
\ee
where $n^\mu$ is a unit vector perpendicular to the hypersurface $\Sigma$, $\gamma$ is the induced metric on this hypersurface, and $d^3y$ is the corresponding volume element. The angular momentum is obtained from the Pauli-Lubanski four-vector 
\be
\label{L-def}
L_\mu = -\frac{1}{2}\epsilon_{\mu\nu\rho\sigma}J^{\nu\rho}u^\sigma ,
\ee
where $u^\mu$ is the four-vector that denotes the rest frame of the system. 

As a check one can easily verify that Eqs.~(\ref{M-def}, \ref{J1}, \ref{L-def}) reduce to the usual definition of angular momentum in Minkowski space. We denote indices for spatial variables in Minkowski space with Greek characters from the beginning of the alphabet, for example $\alpha \in(1,2,3)$ and $x^\alpha$ is a component of the vector $(x,y,z)$. We use $n^\mu = (1,0,0,0)$ so that $\Sigma$ is a hypersurface of constant $t$ and work in the rest frame with $u^\mu = (1,0,0,0)$. Equation (\ref{L-def}) becomes
\be
\label{L-mink}
L^\alpha_{\rm mink} 
= -\frac{1}{2}\epsilon^{\alpha\beta\gamma}J^{\beta\gamma} 
= -\frac{1}{2} \epsilon^{\alpha\beta\gamma} 
\int d^3 \vec x \,(T^{0\beta}x^\gamma - T^{0\gamma}x^\beta) 
= \epsilon^{\alpha\beta\gamma} \int d^3 \vec x \, x^\beta P^\gamma ,
\ee
where $d^3 \vec x$ represents the spatial volume element in Minkowski space, and we have written the Poynting vector $P^\alpha\equiv T^{0\alpha}$.

In Milne coordinates we again use $n^\mu = (1,0,0,0)$ so that the tensor
\be
J^{\nu\rho} = \tau \int d\eta\,d^2\! \vec R \, M^{0\nu\rho}
\ee
is defined on a hypersurface of constant $\tau$. In the rest frame determined by the four-velocity $u^\mu = (1,0,0,0)$ one finds
\be
\label{L-def-2}
L_\mu = \frac{1}{2}\tau \, \epsilon_{0\mu\nu\rho} 
\int d\eta \int d^2\! \vec R \, (T^{0\nu}R^\rho - T^{0\rho}R^\nu).
\ee

Since the system is boost invariant we consider the angular momentum per unit rapidity and Eq.~(\ref{L-def-2}) leads to 
\be
\label{L-def-3}
\frac{dL_\mu}{d\eta} = \frac{1}{2}\tau \, \epsilon_{0\mu\nu\rho} 
\int d^2\! \vec R \, (T^{0\nu}R^\rho - T^{0\rho}R^\nu).
\ee
We note that although the right side of  equation (\ref{L-def-3}) is independent of rapidity, our calculation is only meaningful close to mid-rapidity where boost invariance is a good approximation.

\begin{figure}
\begin{center}
\includegraphics[width=9cm]{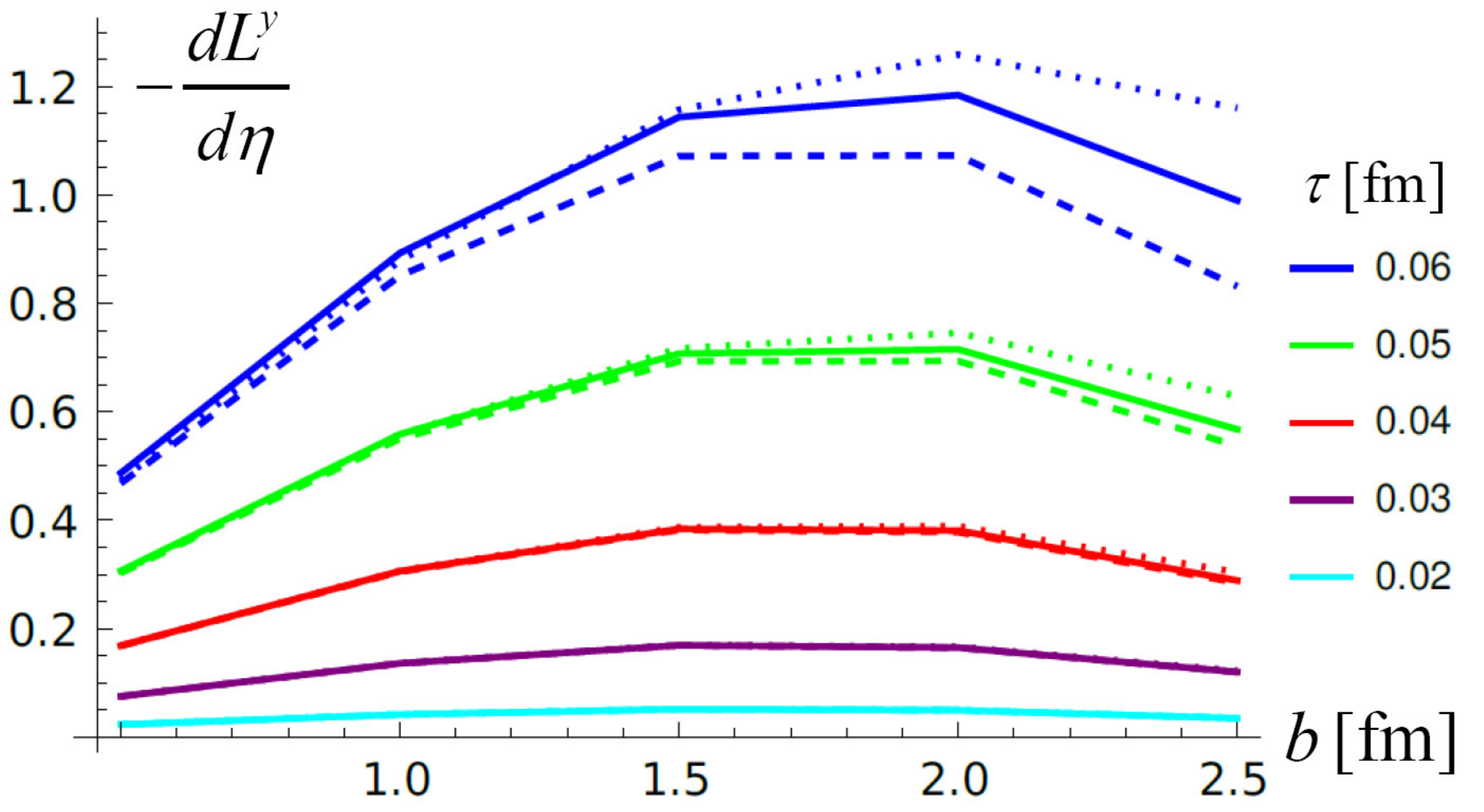}
\end{center}
\vspace{-5mm}
\caption{Angular momentum versus impact parameter at different times to fourth (dotted lines), sixth (dashed lines) and eighth (solid lines) order of the proper time expansion. }
\label{fig-L-impact}
\end{figure}

The integral over the transverse plane in equation (\ref{L-def-3}) can be simplified using symmetry considerations. The distributions of colour charges that we use are even under the transformation $R^y \to -R^y$.  We will consider symmetric displacements of the ions relative to the collision axis ($\vec b_1 = -\vec b_2$) so that the transformation $R^x \to -R^x$ interchanges the distributions for the first and second ions. Using these symmetries one can show that each component of the energy-momentum tensor in Milne coordinates is either even or odd under the $R^x$ and $R^y$ parity transformations. Using the symmetry relations one shows that the only non-zero component of the angular momentum per unit rapidity is, as expected, the $y$ component, which is equal to
\be
\label{L-def-4}
\frac{dL^y}{d\eta} = -\tau^2 \int d^2\! \vec R \, R^x T^{0\eta} .
\ee

In all calculations we displace the ion moving in the positive $z$-direction a distance $b/2$ in the positive $x$-direction, and the ion that is moving in the negative $z$-direction is shifted the same amount in the negative $x$-direction. The collision therefore produces angular momentum in the negative $y$-direction.

In Fig.~\ref{fig-L-impact} we show the glasma angular momentum at five different impact parameters from $b=0.5$ fm to $b=2.5$ fm and at five different proper times from $\tau = 0.02$ fm to $\tau = 0.06$ fm. The calculation is done at fourth, sixth and eighth order of the proper time expansion, which demonstrates the convergence of the expansion at the times that are shown. The shape of the curve shows that the angular momentum initially increases with impact parameter, peaking at  $b \approx 2.0-2.5$ fm, and then slowly decreases. This behaviour is physically reasonable: the initial increase is caused by the increasing asymmetry of the collision, and the eventual dropoff occurs when the centers of the nuclei become so separated that the overlap region is significantly reduced. 

The shape of our graph of the angular momentum versus impact parameter in Fig.~\ref{fig-L-impact} matches the basic form of the results in Refs. \cite{Gao:2007bc,Becattini:2007sr}, where the authors have calculated the angular momentum of a system of colliding ions. However, the values of the angular momentum obtained in \cite{Gao:2007bc,Becattini:2007sr} are of order $10^5$ at RHIC energies, and even larger at LHC energies, and are thus five to six orders of magnitude larger than our results. Our calculation shows that the glasma carries only a very small imprint of the primordial angular momentum, which means that the majority of the angular momentum is carried by valence quarks. This result, which shows that the idea of a rapidly rotating glasma is not relevant at collider energies, is in agreement with the measured global polarization of lambda hyperons which is smaller than 0.5\%  at RHIC \cite{Adam:2018ivw} and consistent with zero at LHC \cite{Acharya:2019ryw}.

The integral over $R^x$ in Eq.~(\ref{L-def-4}) is taken up to $R^x_{\rm max} = 5.9$ fm. We note that the dominant contribution to the angular momentum comes from the parts of the nuclei that are farthest from the collision centre, with respect to which angular momentum is calculated. These are the regions where the gradient expansion we use is least to be trusted. Our results for the angular momentum therefore do depend on the upper limit of the integral, and should only be considered order of magnitude estimates for the glasma angular momentum. In Ref.~\cite{Carrington:2021qvi} we give a more detailed analysis of the extent to which our results for the angular momentum of the glasma depend on the integration region that is used to do the calculation. These considerations do not affect our main result that only a small fraction of the very large angular momentum of the incoming nuclei is transferred to the glasma.

\section{Jet quenching}
\label{sec-jet-quenching}

In the second part of this review article we discuss the transport of hard probes through a glasma. We start with a Fokker-Planck equation whose collision terms encode information about glasma dynamics through correlators of strong electric and magnetic fields. 

\subsection{Fokker-Planck equation}
\label{FP-eq}

The Fokker-Planck equation has been frequently employed to study the transport of heavy quarks through a thermalized quark-gluon plasma, see for example \cite{Moore:2004tg,Svetitsky:1987gq,vanHees:2004gq,Mustafa:2004dr}. Our aim is to study the transport of both heavy quarks and high-$p_T$ light partons through glasma in the earliest period of its temporal evolution. More specifically, we focus on the situation where the hard probes interact with the soft classical gluon fields of the glasma, and not with quasi-particles, which emerge at later stages. 

The formulation of the method and the derivation of the Fokker-Planck equation that we use is presented in detail in  Ref.~\cite{Mrowczynski:2017kso} but the main points are reviewed below. Although the original derivation was presented in the context of heavy quarks traversing a glasma, the formalism can also be used to study relativistic light partons, as long as the diffusion approximation is applicable. For heavy quarks the method can be used for a broad range of velocities, and therefore provides much richer information about their spectra than is the case for light high-energy partons.

When a heavy quark is embedded in a glasma it is subject to stochastic processes due to the action of colour forces. The corresponding distribution function $Q(t,\vec{r},\vec{p})$ can  therefore be decomposed into regular and fluctuating components as follows\footnote{As in Sec.~\ref{sec-angular}, we denote three-vectors  as $\vec{x}=(x^1,x^2,x^3)$ and they are indexed by $\alpha,\beta \in (1,2,3)$ or $\alpha,\beta =x,y,z$.}
\be
\label{reg-fluc}
Q(t,\vec{x},\vec{p}) = \langle Q(t,\vec{x},\vec{p}) \rangle 
+ \delta Q(t,\vec{x},\vec{p}) ,
\ee
where $t$ is time, $\vec{x}$ is position, $\vec{p}$ is momentum and $\langle \cdots \rangle$ denotes a statistical ensemble average over events in a relativistic heavy-ion collision. The regular contribution, denoted $\langle Q(t,\vec{x},\vec{p}) \rangle$, is assumed to be colour neutral and gauge independent, and is expressed as   
\be
\label{whiteness}
\langle Q(t,\vec{x},\vec{p}) \rangle  = 
n (t,\vec{x},\vec{p}) \, \mathds{1},  
\ee
where $\mathds{1}$ is a unit matrix in colour space. We use $\delta Q(t,\vec{x},\vec{p})$ to denote the fluctuating part and we assume that $\langle \delta Q \rangle =0$. It is also assumed that the regular part is a slowly varying function of time and space and is much larger than the fluctuating part. With these conditions, starting from a Vlasov-type equation, one is able to obtain a transport equation in the Fokker-Planck form~\cite{Mrowczynski:2017kso}, which reads
\be
\label{F-K-eq}
\Big({\cal D} - \nabla_p^\alpha  X^{\alpha\beta}(\vec{v}) \nabla_p^\beta - \nabla_p^\alpha  Y^\alpha (\vec{v}) \Big) n(t, \vec{x},\vec{p}) = 0,
\ee
where $\vec{v}=\vec{p}/E_p$ is the velocity of the quark with $E_p=\sqrt{\vec{p}^2 + m^2_Q}$. We also use ${\cal D} \equiv  \frac{\partial}{\partial t} + \vec{v}\cdot \nabla $ for the substantial, or material, derivative. The collision terms entering the Fokker-Planck equation~(\ref{F-K-eq}) are given by
\be
\label{Y-def}
Y^\alpha(\vec{v}) \, n(\vec{p}) = \frac{1}{N_c} {\rm Tr} \big[ \big\langle \mathcal{F}^\alpha(t, \vec{x}) 
\delta Q_0(\vec{x}-\vec{v} t,\vec{p})\big\rangle \big],
\ee
and\footnote{There is a typographic error in Eq.~(5) of \cite{Carrington:2022bnv} and Eq.~(2) of \cite{Carrington:2021dvw},  which are analogous to Eq.~(\ref{X-def}). The factor multiplying the integral in these equations is $1/(2N_c)$ and should be $1/N_c$.}
\be
\label{X-def}
X^{\alpha\beta}(\vec{v}) \equiv
\frac{1}{N_c} \int_0^t dt' \: {\rm Tr}\big[ \big\langle \mathcal{F}^\alpha(t, \vec{x}) 
\mathcal{F}^\beta \big(t-t', \vec{x}-\vec{v} t'\big) \big\rangle \big] ,
\ee 
where $\delta Q_0 \equiv \delta Q(t=0,\vec{x},\vec{p})$ is the initial condition. The Lorentz colour force entering the collision terms (\ref{Y-def}) and (\ref{X-def}) is $\vec{\cal F} (t,\vec{x}) \equiv g \big(\vec{E}(t,\vec{x}) + \vec{v} \times \vec{B}(t,\vec{x})\big)$. The electric $\vec{E}(t,\vec{x})$ and magnetic $\vec{B}(t,\vec{x})$ fields are given in the fundamental representation of the SU($N_c$) group. The tensor $X^{\alpha\beta}(\vec{v})$ in Eq.~(\ref{X-def}) can be written in terms of  correlators of the glasma electric and magnetic fields in the adjoint representation as
\ba
\nn
&&X^{\alpha\beta}(\vec{v}) =
\frac{g^2}{2 N_c} \int_0^t dt' \Big[\big\langle E_a^\alpha(t, \vec{x}) E_a^\beta(t-t',\vec{y})\big\rangle 
\\[2mm] \nn
&&~~~~~~~~~~
+ \epsilon^{\beta \gamma \gamma'}v^\gamma 
\big\langle E_a^\alpha(t, \vec{x}) B_a^{\gamma'}(t-t',\vec{y})\big\rangle 
+ \epsilon^{\alpha \gamma \gamma'}v^\gamma 
\big\langle B_a^{\gamma'}(t, \vec{x}) E_a^\beta (t-t',\vec{y})\big\rangle
\\[2mm] \label{tensor-gen}
&&~~~~~~~~~~~~~~~~~~~~~~
+ \epsilon^{\alpha \gamma \gamma'} \epsilon^{\beta \delta \delta'}v^\gamma v^\delta 
\big\langle B_a^{\gamma'}(t, \vec{x}) B_a^{\delta'}(t-t',\vec{y})\big\rangle
\Big], 
\ea
where $\vec{y} = \vec{x} - \vec{v}t'$. For future convenience we also define $\vec{v}=(v_\parallel,\vec{v}_\perp)$ and $v_\perp=|\vec{v}_\perp|$.

The field correlators in Eq.~(\ref{tensor-gen}) are non-local and consequently they are not gauge invariant. As discussed in Ref.~\cite{Mrowczynski:2017kso}, the problem can be remedied by inserting between the two fields the link operator
\be
\label{link-def}
\Omega (t_1,\vec{x}_1|t_2,\vec{x}_2) \equiv {\cal P} \exp\Big[ ig \int_{(t_2,\vec{x}_2)}^{(t_1,\vec{x}_1)} ds_\mu A^\mu_c(s) T^c\Big] ,
\ee
where ${\cal P}$ denotes `left later' path ordering. Since analytic calculations become prohibitively difficult with the link operator included, we use the expression in Eq.~(\ref{tensor-gen}). In Sec.~\ref{sec-gauge-dependence} we give a quantitative estimate of the size of the link operator and argue that the effect of neglecting it is numerically small. The physical reason is the very short time intervals we deal with. 

The equilibrium distribution function $n^{\rm eq}(\vec{p}) \sim \exp (-E{_{\vec{p}}}/T) $, where $T$ is the temperature of the system, should solve the Fokker-Planck equation. This requires a relation between $X^{\alpha\beta} (\vec{v})$ and $Y^\alpha(\vec{v})$ of the form
\ba
\label{XY}
Y^\alpha(\vec{v}) = \frac{v^\beta}{T} X^{\alpha\beta} (\vec{v}),
\ea
where $T$ is the temperature of an equilibrated quark-gluon plasma that has the same energy density as the glasma, or equivalently the temperature the glasma would have, if it equilibrated without expanding. Our calculation gives no information about this temperature, but in Sec.~\ref{sec-res-0} we discuss how to estimate its value. Since the formula (\ref{Y-def}) is difficult to apply to a non-equilibrium system, we use the relation~(\ref{XY}) to determine $Y^\alpha(\vec{v})$. As we will show below, the quantity $Y^{\alpha}(\vec{v})$ is needed to obtain $dE/dx$, but it is not required for a calculation of $\hat q$.

When the system under consideration is translationally invariant, the tensor $X^{\alpha\beta} (\vec{v})$ is independent of the variable $\vec{x}$ present on the right side of Eq.~(\ref{X-def}). In most of this section we consider collisions of nuclei which are infinite and homogeneous in the transverse plane. Such a system is translationally invariant in the transverse plane but  is not completely uniform along the $z$-axis. We therefore expect a weak dependence of $X^{\alpha\beta} (\vec{v})$ on the longitudinal coordinate $z$ which is not explicitly shown on the left side of Eqs.~(\ref{Y-def}) and~(\ref{X-def}). This is discussed in Sec.~\ref{sec-res-b}. In Sec.~\ref{sec-qhat-finite-nuclei} we consider realistic finite nuclei and verify that the momentum broadening coefficient is largely independent of the average transverse coordinate.

The quantities $Y^\alpha(\vec{v})$ and $X^{\alpha\beta} (\vec{v})$ in Eqs.~(\ref{Y-def}) and (\ref{X-def}) are expected to saturate at large times. This can be proven analytically in equilibrium in the long time limit, as a consequence of the fact that the correlators in Eqs.~(\ref{Y-def}) and (\ref{X-def}) are translation invariant. In our inhomogeneous system  time independence occurs due to finite correlation lengths. If the correlator $\big\langle \mathcal{F}^\alpha(t, \vec{x}) \mathcal{F}^\beta (t', \vec{x}') \big\rangle$ vanishes for $|\vec{x}'- \vec{x}| > \lambda_x$ or $|t'- t| > \lambda_t$, the integral (\ref{X-def}) saturates for $ t > \lambda_t$ or $ t > \lambda_x/v$. In practice, the saturation of $X^{\alpha\beta} (\vec{v})$ at long times and its approximate independence on $z$ provide an estimate of the range of validity of the approximations that we use to obtain the  transport coefficients of the glasma. We note that we have not indicated dependence on time on the left side of Eqs.~(\ref{Y-def}) and (\ref{X-def}) since the transport coefficients that we will calculate are only meaningful when at least approximate saturation is observed. 

The physical interpretation of the collision terms $Y^\alpha(\vec{v})$ and $X^{\alpha\beta}(\vec{v})$ is easy to understand. As discussed in the textbook \cite{Kampen:1987}, they determine the average momentum change per unit time, and the correlation of momentum changes per unit time, as follows
\ba
\label{Dpi-final}
\frac{\langle \Delta p^\alpha \rangle}{\Delta t} 
&=& - Y^\alpha(\vec{v}) ,
\\[2mm]
\label{Dpi-Dpj-final}
\frac{\langle \Delta p^\alpha \Delta p^\beta \rangle}{\Delta t} 
&=& X^{\alpha\beta}(\vec{v}) + X^{\beta\alpha}(\vec{v}) .
\ea
The collisional energy loss $dE/dx$ and the transverse momentum broadening parameter $\hat{q}$ of a heavy quark in a glasma are obtained from the results in Eqs.~(\ref{Dpi-final}) and (\ref{Dpi-Dpj-final}) using the equations 
\ba
\label{enn}
\frac{dE}{dx} &=& \frac{v^\alpha}{v} \frac{\langle \Delta p^\alpha \rangle}{\Delta t} ,\\
\label{qq}
\hat q &=& \frac{1}{v} \Big( \delta^{\alpha\beta} -\frac{v^\alpha v^\beta}{v^2} \Big)
\frac{\langle \Delta p^\alpha \Delta p^\beta \rangle}{\Delta t} ,
\ea
where $v=|\vec{v}|$. Equations~(\ref{XY}), (\ref{enn}) and (\ref{qq}) give
\ba
\label{e-loss-X-Y}
\frac{dE}{dx} &=& - \frac{v}{T} \frac{v^\alpha v^\beta}{v^2} X^{\alpha\beta}(\vec{v}), \\
\label{qhat-X-T}
\hat{q} &=&  \frac{2}{v} \Big(\delta^{\alpha\beta} - \frac{v^\alpha v^\beta}{v^2}\Big) X^{\alpha\beta}(\vec{v}).
\ea

Using the techniques of the previous sections, we can now calculate all of the field correlators that enter the tensor $X^{\alpha\beta}(\vec{v})$ in Eq.~(\ref{tensor-gen}). When the glasma system is translationally invariant in the transverse plane the correlator (\ref{B-res}), which is the building block to construct the field correlations from Eq.~(\ref{tensor-gen}), takes the form
\be
\label{Ai0-Aj0-C1-C2}
B^{ij}(r)
= \delta^{ij}  C_1 (r) - \hat{r}^i \hat{r}^j C_2 (r) ,
\ee
where $r\equiv|\vec{x}_\perp - \vec{y}_\perp|$ and $\hat{r}^i \equiv r^i/r$. The functions $C_1(r)$ and $C_2(r)$ are
\be
\label{C1-def}
C_1(r) \equiv  \frac{m^2 K_0 (mr)}{g^2 N_c  \big( mr K_1(mr) - 1\big) }
\bigg\{ \exp\bigg[\frac{g^4 N_c \mu \big( mr K_1(mr) - 1\big)  }{4\pi m^2 }\bigg] -1 \bigg\} ,
\ee
\be
\label{C2-def}
C_2(r) \equiv  \frac{m^3 r \, K_1(mr) }{g^2 N_c  \big( mr K_1(mr) - 1\big) }
\bigg\{ \exp\bigg[\frac{g^4 N_c \mu \big( mr K_1(mr) - 1\big)  }{4\pi m^2 }\bigg] -1 \bigg\}.
\ee
At lowest order in the $\tau$ expansion, the correlators of the electric and magnetic fields are given by 
\ba
\nn
\langle E^{z(0)}_a(\vec{x}_\perp) \, E^{z(0)}_b(\vec{y}_\perp)\rangle  &=&  g^2 N_c \delta^{ab} 
\Big( 2 C_1^2 (r) - 2 C_1 (r) \, C_2 (r) + C_2^2 (r) \Big) ,
\\[2mm]
\label{1-st-field-corr}
\langle B^{z(0)}_a(\vec{x}_\perp) \, B^{z(0)}_b(\vec{y}_\perp)\rangle  &=&  g^2 N_c \delta^{ab} 
\Big( 2 C_1^2 (r) - 2 C_1 (r) \, C_2 (r) \Big),
\\[2mm]
\nn
\langle E^{z(0)}_a(\vec{x}_\perp) \, B^{z(0)}_b(\vec{y}_\perp)\rangle  &=& 0 .
\ea
All higher order correlators are given by similar expressions involving the functions $C_1(r)$ and $C_2(r)$ and their derivatives. The tensor $X^{\alpha\beta}(\vec{v})$ also involves correlators of fields at the same point, and we treat these 
as two-point correlators. 

We comment that although collisional energy loss and the momentum broadening coefficient are obtained from the two specific projections of the tensor $X^{\alpha\beta}(\vec{v})$ in Eqs.~(\ref{e-loss-X-Y}) and (\ref{qhat-X-T}), we have calculated all of the electric and magnetic field correlators. These expressions could be used in other calculations. For example, one could solve the Fokker-Planck equation and determine how the distribution functions of hard probes evolve in time.

\subsection{Physical picture}
\label{sec-res-0}

Heavy and high-$p_T$ probes are produced at the very first moments of the collision and then propagate through the evolving glasma. When the magnitude of the velocity of the probe $v$ is close to one it is highly relativistic, and could be a heavy or light quark, or gluon. When $v$ is significantly less than one, the probe is necessarily a heavy quark. We will study only the behaviour of quark probes, but the tensor $X^{\alpha\beta}(\vec v)$ for a gluon, and consequently $\hat q$ and $dE/dx$, could be obtained from Eq.~(\ref{X-def}) by multiplying by a factor $(N_c^2 - 1)/2N_c^2$, which equals $4/9$ for $N_c=3$. 

Because experiments at RHIC and the LHC focus on hard probes from the momentum space mid-rapidity region, $y\in (-1,+1)$, we are primarily interested in the transport properties of probes moving mostly perpendicularly to the beam axis. The momentum-space rapidity $y$ is related to the longitudinal component of the probe's velocity by $y=\frac{1}{2} \ln \frac{1+v_\parallel}{1-v_\parallel}$, and therefore the values $y=\pm 1$ correspond to $v_\parallel = \pm 0.76$, and the mid-rapidity value $y=0$ corresponds to strict transverse motion, $v_\parallel=0$. We use the parameter $v_\parallel$ instead of $y$ to quantify deflection from transverse motion, and we consider quarks with $0\leq v_\parallel < 0.76$.
 
\begin{figure}[t]   
\centering
\includegraphics[scale=0.4]{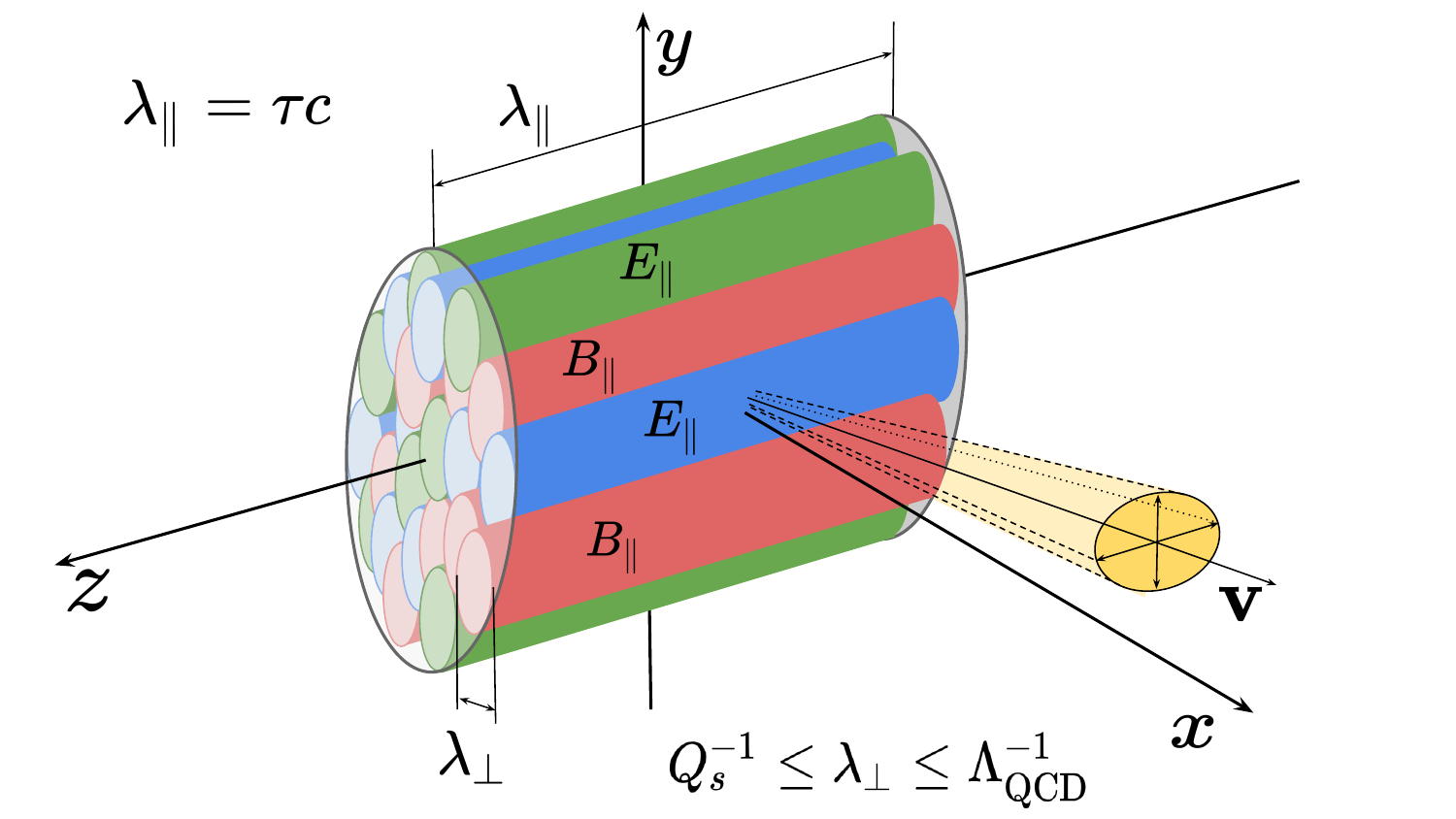}  
\caption{Cartoon of the zeroth order glasma fields and a probe moving mostly transverse to the collision axis.
\label{tubes}}  
\end{figure}

An idealized picture of a probe emerging from the glasma at very early proper times is shown in Fig.~\ref{tubes}, 
where the glasma fields at zeroth order in the proper time expansion are represented by coloured flux tubes. At this order the electric and magnetic fields are purely longitudinal and static. There are two qualitatively different correlation lengths, which we will denote $\lambda_\parallel$ and $\lambda_\perp$. The longitudinal correlation length $\lambda_\parallel$ is proportional to the distance between the nuclei and can be identified with the proper time $\tau$. The transverse correlation length $\lambda_\perp$ can be inferred from the correlators (\ref{1-st-field-corr}). Qualitatively the transverse correlation length obeys $Q_s^{-1} \leq\lambda_\perp \leq \Lambda^{-1}_{\rm QCD}$. 

The collisional energy loss and the momentum broadening parameter are both built up during the time that the probe spends within the domain of correlated fields. At zeroth order this time is determined by the transverse correlation length and the orientation and magnitude of the probe’s velocity. The transport coefficients will saturate when the probe leaves the region of correlated fields\footnote{We note that we consider only the glasma phase, where coherent fields are present and correlation lengths are sizable. We do not consider later kinetic or hydrodynamic stages where further broadening of the momenta of hard probes occurs due to scattering on plasma constituents.}. These simple arguments indicate that the Fokker-Planck methodology we are using might be well suited to describe the problem of a hard probe moving through a glasma, at least at very early times.

The simple picture presented in Fig.~\ref{tubes} is valid at zeroth order, but at later times it does not accurately describe the glasma. As $\tau$ increases, transverse electric and magnetic fields develop, and the glasma fields can grow or decrease rapidly. Higher and higher orders in the $\tau$ expansion are needed to describe the glasma fields, as $\tau$ increases. Our method will work  if saturation is reached before the $\tau$ expansion that is used to calculate the field correlators breaks down.

In addition to determining how long a probe spends in the region of correlated fields, the probe's velocity affects the transport coefficients in another way. To see this we look at Eqs.~(\ref{X-def}), (\ref{e-loss-X-Y}) and (\ref{qhat-X-T}), and use the form of the Lorentz force. At zeroth order the integrand that gives collisional energy loss is proportional to
\be
\label{struc-a}
v^\alpha v^\beta {\cal F}^{\alpha(0)} {\cal F}^{\beta(0)}= 
g^2 v^2_\parallel E^{z(0)}(\vec{x}_\perp) E^{z(0)}(\vec{x}'_\perp) ,
\ee
and the integrand for momentum broadening is proportional to 
\ba
\nn
&&\Big(\delta^{\alpha\beta}-\frac{v^\alpha v^\beta}{v^2}\Big){\cal F}^{\alpha(0)} {\cal F}^{\beta(0)}
= g^2\frac{v_\perp^2}{v^2} \Big( E^{z(0)}(\vec{x}_\perp) E^{z(0)}(\vec{x}'_\perp)
\\[2mm] \label{struc-b}
&&~~~~~~~~~~~~~~~~~~~~~~~~~~~~~~~~~~~~~~~~
+ v^2 B^{z(0)}(\vec{x}_\perp) B^{z(0)} (\vec{x}'_\perp)\Big).
\ea
We see that at zeroth order the collisional energy loss is caused by the electric field and vanishes when the probe moves in the transverse direction. In contrast, zeroth order momentum broadening is caused by both electric and magnetic fields, and is maximal when the probe moves transversely. We will show that at higher orders the same behaviour is observed: for fixed $v$, when $v_\perp$ increases and $v_\parallel$ decreases, one finds that the collisional energy loss decreases and the momentum broadening increases.

We emphasize that the arguments presented in this section give a qualitative description of the behaviour of the glasma at very early times, corresponding to the lowest orders of the $\tau$ expansion. Collisional energy loss and momentum broadening at higher orders in the $\tau$ expansion require calculations to understand the full picture.

\subsection{Numerical results for $\hat q$ and $dE/dx$}
\label{sec-numer-qhat-dEdx}

All our results are calculated for $N_c=3$ and $g=1$. We use $Q_s=2$ GeV and $m=0.2$ GeV except Sec.~\ref{sec-res-d} where we consider different values of $Q_s$ and $m$. We work at mid-spatial-rapidity, or $\eta=z=0$, except Sec.~\ref{sec-res-b} where the $\eta$ dependence of $\hat q$ is studied.

The correlators of chromodynamic fields that enter the tensor $X^{\alpha\beta}(\vec{v})$ are restricted to the forward light-cone region, where the glasma description is valid. To take this condition into account the integrand in Eq.~(\ref{tensor-gen}) is multiplied by
\ba
\label{step}
\Theta(t^2-z^2) \Theta((t-t')^2-(z-v_\parallel t')^2).
\ea 
If we use $v_\parallel=0$ and look at $z=0$, both step functions can be ignored as they are always unity. When $v_{\parallel}$ is nonzero the second step function in (\ref{step}) has the effect of reducing $\hat q$ and $dE/dx$. In almost all calculations, we choose $z=0$ so that the first step function plays no role. The exception is Sec.~\ref{sec-res-b} where we study the dependence of our results on spatial rapidity. 

As explained in Sec.~\ref{FP-eq}, in order to calculate $dE/dx$ we need the temperature $T$ of an equilibrated quark-gluon plasma whose energy density is the same as the energy density of the glasma. The energy density of an equilibrium free quark-gluon plasma equals
\be
\label{en-den-QGP}
\varepsilon_{\rm QGP} = \frac{\pi^2}{60} \big( 4(N_c^2 -1) + 7 N_f N_c \big)  T^4 ,
\ee
where only quarks of $N_f$ flavours with masses much smaller than the temperature should be included. The effective temperature of the glasma can therefore be estimated from the glasma energy density. 

\begin{figure}[t]
\begin{center}
\includegraphics[width=10cm]{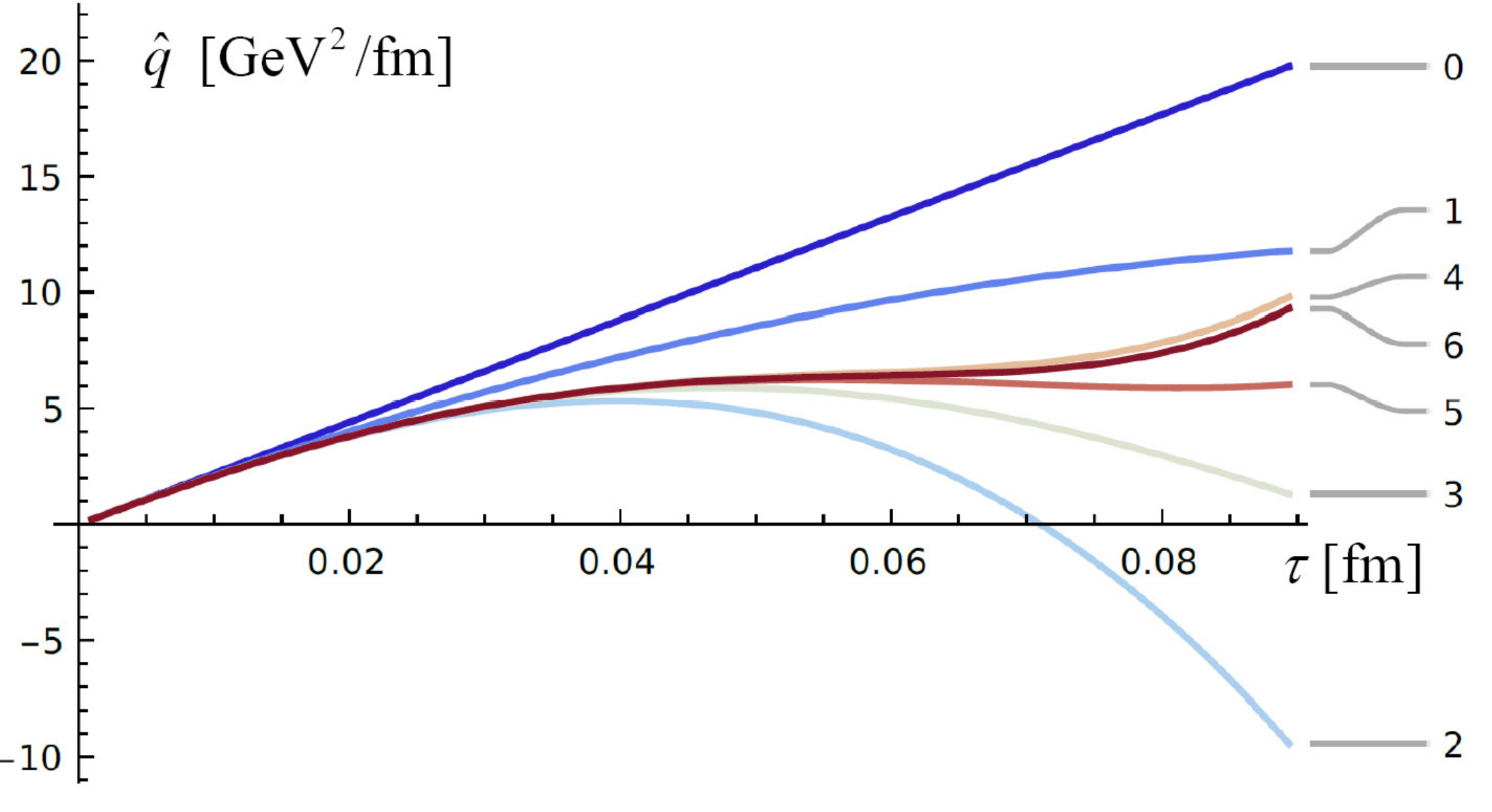}
\end{center}
\vspace{-5mm}
\caption{The momentum broadening coefficient $\hat q$ as a function of $\tau$ at different orders of the proper time expansion. The seventh order result cannot be seen because it lies directly under the sixth order one.}
\label{fig-qhat-time} 
\end{figure}

\subsubsection{Time dependence of $\hat q$ and $dE/dx$}
\label{sec-res-a}

\begin{figure}[t]
\centering
\includegraphics[width=9cm]{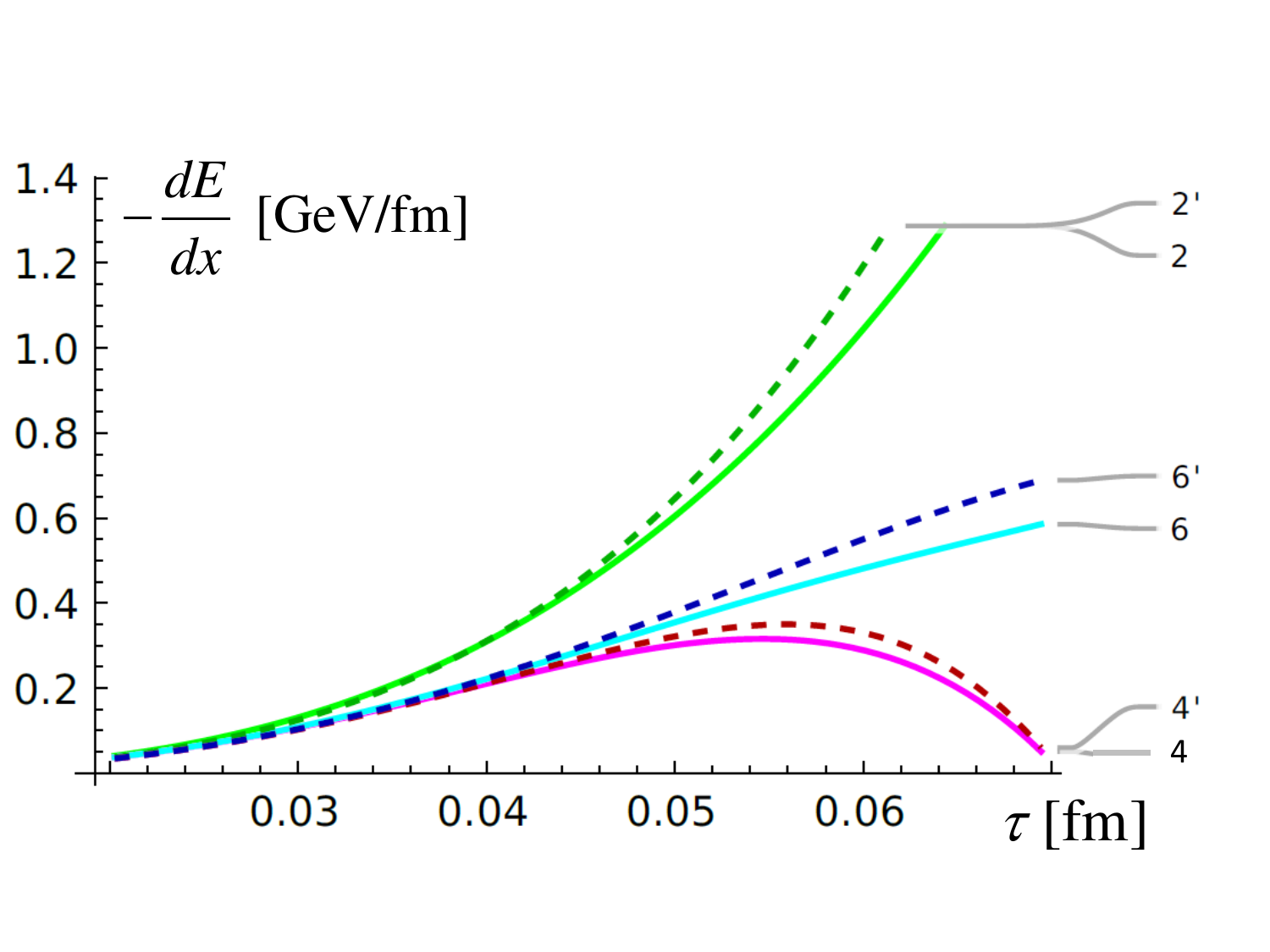}
\vspace{-7mm}
\caption{The energy loss as a function of $\tau$ at different orders of the proper time expansion up to sixth order. The dashed lines are divided by an effective temperature obtained from the glasma energy density (see text for details) and the lighter coloured solid curves are made with $T=1$ GeV. }
\label{fig-eloss-time}
\end{figure}

The momentum broadening coefficient $\hat q$ and the collisional energy loss $dE/dx$, both for an  ultra-relativistic hard probe moving perpendicularly to the beam axis with $v=v_\perp=1$, are shown as a function of $\tau$ in Figs.~\ref{fig-qhat-time} and \ref{fig-eloss-time}. The dependence of both transport coefficients on the order of the $\tau$ expansion illustrates the convergence of the expansion. Taking into account higher order contributions in the $\tau$ expansion extends the range of validity of the result. At very early times all orders of the $\tau$ expansion agree well. 

In Fig.~\ref{fig-qhat-time} we observe that when all terms up to order $\tau^7$ are included the time evolution of $\hat q$ shows initial growth, and then flattening, followed by more rapid growth (which appears after $\tau \sim 0.09$ fm). The region where $\hat q$ flattens shows saturation. The rapid increase of $\hat q$ at later times is not physical, but reflects the breakdown of the proper time expansion. At order $\tau^7$, the highest value of $\hat q$ that is obtained before the proper time expansion breaks down is about 6 GeV$^2$/fm.  The coefficient $\hat q$ was also calculated in Ref.~\cite{Ipp:2020nfu} using real time QCD simulations. The time evolution of $\hat q$ found in this work is qualitatively similar to our finding. Our result is smaller, but still of comparable size. 

Recently calculations of $\hat q$ have been done using a kinetic theory description of an anisotropic quark-gluon plasma \cite{Boguslavski:2023alu,Boguslavski:2023waw}, which is valid between the very early times where the glasma exits and the onset of hydrodynamics. The results of this calculation smoothly connect the two regimes, and support the idea that the pre-equilibrium phase plays an important role in jet quenching.

\begin{figure}[b]
\centering
\includegraphics[width=7.3cm]{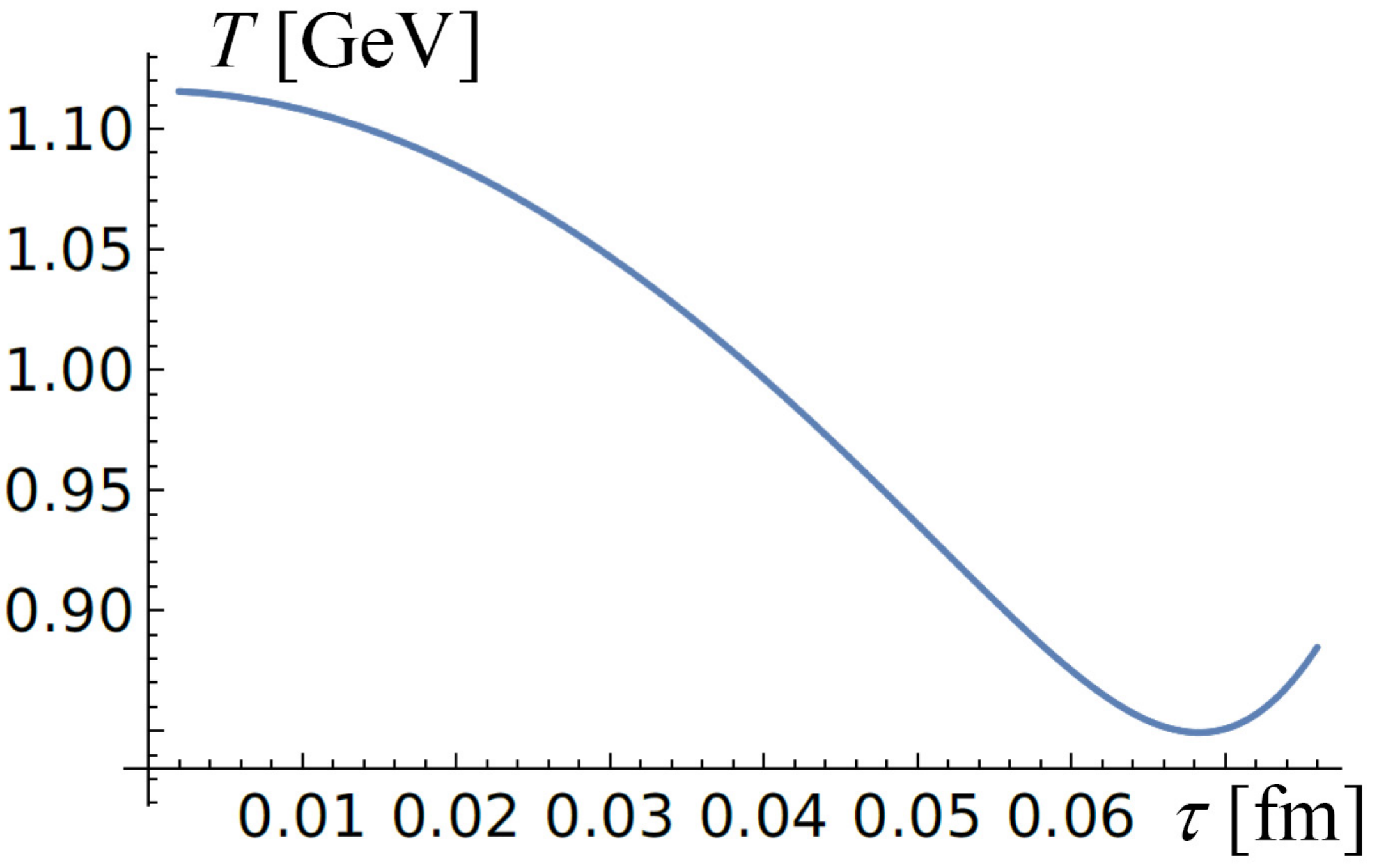}
\vspace{-2mm}
\caption{The effective temperature as a function of the proper time, determined by comparison with an equilibrium system with the same energy density. }
\label{fig-Teff-time}
\end{figure}

The collisional energy loss $-dE/dx$ of a hard parton moving with $v=v_\perp=1$ is shown as a function of $\tau$ in Fig.~\ref{fig-eloss-time}. In order to calculate $dE/dx$ we need the temperature $T$ of an equilibrated quark-gluon plasma whose energy density is the same as the energy density of the glasma. Using the formula for the energy density of an equilibrium noninteracting quark-gluon plasma, the effective temperature of the glasma can be estimated from the glasma energy density. Figure~\ref{fig-Teff-time} shows the temperature obtained from Eq.~(\ref{en-den-QGP}) with $N_f=2$ and the eighth order energy density from a collision of homogeneous nuclei. In Fig.~\ref{fig-eloss-time} the curves for which the order of the expansion is indicated with a prime are obtained using this temperature. The lighter curves are obtained using a constant value $T=1$ GeV, which is not as well motivated from a physics point of view but has the advantage of not mixing the dependence of the two calculations, $-dE/dx$ and ${\cal E}$, on the proper time expansion. The figure shows that the results at different orders converge well up to $\tau \approx 0.06$ fm for both calculations. 

The behaviour of the collisional energy loss $dE/dx$ is very different from what is seen from the momentum broadening parameter $\hat q$. Only the terms at $\tau^2$, $\tau^4$ and $\tau^6$ order contribute to the final result. The zeroth and odd orders vanish because they are proportional to some power of $v_\parallel$, which is zero in the case shown in Fig.~\ref{fig-eloss-time}. The collisional energy loss increases up to around $\tau=0.05-0.06$ fm and for larger times the expansion rapidly breaks down. 

\begin{figure}[t]  
\centering
\includegraphics[scale=0.27]{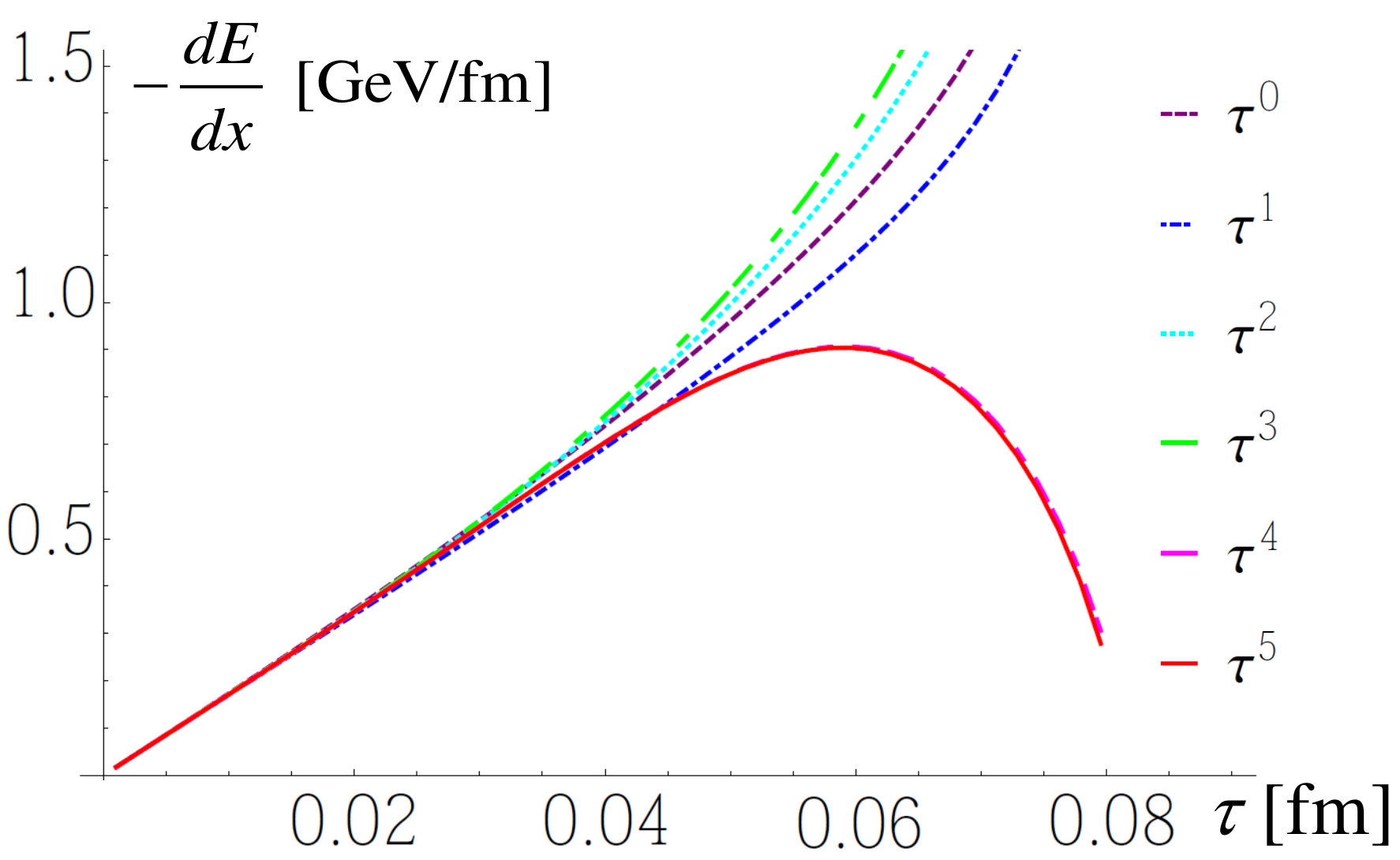}  
\vspace{-2mm}
\caption{Time evolution of $dE/dx$ for $v=1$ and $v_\parallel=v_\perp=1/\sqrt{2}$.
\label{eloss-vel}}
\end{figure}

From the discussion in Sec.~\ref{sec-res-0} we know that the collisional energy loss is much more sensitive to the value of the longitudinal component of the velocity than $\hat q$ is, because the leading order contributions are proportional to $v_\parallel$. In this sense the case of purely transverse motion in Fig.~\ref{fig-eloss-time} might not represent typical behaviour. We therefore show $dE/dx$ in Fig.~\ref{eloss-vel}, also for $v=1$, but now with $v_\perp=v_\parallel=1/\sqrt{2}$. The shape of $dE/dx$ is not significantly different from the one shown in Fig.~\ref{fig-eloss-time}, but all orders in the $\tau$ expansion contribute to the final result. The collisional energy loss is noticeably bigger and equals approximately 0.9 GeV/fm at its maximum, at around $\tau = 0.05$ fm. The $\tau$ expansion breaks down soon after this point. The absence of clear evidence of saturation indicates that our results for collisional energy loss should be considered order of magnitude estimates only.

\subsubsection{Dependence of $\hat q$ on velocity and space-time rapidity}
\label{sec-res-b}

\begin{figure}[t]
\centering
\includegraphics[scale=0.27]{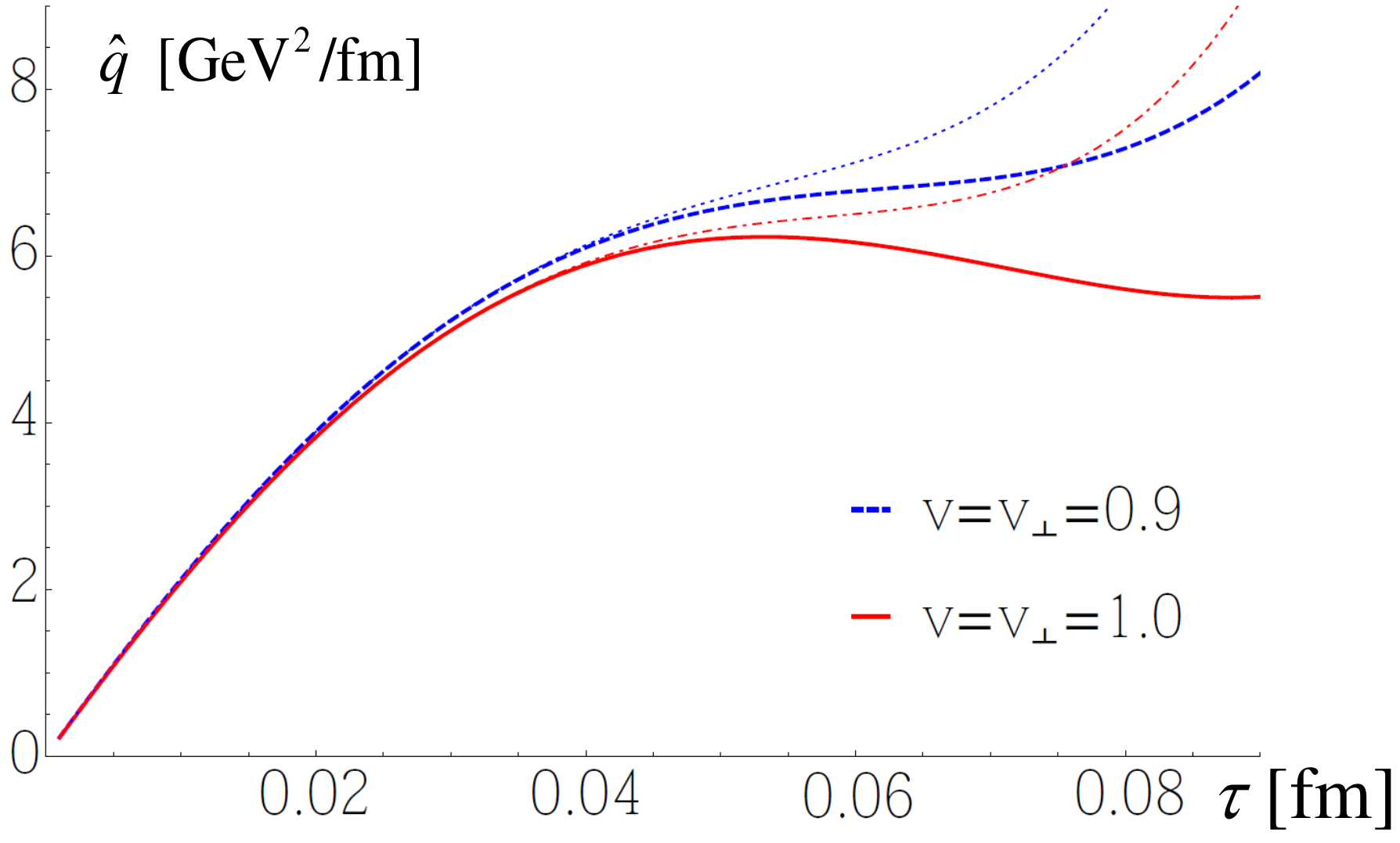}  
\vspace{-2mm} 
\caption{The time evolution of $\hat q$ at order $\tau^5$ for $v=v_\perp=1$ (solid, red) and $v=v_\perp=0.9$ (dashed, blue). The dotted and dashed-dotted lines represent the  corresponding results at order $\tau^4$.
\label{fig-qhat-velocity}}    
\end{figure}  

We first explore the dependence of $\hat q$ on the probe's velocity. We want to understand the relative importance of the two velocity dependent effects discussed in Sec.~\ref{sec-res-0}: the amount of time the probe spends in the region of correlated fields, and the dependence of the Lorentz force on the direction of the probe's velocity. 

\begin{figure}[b]
\centering
\includegraphics[scale=0.27]{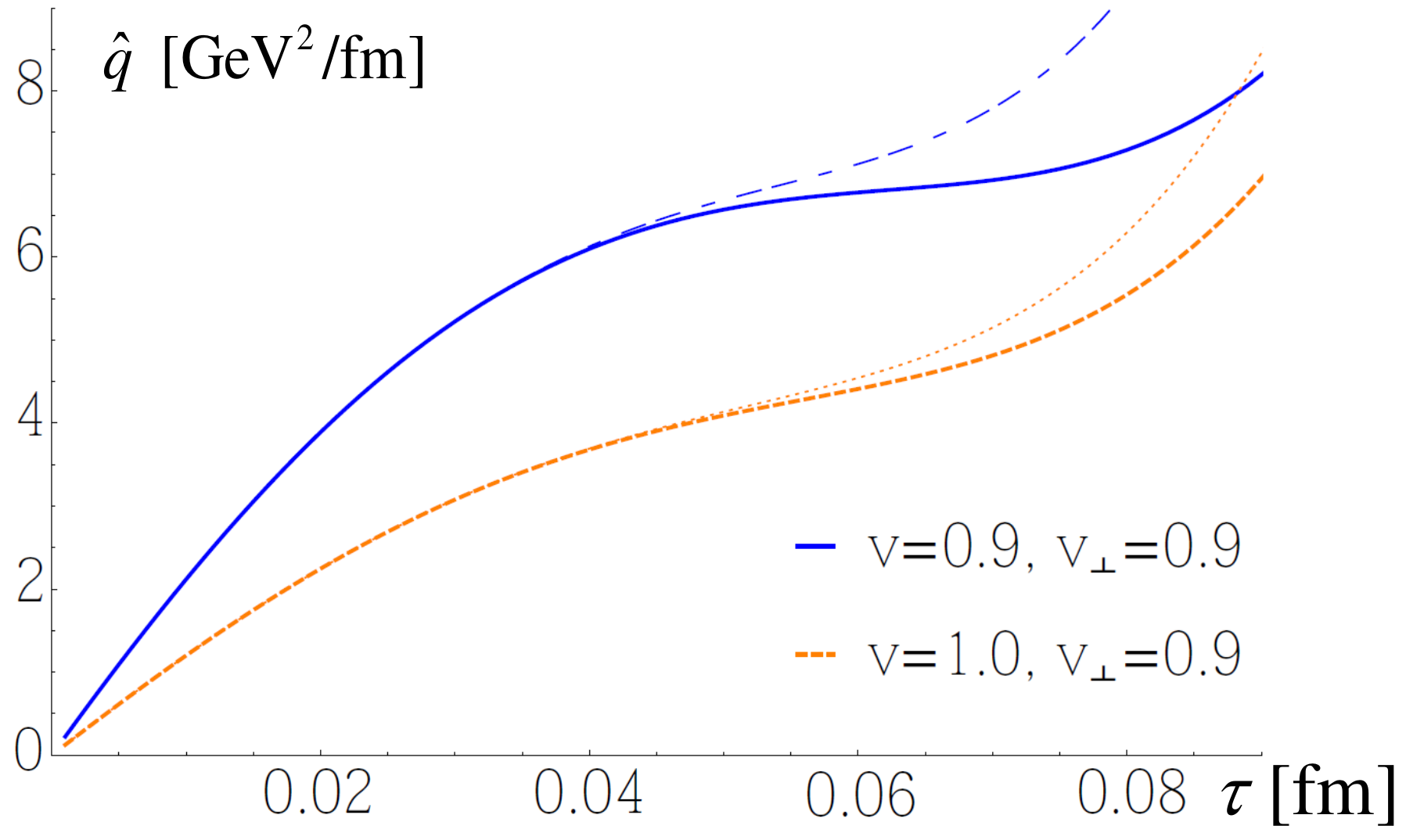} 
\vspace{-2mm} 
\caption{The time evolution of $\hat q$ with $v=0.9$, $v_\perp=0.9$ and $v=1$, $v_\perp=0.9$.
\label{fig-qhat-velocity-1}}
\end{figure}

In Fig.~\ref{fig-qhat-velocity} we show the dependence of $\hat q$ on the speed of a hard probe when the probe moves in the transverse direction. The results calculated at order $\tau^4$ and $\tau^5$ are shown to indicate the region of $\tau$ where the expansion converges. For both values of $v=v_\perp$ the result for $\hat q$ can be trusted to approximately $0.07-0.08$ fm. We observe that at very early times the momentum broadening coefficient is largely independent of $v=v_\perp$, but differences appear at longer times. The value of $\hat q$ for slower quarks flattens and then starts rapidly growing again, whereas $\hat q$ for ultra-relativistic quarks slightly decreases. This shows that when the transverse velocity of the probe increases at fixed $v_\parallel=0$, even though the Lorentz force contribution to $\hat q$ increases (see Eq.~(\ref{struc-b})), the dominant effect is the reduction of the amount of time the probe spends in the domain of correlated fields, which results in a reduction in momentum broadening. This result agrees with the findings of Ref.~\cite{Ipp:2020nfu}, where the momentum broadening parameter of massless quarks is consistently smaller than for larger mass quarks, throughout the whole time evolution. 

\begin{figure}[t]
\centering
\includegraphics[scale=0.27]{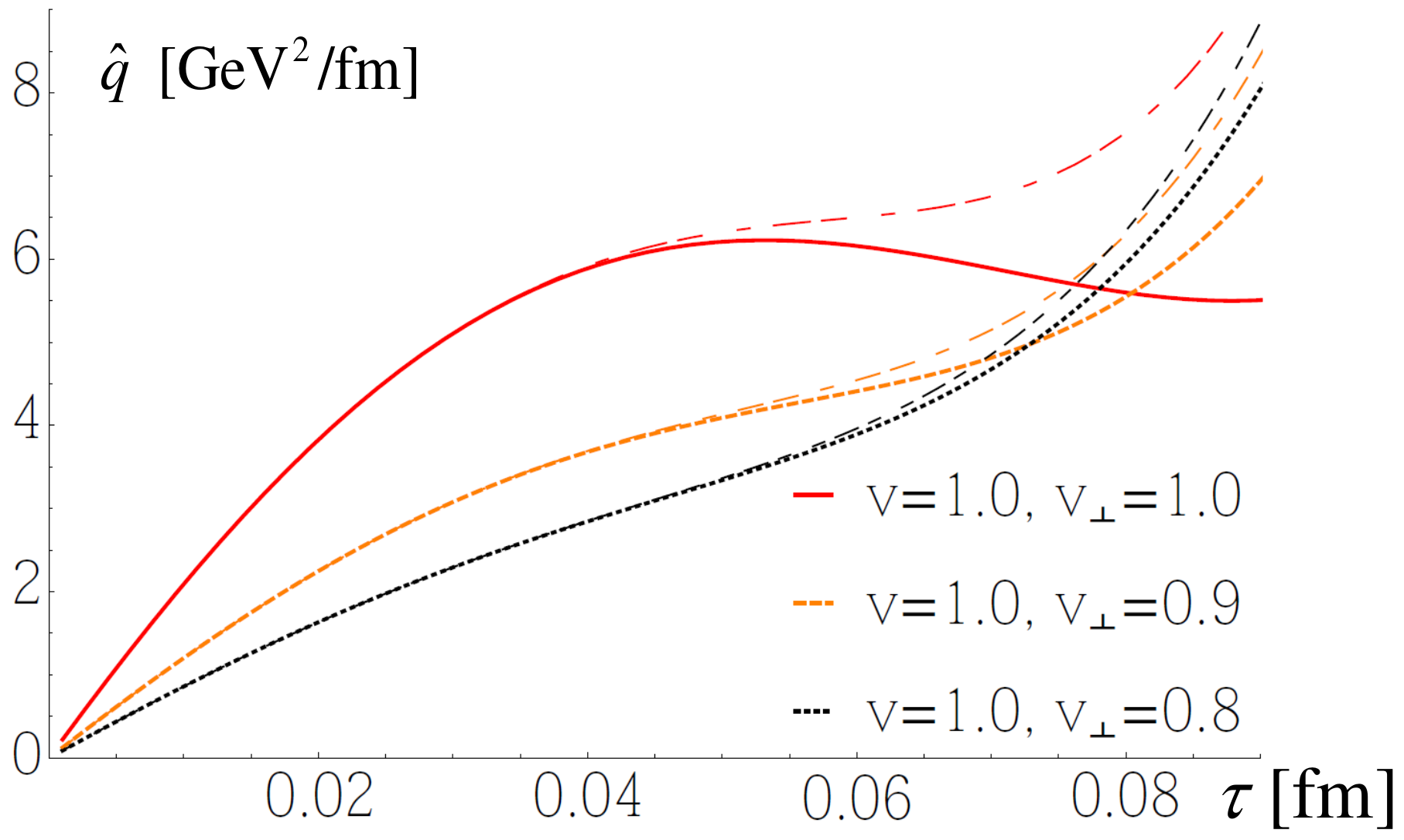}  
\vspace{-2mm}
\caption{The time evolution of $\hat q$ with $v=1$ and $v_\perp=1$, $v_\perp=0.9$, $v_\perp=0.8$.
\label{fig-qhat-velocity-3}}
\end{figure}

In Figs.~\ref{fig-qhat-velocity-1} and \ref{fig-qhat-velocity-3} we show the momentum broadening parameter for several cases with non-zero $v_\parallel$. Figure~\ref{fig-qhat-velocity-1} shows $\hat q$ for $v_\perp=0.9$ and two values of $v$. When the perpendicular component of the velocity $v_\perp$ is fixed, both probes spend the same amount of time in the region of correlated fields and $\hat q$ is affected by the velocity of the probe only because the Lorentz force is velocity dependent. Equation~(\ref{struc-b}) shows that, at very early times, $\hat q$ decreases when $v$ increases at fixed $v_\perp$ because the contribution from electric fields decreases. Figure~\ref{fig-qhat-velocity-1} shows that this effect is also seen at later times.

Figure~\ref{fig-qhat-velocity-3} shows $\hat q$ with $v=1.0$ for two different values of $v_\perp$. When $v$ is fixed and $v_\perp$ varies we also include the effect of changing the amount of time the probe spends in the region of correlation. In Fig.~\ref{fig-qhat-velocity-3} we see that at small times the probe with larger $v_\perp$ has larger $\hat q$, due to the larger Lorentz force, but as $\tau$ increases the curves with large and small $v_\perp$ cross each other. This happens because probes with larger $v_\perp$ escape from the region of correlated fields before the glasma fields become very large, but probes with smaller $v_\perp$ (and larger $v_\parallel$) remain in the domain of correlated fields for a longer time and eventually interact with very large fields. 

We also note that a comparison of Figs.~\ref{fig-qhat-velocity}, \ref{fig-qhat-velocity-1} and \ref{fig-qhat-velocity-3} shows that when $v_\perp=v$ the $\tau$ expansion breaks down at approximately the same point that the saturation regime disappears, but when $v_\perp \approx 0.9 v$, or smaller, the $\tau$ expansion converges fairly well even though no significant saturation regime is observed. This suggests that including higher orders in the $\tau$ expansion could extend the region of saturation when $v_\perp=v$.

\begin{figure}[t]
\centering
\includegraphics[scale=0.27]{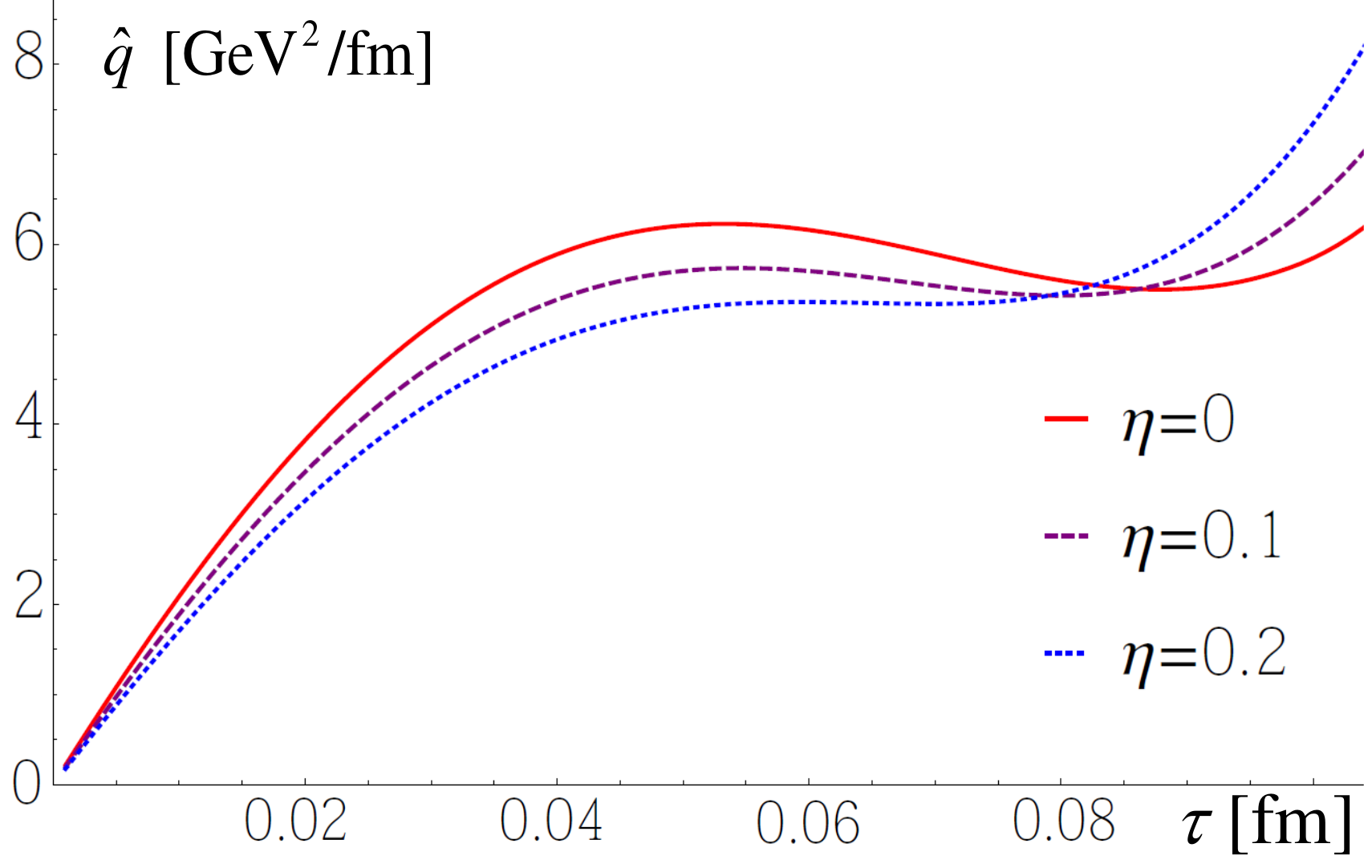} 
\vspace{-2mm} 
\caption{The time evolution of $\hat q$ for $v = v_\perp = 1$ and $\eta=0$, $\eta=0.1$, $\eta=0.2$.
\label{fig-qhat-rapidity-1}}
\end{figure}

\begin{figure}[b]
\centering
\includegraphics[scale=0.27]{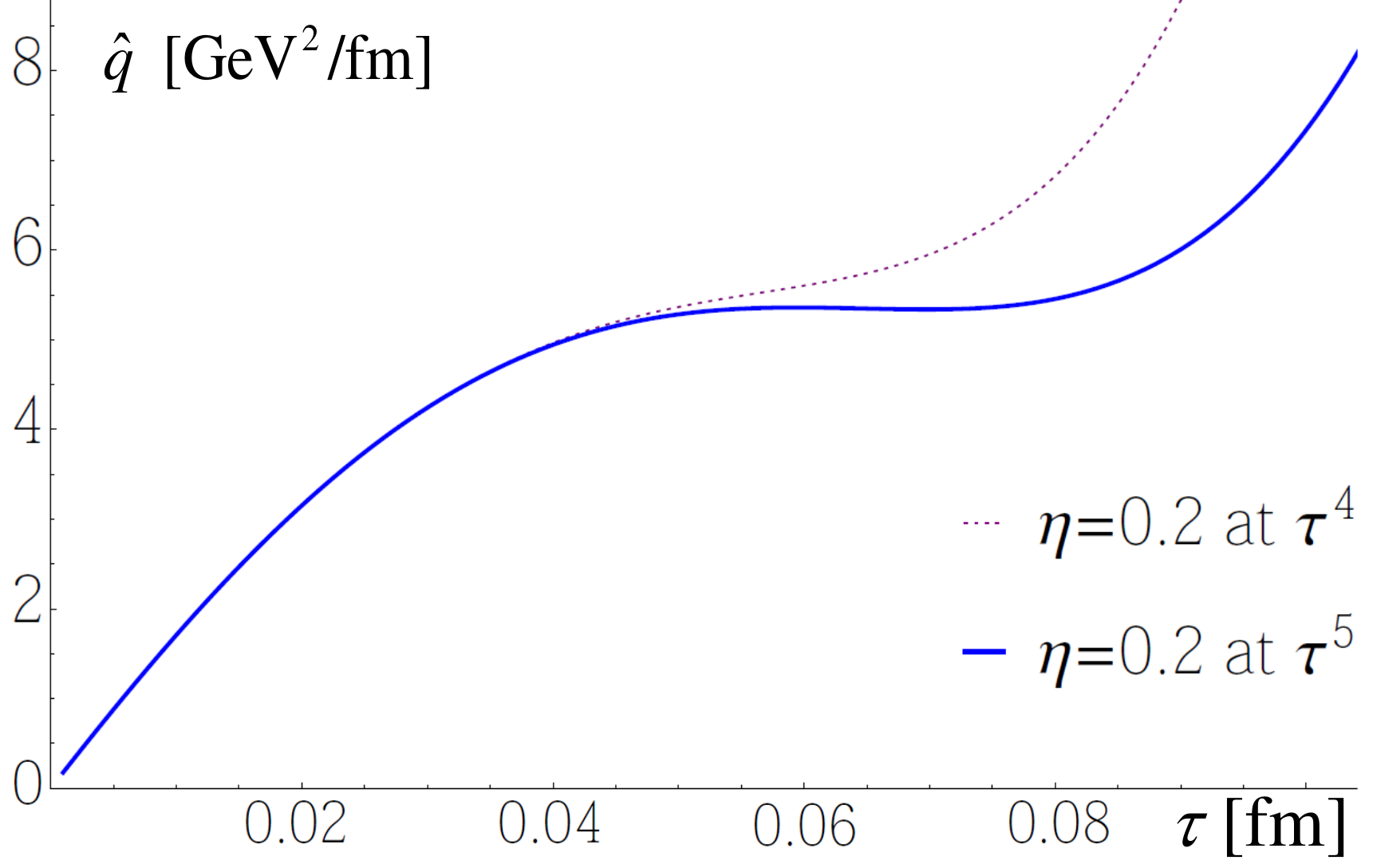}    
\vspace{-2mm}
\caption{The time evolution of $\hat q$ for $v = v_\perp = 1$ and $\eta=0.2$ at $\tau^4$ and $\tau^5$ order.
\label{fig-qhat-rapidity-2}}
\end{figure}

We now explore the dependence of the evolution of $\hat q$ on the spatial rapidity, $\eta$, which is related to the initial position of the probe on the $z$-axis. Figure~\ref{fig-qhat-rapidity-1} shows the time evolution of $\hat q$ for $\eta=0$, $\eta=0.1$ and $\eta=0.2$. In Fig.~\ref{fig-qhat-rapidity-2} we display the results for $\hat q$ with $\eta=0.2$ computed at order $\tau^4$ and order $\tau^5$. The figure indicates that our results can be trusted to $\tau \sim 0.06$ fm.

We consider only small values of $\eta$ because our approach is expected to work best in the mid-spatial-rapidity region, where the CGC approach that we use is most reliable. The momentum broadening parameter depends only weakly on $\eta$ in the region where the curves flatten. Figs.~\ref{fig-qhat-rapidity-1} and \ref{fig-qhat-rapidity-2} show that at least until the region of approximate saturation ends, the result for $\hat q$ is largely independent of spatial rapidity. This result verifies that there is a range of proper times for which the boost invariant ansatz that was used to calculate the glasma correlators is compatible with the approximations that were used to derive the Fokker-Planck equation. 

\subsubsection{Dependence on IR and UV energy scales}
\label{sec-res-d}

The UV scale $Q_s$ and the IR regulator $m$ enter our calculation as parameters that are related to the saturation and confinement scales. We remind the reader that we must stay below $Q_s$, or the assumption that the glasma is composed of classical gluon fields breaks down, and above $m$, so that we do not enter the regime where non-perturbative effects become dominant. The numerical values of these scales cannot be precisely determined within the formalism we are using. The energy-scale interval from $m$ to $Q_s$, where the CGC approach we use is valid, is rather narrow and it is therefore expected that calculated quantities show some scale dependence.

\begin{figure}[t]
\centering 
\includegraphics[scale=0.27]{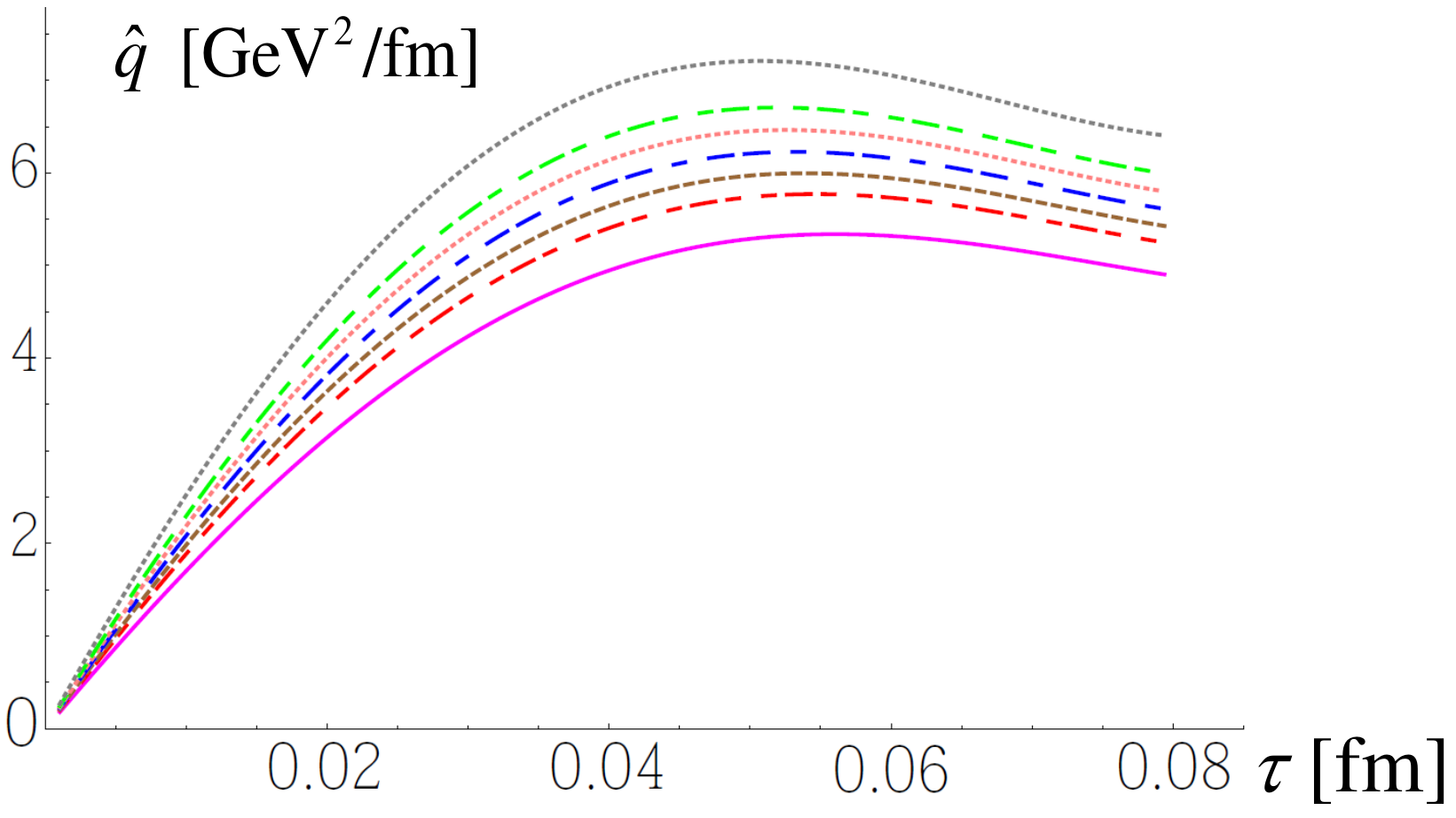}  
\vspace{-2mm}      
\caption{The momentum broadening coefficient $\hat q$ versus $\tau$ for different values of $Q_s$ with $Q_s/m=10$. The curves are labelled with different colours and patterns from $Q_s=1.9$ GeV (magenta solid line) to $Q_s=2.1$ GeV (grey dotted line).
\label{fig-qhat-ratio-Qs-m}} 
\end{figure}

The typical form of this dependence is shown in Fig.~\ref{Qs_mass} for one of the quantities we calculated from the energy-momentum tensor.  In Ref.~\cite{Carrington:2022bnv} the dependence of $\hat{q}$ on $Q_s$ and $m$ was shown to be stronger than what is seen in Fig.~\ref{Qs_mass}, but this is not inherently surprising since there is no reason that different quantities should not have different scale dependence. In the calculation of the momentum broadening parameter, the scale dependence enters both through the two-point correlator and the regularization scheme. It is therefore not unexpected that the scale dependence is stronger than for quantities calculated from the energy-momentum tensor. We have found that the dependence of $\hat{q}$ on $Q_s$ and $m$ can be mostly controlled by holding fixed the ratio $Q_s/m$. This is shown in Fig.~\ref{fig-qhat-ratio-Qs-m}.

The dependence of $\hat q$ on the saturation scale can also have a physical interpretation. Since the saturation scale $Q_s$ increases with the collision energy ($Q_s\sim 1-2$~GeV at RHIC and $Q_s \sim 2-3$~GeV at LHC \cite{Iancu:2003xm}) our calculations predict the growth of $\hat q$ with the collision energy. A reduction in $\hat q$ at  RHIC energies when compared to  LHC energies was found by the JET Collaboration \cite{JET:2013cls}, see also Ref.~\cite{Andres:2016iys}.

\subsubsection{Regularization dependence}
\label{sec-res-c}

\begin{figure}[t]
\centering 
\includegraphics[scale=0.27]{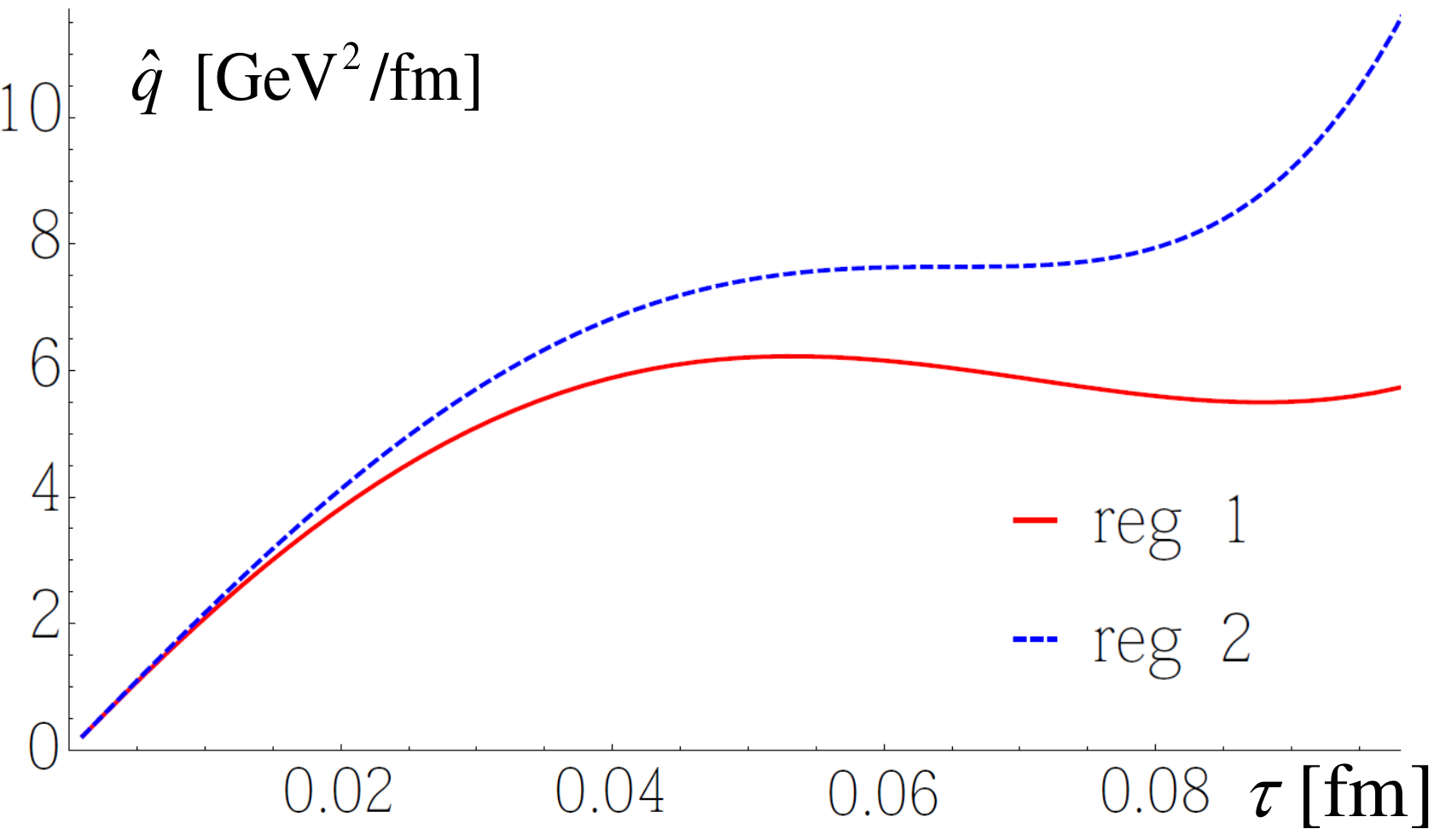}
\vspace{-2mm}    
\caption{The time evolution of $\hat q$ at order $\tau^5$ for $v=v_\perp=1$ regularized in two different ways (see Eqs.~(\ref{regularization1}) and (\ref{regularization2})). 
\label{fig-qhat-reg}}
\end{figure}

All of the results we have obtained for transport coefficients require a regularization procedure. The tensor $X^{\alpha\beta}(\vec{v})$ in Eq.~(\ref{tensor-gen}) is calculated from a time integral with lower limit $t'=0$, which corresponds to the point where $r=0$. The integrand depends on the functions $C_1(r)$ and $C_2(r)$ defined in Eq.~(\ref{C1-def}), and their derivatives, but $C_1(r)$ diverges as $r$ approaches zero. We note that this divergence is a natural consequence of the fact that the classical CGC approach breaks down at small distances. 

We use two different methods to regularize the divergence to verify that our results are largely independent of the regularization method. To explain this, we write the integrand for either momentum broadening or collisional energy loss as a function of the form $f(t',r,z)$, so that the transport coefficient is obtained from the integral $\int_0^t dt'\, f(t',r,z)$  (see Eqs.~(\ref{e-loss-X-Y}, \ref{qhat-X-T}) and (\ref{tensor-gen})). The first regularization method we use is to cut off the singular part of the integrand at a distance $r_s=Q_s^{-1}$ by defining the regularized function 
\be
\label{regularization1}
f^{\rm reg \, 1}(t',r,z) \equiv \Theta(r_s-r)  f(t',r_s,z) + \Theta(r-r_s) f(t',r,z) .
\ee
The transport coefficient is computed as $\int_0^t dt' f^{\rm reg \, 1}(t',v_\perp t',v_\parallel t')$. This method of regularization was used in all results outside of this section. 

The second regularization method is to subtract the leading order $\mathcal{O}(1/r)$ divergences before multiplying by the step function. We can represent this by defining $\tilde f (t',r,z)= f(t',r,z) -a/r$ with $a=\lim_{r\to 0} rf(t',r,z)$ and writing
\be
\label{regularization2}
f^{\rm reg \, 2}(t',r,z) \equiv \Theta(r_s-r)  \tilde f(t',r_s,z) + \Theta(r-r_s) f(t',r,z) .
\ee 
The transport coefficient is given as $\int_0^t dt' f^{\rm reg \, 2}(t',v_\perp t',v_\parallel t')$. 

The results of the two different regularization methods are depicted in Fig.~\ref{fig-qhat-reg} where the fifth order results are shown. The figure shows that the dependence on the regularization is fairly weak.

\subsubsection{$\hat{q}$ in collisions of finite nuclei}
\label{sec-qhat-finite-nuclei}

\begin{figure}[t]
\begin{center}
\includegraphics[width=8cm]{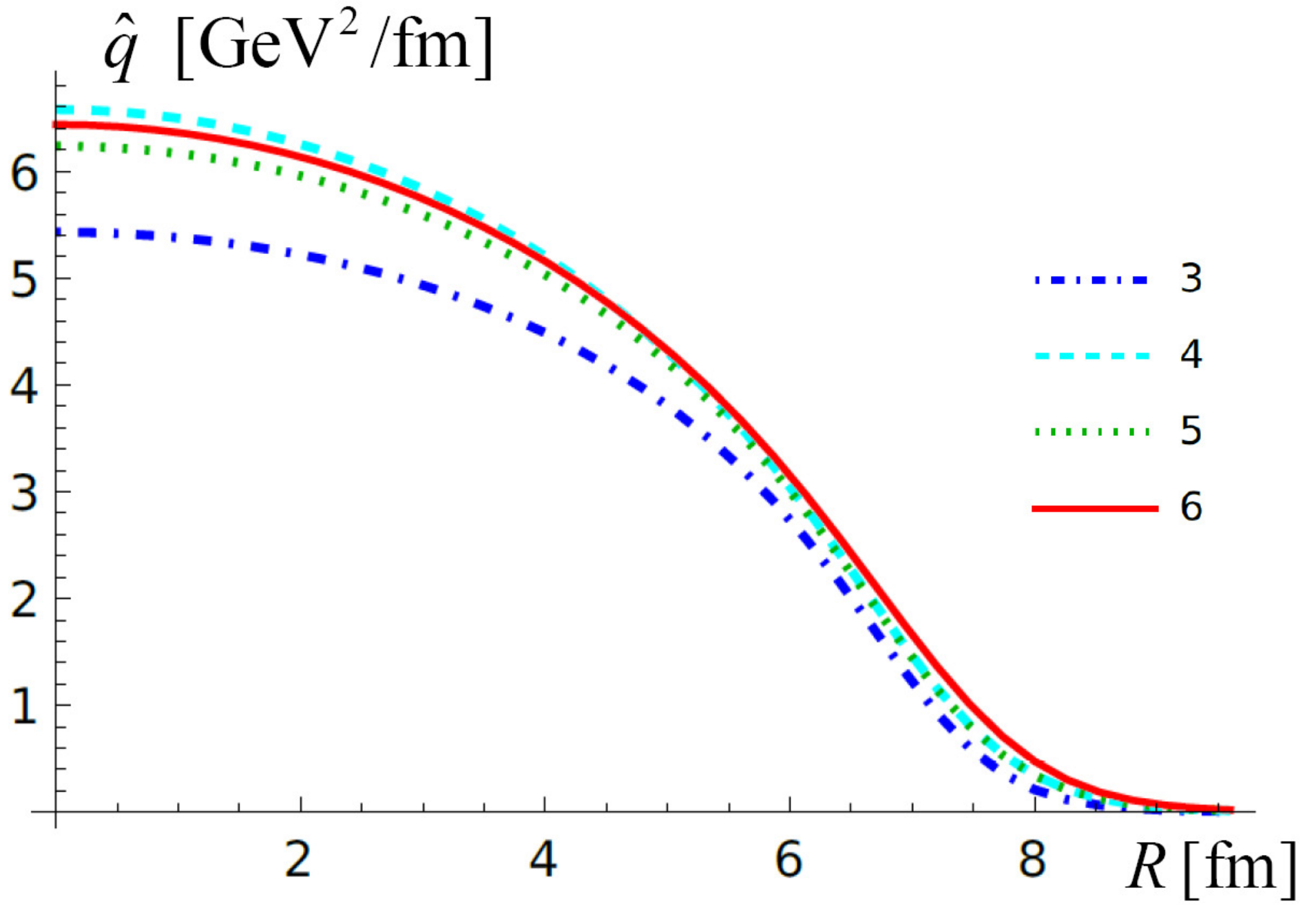}
\end{center}
\vspace{-7mm}
\caption{$\hat q$ versus $R$ for central collisions at $\tau=0.04$ fm and different orders of the proper time expansion from 3 to 6. }
\label{fig-qhat-R}
\end{figure}

So far we have considered only the transport coefficients $\hat q$ and $dE/dx$ for the case of a homogeneous glasma, which means that the incoming nuclei are assumed to be infinitely extended and homogeneous in the plane transverse to the beam direction. A realistic modeling of jet quenching in relativistic heavy-ion collisions requires treating nuclei as finite objects of varying density. We have generalized our previous calculations of $\hat q$ and $dE/dx$ so that the glasma under consideration is produced in collisions of finite nuclei with a Woods-Saxon density distribution. The field correlators are computed using a first order gradient expansion. The correlator $\langle \beta_n^i(\vec x_\perp) \, \beta_n^j(\vec y_\perp) \rangle$ is expanded around $\vec R \pm \vec b/2$ where $\vec R = \frac{1}{2}(\vec x_\perp + \vec y_\perp)$ and only the first two terms of the expansion are included. We note that the correlator $\langle \beta_n^i(\vec x_\perp) \, \beta_n^j(\vec y_\perp) \rangle$  is independent of $\vec R$ if the system is  translationally invariant in the transverse plane.
 
Figure~\ref{fig-qhat-R} shows $\hat q$ versus $R$ for central collisions ($b=0$) at $\tau=0.06$ fm at different orders of the proper time expansion up to sixth order. The calculation is done for a lead nucleus with radius $R_A= 7.4$ fm. In the outer part of the system the charge density drops rapidly to zero and the gradient expansion is not reliable. In the region $0<R\lesssim 4$ fm, which covers most of the glasma's volume, $\hat q$ depends on $R$ only weakly.  This means that the assumption of translation invariance in the transverse plane is valid to good accuracy in this domain.

\subsubsection{Gauge dependence}
\label{sec-gauge-dependence}

In this section we consider the issue of the gauge dependence of our results. As discussed in Sec. \ref{FP-eq} the tensor defined in Eq. (\ref{X-def}) is not gauge invariant due to the absence of the link operator (\ref{link-def}) in the field correlators. We can obtain an estimate of the size of the effect of this approximation by computing the statistical ensemble average of the operator in Eq. (\ref{link-def}). The crucial observation is that the magnitude of the tensor (\ref{X-def}) is saturated after a time of approximately 0.06 fm. Consequently, the time interval covered by the link operator is very short and thus the operator can be approximated by the first three terms in the expansion of the exponential function in (\ref{link-def}). Since the ensemble average of a single potential vanishes, we obtain
\be
\label{ave-Omega-approx}
\frac{1}{N_c^2-1} \,\big\langle \Omega (t,\vec{x}|t-t', \vec{x}-\vec{v} t')\big\rangle
 = 1 - \frac{g^2 N_c}{2(N_c^2-1)}
\big\langle A_a^\mu(x)  \, A_a^\nu(x) \big\rangle \Delta s_\mu \Delta s_\nu .
\ee

We compute the expression (\ref{ave-Omega-approx}) in the zeroth order of the proper time expansion. Using the correlator (\ref{corr-res-basic}) with $\Lambda=Q_s$ and assuming that the velocity $\vec{v}$ of the hard probe is perpendicular to the collision axis, one obtains
\ba
\nn
&& \frac{1}{N_c^2-1} \,\big\langle \Omega (t,\vec{x}|t-t', \vec{x}- \vec{v} t')\big\rangle 
\\[2mm] \label{ave-Omega-approx-final}
&&~~~~~~~~~~~~
= 1 -  \frac{N_c Q_s^2}{8\pi} 
\bigg[ \ln\Big(\frac{Q_s^2}{m^2} +1 \Big) -  \frac{Q_s^2}{Q_s^2 + m^2} \bigg]  \vec{v}^2 t'^2 .
\ea
With $N_c = 3$, $Q_s=2$~GeV, $m=200$~MeV, $\vec{v}^2 = 1$ and $t' = 0.06$~fm, we find 
\be
\label{ave-link-final}
1 - \frac{1}{N_c^2-1} \,\big\langle \Omega (t,\vec{x}|t-t', \vec{x}-\vec{v} t')\big\rangle = 0.16 .
\ee
This estimate most likely provides an upper limit, for two reasons. First, we have computed the potential correlator at the zeroth order of the proper-time expansion which corresponds to the strongest fields. Second, the correlator (\ref{ave-Omega-approx}) has been taken at the minimal distance where its size is maximal.

Equation (\ref{ave-link-final}) shows that the ensemble average of the link operator per colour degree of freedom differs from unity by 0.16, which is not a big number. This result supports the idea that neglecting the link operator in the collision term of the Fokker-Planck equation does not invalidate our results.

\subsection{Glasma impact on jet quenching}
\label{sec-res-add}

We have found that in the glasma phase the momentum broadening parameter $\hat q$ can be as large as $\hat{q} \approx 6 ~ {\rm GeV^2/fm}$. The value of $\hat{q}$ in equilibrium quark-gluon plasma for a hard quark of $p_T > 40$ GeV is $2 < \hat{q}/T^3 < 4$ where $T$ is the plasma temperature, as inferred from experimental data by the JETSCAPE Collaboration \cite{JETSCAPE:2021ehl}. In the discussion below we take $\hat{q} = 3T^3$. Since the temperature of the plasma produced at the LHC evolves from roughly 450 to 150 MeV \cite{Shen:2011eg}, the momentum broadening coefficient varies from $\hat{q} \approx 1.0~{\rm GeV^2/fm}$ to $\hat{q} \approx 0.05~{\rm GeV^2/fm}$, which is much smaller than the value $\hat{q} \approx 6 ~ {\rm GeV^2/fm}$ for the glasma that we have obtained. However, since the pre-equilibrium phase exists for less than 1 fm, it is not clear if the glasma contributes significantly to the total momentum broadening that the probe experiences when it moves through the system.

The radiative energy loss per unit length of a high-energy parton traversing a medium of length $L$ is proportional to the total accumulated transverse momentum broadening, denoted  $\Delta p_T^2$.
In the case of a static medium, where $\hat q$ is constant, we have $\Delta p_T^2 = \hat{q}L$. When the plasma is not static and $\hat{q}$ is time dependent, the transverse momentum broadening is 
\be
\label{total-broadening}
\Delta p_T^2 = \int_0^L dt \, \hat{q}(t),
\ee
where the probe is assumed to move with the speed of light.  

\begin{figure}
\centering
\includegraphics[width=8cm]{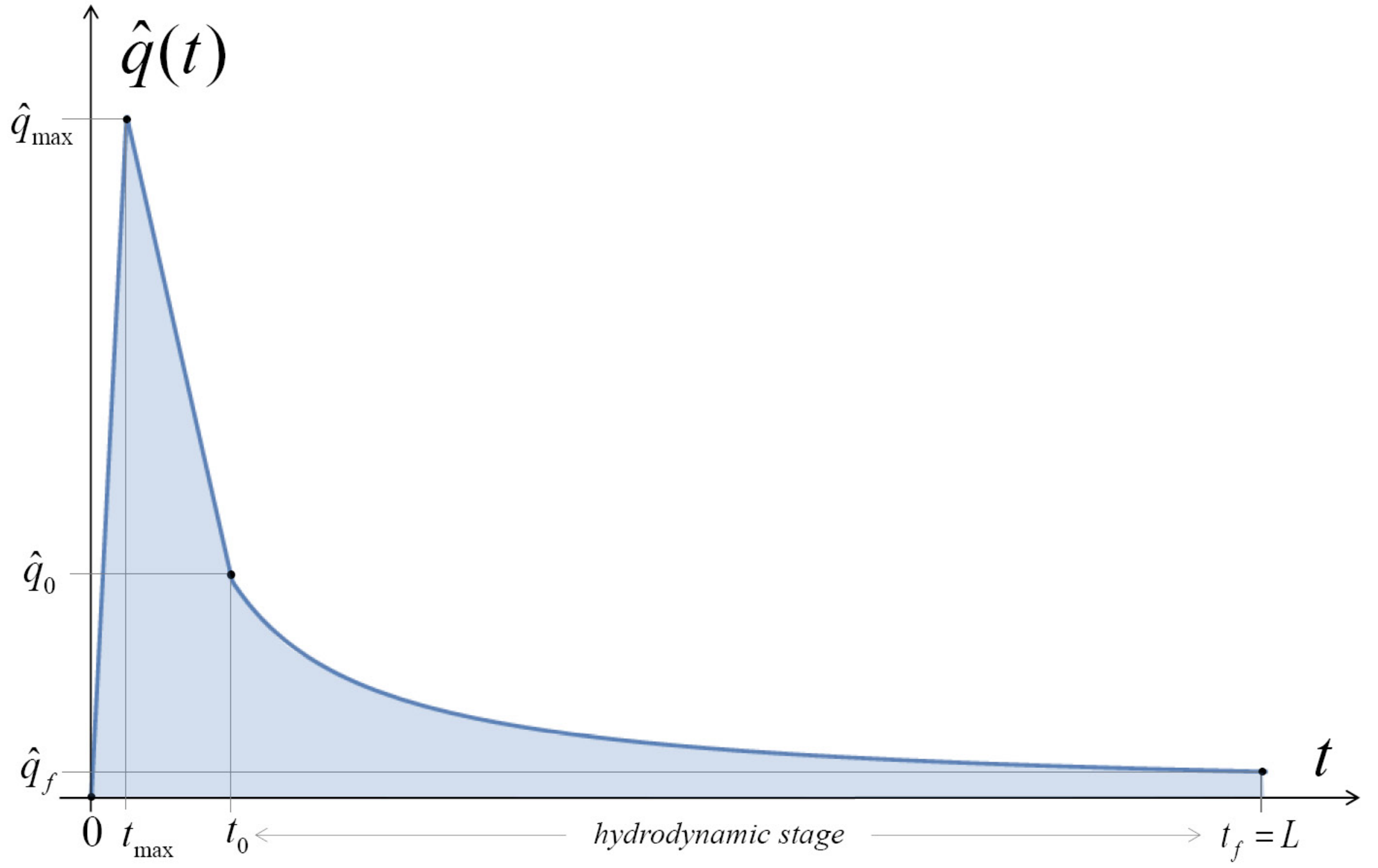}
\caption{Schematic representation of the temporal evolution of $\hat{q}(t)$. } 
\label{Fig-qhat-time}
\end{figure}

Figure~\ref{Fig-qhat-time} is a schematic representation of the time dependence of the momentum broadening coefficient throughout the whole history of the probe's journey across the deconfined matter produced in a relativistic heavy-ion collision. The first part of the figure shows the rapid growth of $\hat{q}(t)$ to a maximal value $\hat{q}_{\rm max} \approx 6~{\rm GeV^2/fm}$ at $t_{\rm max} \approx 0.06$ fm. This is a rough description of the evolution of $\hat q$ in the glasma phase from our calculation. The value of $\hat q(t)$ subsequently decreases at later times. At $t_0 \approx 0.6$ fm it has the value $\hat{q}_0 \approx 1.4$~GeV$^2$/fm (these numbers are estimates inferred from experimental data and are discussed in more detail below). We comment that the saturation region observed clearly in Fig.~\ref{fig-qhat-time} is not seen in Fig.~\ref{Fig-qhat-time} because different time scales are used in the two figures. The time interval between $t_{\rm max}$ and $t_0$ is beyond the region of validity of the proper time expansion and the rapid decrease of $\hat q$ in this domain is not captured by our calculation, but it is reproduced by the simulations in Ref.~\cite{Ipp:2020nfu}. Using linear interpolation between the points $\hat{q}(0)=0$, $\hat{q}(t_{\rm max})=\hat{q}_{\rm max}$, and $\hat{q}(t_0)=\hat{q}_0$, one finds the following non-equilibrium contribution to the accumulated transverse momentum broadening
\be
\label{non-eq}
\Delta p_T^2\Big|^{\rm non-eq} =  \int_0^{t_0} dt \, \hat{q}(t) 
= \frac{1}{2}\hat{q}_{\rm max} t_0 +  \frac{1}{2}\hat{q}_0(t_0 - t_{\rm max}) .
\ee

At $t>t_0$ we have equilibrated quark-gluon plasma which expands hydrodynamically. 
Using  ideal one-dimensional boost invariant hydrodynamics the temperature decreases as
\be
T = T_0 \Big(\frac{t_0}{t}\Big)^{1/3}. 
\ee
Consequently, the momentum broadening coefficient depends on time as
\be
\hat{q}(t)  = 3 T_0^3 \, \frac{t_0}{t} = \hat{q}_0 \frac{t_0}{t}
\ee
and the equilibrium contribution to $\Delta p_T^2$ is
\be
\label{eq}
\Delta p_T^2\Big|^{\rm eq} =  \int_{t_0}^L dt \, \hat{q}(t) 
= 3 T_0^3 \, t_0 \, \ln \frac{L}{t_0} .
\ee

To estimate the role of the glasma in jet quenching we need, in addition to $\hat{q}_{\rm max}$ and $t_{\rm max}$ which come from our calculation, the following parameters: $T_0$, $t_0$, $\hat q_0$ and $L$. The time $t_0$, which marks the beginning of the hydrodynamic evolution, and the initial temperature $T_0$, which determines the system's initial energy density, are obtained by comparing hydrodynamic models with experimental data on particle spectra and collective flows. 
The initial time cannot be too small as the system should reach, at least approximately, local thermodynamic equilibrium 
for a hydrodynamic approach to be applicable. On the other hand the initial time cannot be too big, because in that case the initial shape of the system would be washed out and hydrodynamics would not be able to reproduce the Fourier coefficients of the collective flow. We use $T_0 = 0.45$ GeV and $t_0 = 0.6$~fm taken from \cite{Shen:2011eg} and \cite{JETSCAPE:2021ehl}, respectively. The momentum broadening parameter is inferred from experimental data on jet quenching through complex modelling of the process of hard probe propagation through the evolving plasma. Using again the results of the JETSCAPE Collaboration \cite{JETSCAPE:2021ehl}, we take $\hat q_0 \approx 3 T_0^3 \approx 1.4$ GeV$^2$/fm. Finally, keeping in mind that the radius of a heavy nucleus like Au or Pb is about 7 fm, we assume that the typical path length of a hard probe in the quark-gluon plasma is $L = 10$ fm. The length scale $L$ is chosen to be slightly bigger than a typical nuclear radius, because the effect of jet quenching is particularly evident when the point of the jet production is close to the system's surface. In this case, one jet easily escapes into vacuum while the jet going in the opposite direction propagates through the plasma and, in central collisions, its path can be as long as the diameter of the nucleus. Substituting these values into Eqs.~(\ref{non-eq}) and (\ref{eq}), we find
\be
\label{ratio}
\frac{\Delta p_T^2\Big|^{\rm non-eq}}{\Delta p_T^2\Big|^{\rm eq}} = 0.93.
\ee
We note that this result is not very sensitive to the  parameters $T_0$, $t_0$, $\hat q_0$ and $L$, or the precise shape of the peak in Fig. \ref{Fig-qhat-time}. Equation (\ref{ratio}) shows that the non-equilibrium phase gives a contribution to the radiative energy loss which is comparable to that of the equilibrium phase. The conclusion is that the glasma plays an important role in jet quenching, which contradicts the commonly made assumption that the contribution of the glasma phase to momentum broadening is negligible.

\section{Summary, conclusions and outlook}
\label{sec-conclusions}

We have used a CGC approach and an expansion in proper time to study various characteristics of glasma from the earliest phase of relativistic heavy-ion collisions. We have considered two classes of glasma characteristics: those which are encoded in the glasma energy-momentum tensor and those which determine glasma transport properties. The first class includes: the energy density, transverse and longitudinal pressures, Poynting vector and angular momentum. The transverse momentum broadening of a hard probe and its collisional energy loss belong to the second class. Our results, which are obtained up to sixth or eighth order in the $\tau$ expansion, are reliable to about $\tau=0.06 - 0.07$ fm. The results have complicated structure but are analytic and thus free of the numerical artifacts that are an inherent part of the simulations which are usually used in studies of glasma. 

Our general conclusion is that the glasma that exists during the earliest phase of relativistic heavy-ion collisions plays an important role in the temporal evolution of the system produced in the collisions. This is important because phenomena observed in the final state have their origins in the glasma. The most important specific findings of our study are the following.

\begin{itemize}

\item The glasma energy density and transverse pressure decrease with time while the longitudinal pressure increases. Consequently, the glasma anisotropy decreases. This behaviour marks the beginning of glasma evolution towards thermodynamic equilibrium. 

\item Initially the Poynting vector is zero but the collective flow, both radial and azimuthally asymmetric, develops very rapidly, and consequently should contribute to the observed collective flow of final state particles. 

\item A growth of glasma elliptical flow is associated with a decrease of the spatial eccentricity of the system which strongly resembles hydrodynamic behaviour. This result might partially explain why a hydrodynamic description works well even at very early times when the system is far from equilibrium. 

\item Only a small fraction of the huge angular momentum carried by the valence quarks is transmitted to the glasma. This contradicts the picture of a rapidly rotating glasma but agrees with measurements of global polarization of lambdas, which is zero or almost zero at RHIC and LHC energies \cite{Adam:2018ivw,Acharya:2019ryw}. 

\item The momentum broadening coefficient $\hat{q}$ and collisional energy loss $dE/dx$ in the glasma are significantly bigger than those in the hydrodynamic phase of matter produced in heavy-ion collisions. This indicates that the short lived glasma state plays an important role in  jet quenching and should not be neglected. 

\end{itemize}

We computed a variety of glasma characteristics, but the approach we developed can be used to study other properties of  glasma. 

\begin{itemize}

\item We argued that the temporal evolution of glasma mimics hydrodynamic behaviour. It would be interesting to explore this further with a study of other glasma characteristics. 

\item We mostly considered collisions of heavy nuclei for which our approach is most reliable. There is a lot of interest in asymmetric collisions, and collisions involving smaller systems which are known to reveal some similarities with those of heavy nuclei, in particular concerning collective phenomena. The approach we developed can be used to study the dependence of different glasma properties on the sizes of the colliding nuclei.

\item We found that the global angular momentum of the glasma system transverse to the reaction plane is very small, which agrees with the small or vanishing global polarization of lambdas measured at RHIC and LHC \cite{Adam:2018ivw,Acharya:2019ryw}. However, there are interesting new experimental results on the non-vanishing polarization of lambdas along the beam direction \cite{STAR:2019erd,STAR:2019erd}. The glasma angular momentum in the beam direction is therefore of physical interest. This calculation is currently in progress.

\item We computed the two-point field correlators which determine the momentum broadening and collisional energy loss of a hard probe traversing the glasma. Using the same method one can derive the complete collision terms of the Fokker-Planck equation. A numerical solution of the equation can then be obtained. 

\end{itemize}

Several aspects of our approach to study glasma dynamics could be improved. Below we list a few examples starting with the most ambitious. 

\begin{itemize}

\item An important simplification of the CGC approach we used is the assumption of boost invariance. It would very desirable to go beyond this simplification and include some of the effects of the longitudinal dynamics. However, this is a difficult task that requires a modification of the ansatz (\ref{ansatz}) that determines the structure of the gauge potential. 

\item We used the Glasma Graph Approximation to perfom averaging over the colour configurations of the incoming nuclei. We found a way to estimate the validity of the approximation, as discussed in Sec.~\ref{sec-glasma-graph}. However, a more accurate assessment would involve testing the stability of our results when the approximation is relaxed, which could be done using an approach like that in Refs.~\cite{Fukushima:2007dy,FillionGourdeau:2008ij,Albacete:2018bbv}. 

\item The Fokker-Planck equation we used violates  gauge invariance because link operators of the form shown in equation (\ref{link-def}) are not included in the collision terms. In Sec.~\ref{sec-gauge-dependence} we argued that numerically the effect is not large because the expectation value of the link operator is not much different from unity. The link operators can be included using the machinery that we have developed to implement the proper time expansion, at least at lower orders. 

\item Our description of colliding nuclei breaks down at their edges because we use a gradient expansion to calculate correlation functions. Our calculation of the energy-momentum tensor includes terms to second order in the gradient expansion, and the calculation of the momentum broadening parameter includes zeroth and first order terms only. Many of the quantities we calculated are largely insensitive to higher order corrections, but quantities that characterize transverse anisotropy and glasma angular momentum depend much more strongly on the gradient expansion. The inclusion of higher order terms in the gradient expansion is a straightforward task that will  extend the region of validity of our results. 

\end{itemize}

\section*{Acknowledgments}

This work was partially supported by the National Science Centre, Poland under grant 2018/29/B/ST2/00646, and by the Natural Sciences and Engineering Research Council of Canada under grant SAPIN-2017-00028. 

\appendix

\section{Notation}
\label{AppendixA}

\vspace{-5mm}

\subsubsection*{Indices}

Greek letters from the second half of the alphabet $\mu, \nu, \rho, \dots$ label components of four-vectors and letters from the first half of the alphabet $\alpha, \beta, \gamma, \dots$ label components of spatial three-vectors denoted as $\vec{x}$. Components $x, y$ of three vectors which are transverse to the collision axis $z$ are indexed with Latin letters $i,j,k, \dots$. These transverse two vectors are denoted $\vec x_\perp$. The Latin letters  from the beginning of the alphabet $a,b, c \dots$ label colour components of elements of the SU($N_c$) gauge group in the adjoint representation.

\subsubsection*{Coordinates}

We use Minkowski, light-cone and Milne coordinates in different parts of the calculation, and these  coordinates are written $(t,z,\vec x_\perp)$, $(x^+,x^-,\vec x_\perp)$ and $(\tau,\eta,\vec x_\perp)$.  We use the conventional definitions
\ba
&& x^\pm \equiv\frac{t \pm z}{\sqrt{2}} ,
\\
&& \tau \equiv \sqrt{t^2-z^2}=\sqrt{2x^+ x^-} ,
\\
&& \eta \equiv \frac{1}{2} \ln \Big( \frac{x^+}{x^-} \Big) .
\ea

We define the relative and average transverse coordinates
\be
\label{rR-def}
\vec r \equiv \vec x_\perp-\vec y_\perp 
\text{~~~~and~~~~}
\vec R \equiv \frac{1}{2}\left(\vec y_\perp+\vec x_\perp\right) .
\ee
We will write unit vectors as $\hat r \equiv \vec r/|\vec r|=\vec r/r$ and $\hat R \equiv \vec R/|\vec R|=\vec R/R$ and use standard notation for derivatives like
\bea
\partial^i_{x} \equiv   - \frac{\partial}{\partial x_\perp^i} 
\text{~~~~and~~~~}
\partial^i_{R} \equiv  - \frac{\partial}{\partial R^i}\,.
\eea
In light-cone coordinates we have 
\be
\partial^{+} = \frac{\partial}{\partial x^-} 
\text{~~~~and~~~~} 
\partial^{-} = \frac{\partial}{\partial x^+}.
\ee
We note that the chain rule gives
\be
\label{chain}
-\partial^i_{x}  = \frac{\partial}{\partial r^i} +\frac{1}{2} \frac{\partial}{\partial R^i} 
\text{~~~~and~~~~}
-\partial^i_{y} = -\frac{\partial}{\partial r^i} +\frac{1}{2} \frac{\partial}{\partial R^i} .
\ee

The metric tensors in these three coordinate systems are 
$$
g_{\rm mink} = (1,-1,-1,-1)_{\rm diag}
$$ 
and 
\ba
\label{metric}
g_{\rm lc} = \left(
\begin{array}{cccc}
 0 & 1 & 0 & 0 \\
 1 & 0 & 0 & 0 \\
 0 & 0 & -1 & 0 \\
 0 & 0 & 0 & -1 \\
\end{array}
\right) \! ,
~~~
g_{\rm milne} = \left(
\begin{array}{cccc}
 1 & 0 & 0 & 0 \\
 0 & -\tau^2 & 0 & 0 \\
 0 & 0 & -1 & 0 \\
 0 & 0 & 0 & -1 \\
\end{array}
\right) \! .
\ea
The coordinate transformations are
$x^\mu_{\rm mink} = M^\mu_{~\nu} \, x^\nu_{\rm lc}$
with 
\ba
\label{lc2mink}
M^\mu_{~\nu} = \frac{dx^\mu_{\rm mink}}{dx^\nu_{\rm lc}} = 
\left(
\begin{array}{cccc}
 \frac{1}{\sqrt{2}} & \frac{1}{\sqrt{2}} & 0 & 0 \\
 \frac{1}{\sqrt{2}} & -\frac{1}{\sqrt{2}} & 0 & 0 \\
 0 & 0 & 1 & 0 \\
 0 & 0 & 0 & 1 \\
\end{array}
\right) ,
\ea
and $x^\mu_{\rm mink} = M^\mu_{~\nu} \, x^\nu_{\rm milne}$ with
\ba
\label{milne2mink}
M^\mu_{~\nu} = \frac{dx^\mu_{\rm mink}}{dx^\nu_{\rm milne}} = 
\left(
\begin{array}{cccc}
 \cosh\eta & \tau  \sinh\eta & 0 & 0 \\
 \sinh\eta & \tau  \cosh\eta & 0 & 0 \\
 0 & 0 & 1 & 0 \\
 0 & 0 & 0 & 1 \\
\end{array}
\right) .
\eea

We define a four-dimensional gradient operator where the transverse components are derivatives with respect to the average coordinate $\vec R$ defined in equation (\ref{rR-def}). We can transform this gradient operator from Milne to Minkowski coordinates by taking the inverse of the transpose of (\ref{milne2mink}). This gives
\ba
\label{gradient}
\partial_\mu ^{\rm mink} = \left(
\begin{array}{c}
\cosh\eta \,\frac{\partial}{\partial\tau} - \frac{\sinh\eta}{\tau} \,\frac{\partial}{\partial\eta} 
\\
-\sinh\eta  \,\frac{\partial}{\partial\tau} + \frac{\cosh\eta}{\tau} \,\frac{\partial}{\partial\eta} 
\\
-\partial_R^1
\\
-\partial_R^2
\end{array}
\right).
\ea

\subsubsection*{Gauge group}

The generators $t_a$ of the fundamental representation of SU$(N_c)$ group, where $a = 1,2, \dots N_c^2-1$, satisfy
\bea
\label{gen-def}
&& [t_a,t_b] = i f_{abc} t_c , \nn
\\
&& {\rm Tr} (t_a t_b) = \frac{1}{2}\delta_{ab} ,
\\
&& f_{abc} = -2i {\rm Tr}\big(t_a[t_b,t_c]\big) . \nn
\ea

Functions like $A_\mu$, $J_\mu$ and $\rho$ are SU$(N_c)$ valued and can be written as linear combinations of the SU$(N_c)$ generators. In the adjoint representation we write the generators with a tilde as $(\tilde t_a)_{bc} = -i f_{abc}$. 

\subsubsection*{Covariant derivatives}

In Minkowski coordinates the covariant derivative is defined as $D_\mu \equiv \partial_\mu - i  g A_\mu$ in the fundamental representation of SU($N_c$) gauge group. In the adjoint representation this becomes $D_{\mu\, ab} \equiv \delta_{ab}\partial_\mu - g f_{abc}A_{\mu\,c}$. Covariant derivatives in Milne coordinates include both the gauge potential contribution and connection terms. Products of  covariant derivatives acting on a scalar function $\phi$ are written 
\ba
\nn
&& \nabla_\nu \, \phi =D_\nu \, \phi ,
\\ \label{nabla}
&& \nabla_\mu \nabla_\nu \,\phi 
= (D_\mu D_\nu  -\Gamma^\rho_{\mu\nu}D_\rho) \, \phi ,
 \\ \nn
&& \nabla_\rho \nabla_\mu \nabla_\nu \,\phi 
= (D_\rho \nabla_\mu \nabla_\nu  
- \Gamma^\tau_{\rho\mu} \nabla_\tau \nabla_\nu
- \Gamma^\tau_{\rho\nu}\nabla_\mu \nabla_\tau)\,\phi  \,.
\ea
The connection $\Gamma^\rho_{\mu\nu}$ can be calculated from the metric tensor (\ref{metric}) and one easily shows that the only non-zero components are
\be
\label{connect}
\Gamma^0_{11} = \tau
\text{~~~~and~~~~}
 \Gamma^1_{01} = \Gamma^1_{10} = \frac{1}{\tau}.
\ee

\section{Azimuthal distribution}
\label{sec-phi-distribution}

We define here the azimuthal distribution of the flow vector $T^{i0}(\vec x_\perp)$. The azimuthal angle is measured with respect to the  $x$-axis and is written
\be
\label{phi-x-perp}
\varphi(\vec x_\perp) = \tan^{-1}\Big(\frac{T^{0y}(\vec x_\perp)}{T^{0x}(\vec x_\perp)}\Big)  
= \cos^{-1}\bigg(\frac{T^{0x}(\vec x_\perp)}
{\sqrt{\big(T^{0x}(\vec x_\perp)\big)^2 + \big(T^{0y}(\vec x_\perp)\big)^2 }}\bigg) .
\ee
We define the distribution 
\be
\label{phi-distribution}
P(\phi) \equiv \frac{1}{\Omega}  \int d^2 \vec x_\perp \,\delta \big( \phi -\varphi(\vec x_\perp) \big) \,
W(\vec x_\perp) ,
\ee
where we have introduced the weighting function
\be
 W(\vec x_\perp)  \equiv \sqrt{\big(T^{0x}(\vec x_\perp)\big)^2 + \big(T^{0y}(\vec x_\perp)\big)^2} 
 \ee
and the normalization factor
 \be
 \Omega \equiv \int d^2 \vec x_\perp \, W(\vec x_\perp) .
\ee
The distribution $P(\phi)$ can be decomposed into Fourier harmonics as
\be
P(\phi) = \frac{1}{2\pi} \Big(1 +  2\sum_{n=1}^\infty v_n \cos(n \phi ) \Big) ,
\ee
where the coefficients $v_n$ are given by the relation
\be
\label{vn-1}
v_n = \int_0^{2\pi} d\phi \, \cos(n\phi) \, P(\phi)  . 
\ee



\begin{thebibliography}{}

\bibitem{McLerran:1993ni}
L.~D.~McLerran and R.~Venugopalan,
Phys. Rev. D \textbf{49}, 2233 (1994).

\bibitem{McLerran:1993ka}
L.~D.~McLerran and R.~Venugopalan,
Phys. Rev. D \textbf{49}, 3352 (1994).

\bibitem{McLerran:1994vd}
L.~D.~McLerran and R.~Venugopalan,
Phys. Rev. D \textbf{50}, 2225 (1994).

\bibitem{Gelis:2012ri}
F.~Gelis,
Int. J. Mod. Phys. A \textbf{28}, 1330001 (2013). 

\bibitem{JalilianMarian:1996xn}
J.~Jalilian-Marian, A.~Kovner, L.~D.~McLerran and H.~Weigert,
Phys. Rev. D \textbf{55}, 5414 (1997).

\bibitem{Kovchegov:1999yj}
Y.~V.~Kovchegov,
Phys. Rev. D \textbf{60}, 034008 (1999).

\bibitem{Sun:2019fud}
Y.~Sun, G.~Coci, S.~K.~Das, S.~Plumari, M.~Ruggieri and V.~Greco,
Phys. Lett. B \textbf{798}, 134933 (2019).

\bibitem{Boguslavski:2021buh}
K.~Boguslavski, A.~Kurkela, T.~Lappi and J.~Peuron,
JHEP \textbf{05}, 225 (2021).

\bibitem{Ipp:2021lwz}
A.~Ipp, D.~I.~M\"uller, S.~Schlichting and P.~Singh,
Phys. Rev. D \textbf{104}, 114040 (2021).

\bibitem{Avramescu:2023qvv}
D.~Avramescu, V.~B\u{a}ran, V.~Greco, A.~Ipp, D.~I.~M\"uller and M.~Ruggieri,
Phys. Rev. D \textbf{107}, 114021 (2023).

\bibitem{Matsuda:2023gle}
H.~Matsuda and X.~G.~Huang,
Phys. Rev. D \textbf{108}, 114008 (2023).

\bibitem{Fries:2005yc}
R.~J.~Fries, J.~I.~Kapusta and Y.~Li,
Nucl. Phys. A \textbf{774}, 861 (2006).

\bibitem{Fukushima:2007yk}
K.~Fukushima,
Phys. Rev. C \textbf{76}, 021902 (2007).

\bibitem{Fujii:2008km}
H.~Fujii, K.~Fukushima and Y.~Hidaka,
Phys. Rev. C \textbf{79}, 024909 (2009).

\bibitem{Chen:2015wia}
G.~Chen, R.~J.~Fries, J.~I.~Kapusta and Y.~Li,
Phys. Rev. C \textbf{92}, 064912 (2015).

\bibitem{Fries:2017ina}
R.~J.~Fries, G.~Chen and S.~Somanathan,
Phys. Rev. C \textbf{97}, 034903 (2018).

\bibitem{Li:2017iat}
M.~Li,
Phys. Rev. C \textbf{96}, 064904 (2017).

\bibitem{Carrington:2020ssh}
M.~E.~Carrington, A.~Czajka and St.~Mr\'owczy\'nski,
Eur. Phys. J. A \textbf{58}, 5 (2022).

\bibitem{Carrington:2021qvi}
M.~E.~Carrington, A.~Czajka and St.~Mr\'owczy\'nski,
Phys. Rev. C \textbf{106}, 034904 (2022).

\bibitem{Carrington:2020sww}
M.~E.~Carrington, A.~Czajka and St.~Mr\'owczy\'nski,
Nucl. Phys. A \textbf{1001}, 121914 (2020).

\bibitem{Carrington:2022bnv}
M.~E.~Carrington, A.~Czajka and St.~Mr\'owczy\'nski,
Phys. Rev. C \textbf{105}, 064910 (2022).

\bibitem{Carrington:2021dvw}
M.~E.~Carrington, A.~Czajka and St.~Mr\'owczy\'nski,
Phys. Lett. B \textbf{834}, 137464 (2022).

\bibitem{Carrington:2023nty}
M.~E.~Carrington, W.~N.~Cowie, B.~T.~Friesen, St.~Mr\'owczy\'nski and D.~Pickering,
Phys. Rev. C \textbf{108}, 054903 (2023).

\bibitem{Kovner:1995ts}
A.~Kovner, L.~D.~McLerran and H.~Weigert,
Phys. Rev. D \textbf{52}, 3809 (1995).

\bibitem{Kovner:1995ja}
A.~Kovner, L.~D.~McLerran and H.~Weigert,
Phys. Rev. D \textbf{52}, 6231 (1995).

\bibitem{Blaizot:2004wv}
J.~P.~Blaizot, F.~Gelis and R.~Venugopalan,
Nucl. Phys. A \textbf{743}, 57 (2004).

\bibitem{Fukushima:2007dy}
K.~Fukushima and Y.~Hidaka,
JHEP \textbf{06}, 040 (2007).

\bibitem{FillionGourdeau:2008ij}
F.~Fillion-Gourdeau and S.~Jeon,
Phys. Rev. C \textbf{79}, 025204 (2009).

\bibitem{Lappi:2017skr}
T.~Lappi and S.~Schlichting,
Phys. Rev. D \textbf{97}, 034034 (2018).

\bibitem{Albacete:2018bbv}
J.~L.~Albacete, P.~Guerrero-Rodr\'\i{}guez and C.~Marquet,
JHEP \textbf{01}, 073 (2019).

\bibitem{Lappi:2015vta}
T.~Lappi, B.~Schenke, S.~Schlichting and R.~Venugopalan,
JHEP \textbf{01}, 061 (2016). 

\bibitem{Iancu:2003xm} E.~Iancu and R.~Venugopalan, 
in {\it Quark–Gluon Plasma 3}, 
eds. R.C. Hwa and X.-N. Wang
(World-Scientific, Singapore, 2004), p.~249.

\bibitem{Dokshitzer:2001zm}
Y.~L.~Dokshitzer and D.~E.~Kharzeev,
Phys. Lett. B \textbf{519}, 199 (2001).

\bibitem{Dong:2019byy}
X.~Dong, Y.~J.~Lee and R.~Rapp,
Ann. Rev. Nucl. Part. Sci. \textbf{69}, 417 (2019).

\bibitem{Baier:1996sk}
R.~Baier, Y.~L.~Dokshitzer, A.~H.~Mueller, S.~Peigne and D.~Schiff,
Nucl.\ Phys.\  B {\bf 484}, 265 (1997).

\bibitem{Lappi:2007ku}
T.~Lappi,
Eur. Phys. J. C \textbf{55}, 285 (2008).

\bibitem{Lappi:2006hq}
T.~Lappi,
Phys. Lett. B \textbf{643}, 11 (2006). 

\bibitem{Mrowczynski:2017kso}
St.~Mr\'owczy\'nski,
Eur. Phys. J. A \textbf{54}, 43 (2018).

\bibitem{Jankowski:2020itt}
J.~Jankowski, S.~Kamata, M.~Martinez and M.~Spali\'nski,
Phys. Rev. D \textbf{104}, 074012 (2021).

\bibitem{Krasnitz:2002ng}
A.~Krasnitz, Y.~Nara and R.~Venugopalan,
Phys. Lett. B \textbf{554}, 21 (2003). 

\bibitem{Heinz:2013th}
U.~Heinz and R.~Snellings,
Ann. Rev. Nucl. Part. Sci. \textbf{63}, 123 (2013).

\bibitem{Gao:2007bc}
J.~H.~Gao, S.~W.~Chen, W.~T.~Deng, Z.~T.~Liang, Q.~Wang and X.~N.~Wang,
Phys. Rev. C \textbf{77}, 044902 (2008).

\bibitem{Becattini:2007sr}
F.~Becattini, F.~Piccinini and J.~Rizzo,
Phys. Rev. C \textbf{77}, 024906 (2008).

\bibitem{Adam:2018ivw}
J.~Adam \textit{et al.} [STAR],
Phys. Rev. C \textbf{98}, 014910 (2018).

\bibitem{Acharya:2019ryw}
S.~Acharya \textit{et al.} [ALICE],
Phys. Rev. C \textbf{101}, 044611 (2020).

\bibitem{Moore:2004tg} 
G.~D.~Moore and D.~Teaney,
Phys.\ Rev.\ C {\bf 71}, 064904 (2005).
   
\bibitem{Svetitsky:1987gq} 
B.~Svetitsky,
Phys.\ Rev.\ D {\bf 37}, 2484 (1988).
  
\bibitem{vanHees:2004gq} 
H.~van Hees and R.~Rapp,
Phys.\ Rev.\ C {\bf 71}, 034907 (2005).

\bibitem{Mustafa:2004dr} 
M.~G.~Mustafa,
Phys.\ Rev.\ C {\bf 72}, 014905 (2005).

\bibitem{Kampen:1987}
N.G.~van~Kampen,
{\it Stochastic processes in physics and chemistry}
(North-Holland, Amsterdam, 1987).

\bibitem{Ipp:2020nfu}
A.~Ipp, D.~I.~M\"uller and D.~Schuh,
Phys. Lett. B \textbf{810} (2020) 135810.

\bibitem{Boguslavski:2023alu}
K.~Boguslavski, A.~Kurkela, T.~Lappi, F.~Lindenbauer and J.~Peuron,
Phys. Lett. B \textbf{850}, 138525 (2024).

\bibitem{Boguslavski:2023waw}
K.~Boguslavski, A.~Kurkela, T.~Lappi, F.~Lindenbauer and J.~Peuron,
arXiv:2312.00447 [hep-ph].

\bibitem{JET:2013cls}
K.~M.~Burke \textit{et al.} [JET],
Phys. Rev. C \textbf{90}, 014909 (2014).

\bibitem{Andres:2016iys}
C.~Andr\'es, N.~Armesto, M.~Luzum, C.~A.~Salgado and P.~Zurita,
Eur. Phys. J. C \textbf{76}, 475 (2016).

\bibitem{JETSCAPE:2021ehl}
S.~Cao \textit{et al.} [JETSCAPE],
Phys. Rev. C \textbf{104}, 024905 (2021).

\bibitem{Shen:2011eg}
C.~Shen, U.~Heinz, P.~Huovinen and H.~Song,
Phys. Rev. C \textbf{84}, 044903 (2011).

\bibitem{STAR:2019erd}
J.~Adam \textit{et al.} [STAR],
Phys. Rev. Lett. \textbf{123}, 132301 (2019).

\bibitem{ALICE:2021pzu}
S.~Acharya \textit{et al.} [ALICE],
Phys. Rev. Lett. \textbf{128}, 172005 (2022).

\end{thebibliography}
\end{document}